\documentclass[fleqn,usenatbib]{mnras}
\usepackage{multirow}
\usepackage{newtxtext,newtxmath}
\usepackage[T1]{fontenc}
\DeclareRobustCommand{\VAN}[3]{#2}
\let\VANthebibliography\thebibliography
\def\thebibliography{\DeclareRobustCommand{\VAN}[3]{##3}\VANthebibliography}
\usepackage{graphicx}	
\usepackage{longtable}

\newcommand{\bz}{\ensuremath{\langle B_z \rangle}}
\newcommand{\bs}{\ensuremath{\langle \vert B \vert \rangle}}
\newcommand{\sbs}{\ensuremath{s_{\bs}}}
\newcommand{\vmax}{\ensuremath{\vert (V/I) \vert_{\rm max}}}
\newcommand{\abz}{\ensuremath{\vert \langle B_z \rangle \vert}}
\newcommand{\nnz}{\ensuremath{\langle N_z \rangle}}

\newcommand{\snr}{\ensuremath{S/N}}
\newcommand{\vsini}{\ensuremath{v\,\sin\,i}}
\newcommand{\teff}{\ensuremath{T_{\rm eff}}}

\newcommand{\kms}{km\,s\ensuremath{^{-1}}}
\newcommand{\fil}{\ensuremath{\phantom{.10}}}
\newcommand{\fili}{\ensuremath{\phantom{0}}}
\newcommand{\lv}{local 20\,pc volume}
\newcommand{\mg}{}
\newcommand{\ssz}[3]{&\multicolumn{1}{c}{\tiny \ensuremath{\sigma_z \simeq         {#1}}{#2} {({#3}$\times$)}}}
\newcommand{\saz}[2]{&\multicolumn{1}{c}{\tiny \ensuremath{\vert \langle B_z \rangle \vert \la {#1}}{#2}                      }}
\newcommand{\bas}[2]{&\multicolumn{1}{c}{\tiny \ensuremath{\bs \la {#1}}{#2}                      }}
\newcommand{\bbs}[1]{\bf {#1}}
\newcommand{\bbm}[1]{&\multicolumn{1}{c}{\tiny {\ensuremath{\bs \simeq {#1}}\,MG?}}}
\title[New insight into the magnetism of white dwarfs]{New insight into the magnetism of degenerate stars from the analysis of a volume limited sample of white dwarfs}
\author[S.\ Bagnulo \& J.D.\ Landstreet]{
S. Bagnulo$^{1}$ \&
J.D. Landstreet$^{1,2}$
\\
$^{1}$Armagh Observatory \& Planetarium, College Hill, Armagh BT61 9DG, UK\\
$^{2}$University of Western Ontario, London, Ontario N6A 3K7, Canada. \\
}

\date{Accepted June 10; Received June 3; in original form: April 7 2021.}

\pubyear{2021}

\begin{document}
\label{firstpage}
\pagerange{\pageref{firstpage}--\pageref{lastpage}}
\maketitle

\begin{abstract}
Many stars evolve into magnetic white dwarfs, and observations may help to understand when the magnetic field appears at the stellar surface, if and how it evolves during the cooling phase, and, above all, what are the mechanisms that generate it. After obtaining new spectropolarimetric observations and combining them with previous literature data, we have checked almost the entire population of about 152 white dwarfs within 20\,pc from the Sun for the presence of magnetic fields, with a sensitivity that ranges from better than 1\,kG for most of the stars of spectral class DA, to 1\,MG for some of the featureless white dwarfs. We find that 33 white dwarfs of the local 20\,pc volume are magnetic. Statistically, the data are consistent with the possibility that the frequency of the magnetic field occurrence is similar in stars of all spectral classes, except that in the local 20\,pc volume, either DQ stars are more frequently magnetic, or host much stronger fields than average. The distribution of the observed field strength ranges from 40\,kG to 300\,MG and is uniform per decade, in striking contrast to the field frequency distribution resulting from spectroscopic surveys. Remarkably, no fields weaker than 40\,kG are found. We confirm that magnetic fields are more frequent in white dwarfs with higher than average mass, especially in younger stars. We find a marked deficiency of magnetic white dwarfs younger than 0.5 Gyr, and we find that the frequency of the occurrence of the magnetic field is significantly higher in white dwarfs that have undergone the process of core crystallisation than in white dwarfs with fully liquid core. There is no obvious evidence of field strength decay with time. We discuss the implications of our findings in relation to some of the proposals that have been put forward to explain the origin and evolution of magnetic fields in degenerate stars, in particular those that predict the presence of a dynamo acting during the crystallisation phase. 
\end{abstract}



\section{Introduction}
That magnetic fields occur in the latest stages of stellar evolution has been known for more than 50\,years, since the discovery that pulsars are strongly magnetised neutron stars \citep{Wolt64,Paci67,Gold68} and the discovery that white dwarfs (WDs) host strong magnetic fields \citep{Kempetal70,AngLan71-Second,AngLan72}.

When present, the magnetic field is one of the main actors that determines the observable features of a WD, or of a binary system containing a WD, by altering the spectral lines; by changing the physical structure through suppression of convection or by Lorentz forces; and by affecting transport of angular momentum and mixing during accretion and mass-loss phases, and during evolutionary changes to internal structure. However, the origin of WD magnetism is not well understood. Schematically, at least three scenarios have been proposed in the literature. (1) According to the classical `fossil field' theory, the field is a descendant of a field that was present when the star was in a previous evolutionary stage,  perhaps during the red giant (RG) phase or on the main sequence (MS), and its strength was amplified due to flux conservation while the star contracted into a WD. The favoured MS candidates progenitors of the magnetic WDs (MWDs) are the magnetic Ap and Bp stars. (2) According to the `merging scenario', a WD that formed in the merger of a binary pair may become magnetic as a result of a dynamo generated during the merger \citep{Touetal08}; a variant of this scenario predicts that a magnetic field may also be formed during the accretion of rocky debris \citep{Faretal11,Brietal18Isolated}. (3) The field may also be produced by some other internal physical mechanism during the cooling of the WD itself \citep{ValFab99}, possibly a dynamo acting in an unstable WD liquid mantle on the top of a solid core \citep{Iseetal17}. The latter theory also provides an explanation for planetary magnetic fields.

To understand the origin of the magnetic fields it is useful to know if there are clear correlations between the incidence of magnetic fields and other characteristics of the WDs. For instance, we know that the strength of the magnetic fields in the chemically peculiar stars of the upper main decays rapidly with time as soon as the star has reached the main sequence \citep{Lanetal07,Lanetal08}. Does the strength of the magnetic field in WDs also decrease with cooling age? Is there a correlation between the magnetic field and the mass of a WD, or with the chemical composition of its atmosphere, or with binarity? 

Over the last five decades, a lot of effort have been dedicated to the search for and characterisation of MWDs, both as a side-product of large spectroscopic surveys \citep[e.g.][]{Lieetal03,Kepetal13,Holletal15}, and due to specifically dedicated polarimetric surveys \citep[e.g.][]{AngLan70a,Angetal81,SchSmi95,Putney97,Aznetal04,Joretal07,Kawetal07,Lanetal12,BagLan18}.
A number of proposals have been put forward regarding various kinds of trends, for example that magnetic WDs are more massive than non magnetic WDs \citep{Lieetal03};  that cooler WDs are more often magnetic than hotter WDs \citep{LieSio79,FabVal99}; that magnetic WDs are not found in close binary systems with a WD and a main sequence companion \citep{Lieetal05}; that magnetic fields occur more frequently in cool stars with metal lines, whether they are H-rich DAZ WDs \citep{Kawetal19} or He-rich DZ WDs \citep{Holletal15,BagLan19b,Kawetal21}; and that hot DQs, WDs with a surface composition dominated by carbon and oxygen, show an extremely high incidence of magnetic fields \citep{Dufetal13}. Some of these trends, however, have been challenged by the results of other groups; for instance, \citet{Kawetal07} could not confirm a trend of the incidence of magnetic fields with age, and \citet{LanBag20} showed that MWDs in binary WDs with a main sequence companion may be rare, but do exist. 

The reality is that obtaining firm observational conclusions about evolutionary trends of the magnetic field is strongly hampered by various observational biases. Spectropolarimetric surveys, sensitive to the mean longitudinal magnetic field, are photon-hungry by nature, and tend to favour observations of bright stars. Lower mass WDs have different luminosity and absolute magnitude than higher mass WDs of the same age, and cooler (hence older) WDs may be much less bright than hotter and younger WDs. Both high-mass and cool/old WDs might be underrepresented in magnitude-limited surveys \citep{Lieetal03}. Low-resolution spectroscopic surveys, which can go much deeper in magnitude and examine a much larger sample of stars than spectropolarimetric surveys, are sensitive only to magnetic fields stronger than roughly 1--2\,MG, leaving weak fields in WDs undetected, and may struggle to identify some WDs with field stronger than 100 MG, which splits spectral lines into numerous weak components with low contrast relative to the continuum. 

To obtain the clearest possible view of the occurrence of detectable magnetic fields in WDs, we need to look at a sample which has a relatively clear significance and well-defined theoretical selection criteria. The most obvious choice is a volume-limited sample. The sample of all WDs within 20\,pc from the Sun includes about 150 stars: it is small enough to allow us a careful analysis of each individual member, but large enough to provide some meaningful statistical results. Because the evolution of all single stars with initial masses below about 8\,$M_\odot$ ends with collapse to the white dwarf state, as does the evolution of many binary pairs, the current populations of white dwarfs in the \lv\ records a sample of the results of more than 90\% of all completed stellar nuclear evolution at this distance from the centre of the Milky Way. The magnetic fields found in the WDs of this sample provide a relatively well-defined record of the generation and preservation of such fields through the lifetime of the Milky Way. White dwarfs of different ages in the \lv\ provide evidence about the production rate and evolution of both WDs and of their fields as a function of time over the past 10\,Gyr. 

In this paper we therefore try to  characterise the frequency of occurrence of magnetic fields of various strengths among all WDs of the \lv, and to correlate this with other stellar features such as mass, chemical composition, and cooling age. This task requires first an accurate characterisation of the physical parameters of each member of the \lv, which is discussed in Sect.~\ref{Sect_Sample}. Next, we need to check each member of the \lv\ for a magnetic field. Existing observations from the literature are reviewed in Sect.~\ref{Sect_Techniques}. Because literature data did not provide sufficient information, we have carried out a spectropolarimetric survey aimed at acquiring a complete dataset of all WDs within the local 20\,pc volume. Some preliminary results of our survey have been presented by \citet{LanBag19b}, \citet{BagLan19b} and \citet{LanBag20}, mainly to announce the discovery of new MWDs. In this paper, we publish the remaining observations that we have carried out in the last three years, and which complement those previously published in the literature for the sample of all WDs within 20\,pc from the Sun. These new observations are presented in Sect.~\ref{Sect_New_Obs} (observing log and fields values are given in App.~\ref{Sect_Log}). In Sect.~\ref{Sect_Results} we describe the dataset that we have collected for the WDs of the \lv\  and in Sect.~\ref{Sect_Analysis} we perform our analysis, in particular we discuss how the frequency of occurrence and the strength of the magnetic field vary in WDs of different spectral types, effective temperatures, and ages. In Sect.~\ref{Sect_Discussion} we discuss whether the theories that have been put forward to explain the origin of magnetic fields in degenerate stars are consistent with the characteristics of the MWDs of the \lv. Our conclusions are presented in Sect.~\ref{Sect_Conclusions}. This paper contains also a second appendix (App.~\ref{Sect_Stars}) where we discuss the magnetic nature of all individual WDs of the \lv; for some of the WDS, we also discuss the estimates of the stellar parameters and binarity.

\section{Defining the 20\,pc WD sample}\label{Sect_Sample}
Our sample of WDs is taken from the list proposed by \citet{Holletal18}, mainly on the basis of the astrometry reported in the second data release of the Gaia space astrometry mission \citep{gaia16,gaia18}, but also including several nearby WDs missed in the Gaia survey for one reason or another. Practically we have considered all stars of Table~1 of \citet{Holletal18}, and all stars of their Table~4 except for WD\,0454$+$620 and WD\,1443$+$256: the former seems to be outside of the \lv, and the latter was found to be a distant high-velocity G star by \citet{Schetal18}, an identification confirmed by a very small Gaia parallax. The third Gaia data release became publicly available after we had performed all our observations and carried out most of the analysis; nevertheless we checked if the revised parallax values would alter the membership of our sample. We found for instance that WD\,0728$+$642, a star that we had incorporated in our sample even though Gaia DR2 had put it possibly just outside of the \lv\ ($\pi = 49.97 \pm 0.05$), is according to DR3 most likely within 20\,pc from the Sun ($\pi = 50.06 \pm 0.04$), hence full member of the \lv. None of the stars of our sample was found according to DR3 to be outside of the \lv. We also checked that none of the WDs of the catalogue by \citet{Genetal19} that had parallax $\pi \le 50.0$ in DR2 had $\pi > 50$ in DR3, except for  WD\,0728$+$642.

\subsection{Atmospheric parameters}
Most of the WDs of the \lv\ are reasonably well studied, and atmospheric composition, mass, \teff, gravity, and age have been calculated on the basis of detailed modelling of large data sets such as low resolution visible and/or UV spectra, and extensive photometry with a wide range of filters or calibrated spectrophotometry. We have selected most of physical parameter data from the work of \citet{Giaetal12,Limetal15,Subetal17,Bloetal19} and \citet{Couetal19}. When data from these sources were not available (mainly for newly discovered WDs), we have adopted the values tabulated by \citet{Holletal18} or by \citet{Genetal19}, which are derived from the Gaia three-band photometry and parallax of each WD, on the basis of calibrations provided by fitting a large number of well-studied WDs. However, these calibrations do not provide reliable composition choices for newly identified WDs in the volume, and provide no ages. For WDs without age estimate in the literature we have relied on the on-line cooling tables\footnote{\tt http://www.astro.umontreal.ca/\~{}bergeron/CoolingModels} of the Montreal group \citep{Dufetal17}, which are based on computations described by \citet{Treetal11}, using a two-dimensional logarithmic interpolation.

To obtain an idea of the general level of precision of the adopted physical parameters we have compared the results obtained for a number of individual WDs by several different groups. The scatter of such computations suggests that the uncertainties of the physical parameters are about 300\,K for \teff, 0.1 for $\log g$, $0.1\,M_\odot$ for the mass, and 0.3\,dex for the age (or somewhat better for the youngest WDs). 

We should also note that the parameters that we have adopted come from modelling based on the assumption that the magnetic field has no influence on the stellar atmosphere. In fact, a magnetic field of some kG can suppress convection in the outer layers of the MWD. This changes the stratification of the model, and introduces a further uncertainty into the the estimate of the stellar parameters. Radiation magneto-hydrodynamic simulations of the atmosphere of white dwarf stars are discussed in \citet{Treetal15}, and applied to observational data by \citet{Genetal18} to the WD 2105$-$820. For that weakly magnetic star it was found that the best pure radiative model that reproduces the optical spectrum has $\teff = 9982 \pm 170$\,K and $\log g = 8.22 \pm 0.08$, and is also capable to reproduce the star's FUV spectrum, while the best-fitting optical convective model has $\teff = 10389 \pm 153$\,K and  $\log g = 8.01 \pm 0.05$ and fails to reproduce the FUV spectrum. However, the zero-field atmospheric model adopted by \citet{Subetal17} leads to an estimate of $\teff = 9820 \pm 240$\,K and  $\log g = 8.29 \pm 0.04$, consistent with the best-parameters of the radiative model by \citet{Genetal18}. This suggests that the uncertainties introduced by the zero-field approximation are probably of the same order as of those deduced from the scatter seen in literature values. 

\subsection{Multiple systems}
Some of the WDs occur in binary or multiple systems. In this section we discuss these situations.

\subsubsection{Common proper motion systems}
About 30 of the WDs of the \lv\ are in common proper motion visual binary systems, usually with main sequence stars, with separations ranging from about $2\arcsec$ to several arc-minutes. In such systems, the nature of all components is almost always well established. In three cases (DA WD Sirius B, and DQ WDs Procyon B and WD\,0208$-$510), the components are so close as to prevent polarimetric observations of the WD (circular polarimetric measurements are possible only from ground-based facilities). However, HST spectroscopy is available for all three WDs. There are two double degenerate (DD) systems whose members are well separated and may be observed individually. In addition there are a dozen known or suspected unresolved double systems (uDD), all made up of two WDs (a WD plus a main sequence star pair would usually be obvious from the composite spectrum). In only a few of these systems is it clear what the types of WDs the two stars are, and for half of them it is not even really clear whether one or two stars are present. 

The common proper motion pairs do not present any real difficulty for our analysis. Within the 20\,pc volume, all the visual binary pairs are separated by at least about $2{\arcsec}$, which at a distance of the order of 10\,pc corresponds to a separation in the plane of the sky of the order of 20\,AU. Therefore most of the minimum separations are hundreds or thousands of AU. With such a large separation, it is unlikely that a main sequence companion would have had any influence on the evolution of the WD predecessor through the red giant and AGB stages, although the main sequence companion might accrete some of the outflowing gas during mass loss phases preceding the formation of the WD. 

\subsubsection{Unresolved double degenerate systems}\label{Sect_Unresolved}
In contrast, all the unresolved double systems consist of two WDs, and therefore contribute not one but two WDs to the total tally of WDs within the \lv. 
Some of these uDD systems are identified simply because the deduced mass of the object ($\la 0.45 M_\odot$), treated as a single WD, is too small to have formed by single star evolution. However, in some cases, previous works provide conflicting results, in that the parameter determination by one group may result in small deduced mass, while another group finds a mass large enough to be the product of single star evolution \citep[see for example Sec.\,4.2 of][]{Holletal18}; therefore, before accepting the system as uDD, we look for some further observational hint of a secondary, such as discordance between computed and observed H line profiles.

For the systems that we have accepted as uDD, we have also used the available data to make plausible proposals (often just guesses) about the characteristics of the two stars. As these systems largely have photometry in several bands, and the single star models of \citet{Giaetal12} are generally successful in accounting for the energy distributions, we have looked for models in which both stars have similar \teff\ values. We have also started by looking for model pairs with similar radii (i.e. we have looked for models in which the two components make similar contributions to the total system light). Such models are of course rather uncertain, particularly as concerns the secondary in the system. The best studied cases are those of uDDs WD\,0135-052 and of WD\,0727+482 (the latter is marginally visually resolved), for which the orbits of the WD members are known;  the orbits provide the masses of each of the two components, and light ratios, and the other physical parameters of the stars may be reasonably estimated.

All our choices of which systems should be considered as most likely to be composed of two stars (labelled as ``uDD''), and which uDD systems are only suspected DD (labelled as ``uDD?''), are documented in App.~\ref{Sect_Stars}, where we have also described our modelling attempts. In our statistics, we count uDD systems as two WDs, but uDD? systems as single stars.

Finally, we note that for all established uDD systems it may not be clear how to apportion a magnetic measurement to one of the two stars. This issue is discussed in App.~\ref{Sect_Stars}.

A further complication of the uDD systems is that they are relatively close to each other. For example, at a distance of 15 pc, two WDs separated by less than 0.5\arcsec\ are probably only a few AU apart. This is a small enough separation that the evolution of each star is very likely to have affected that of the other star. These WDs may well have evolved rather differently than isolated single stars.

\subsection{The list of WDs of the \lv}
Our final list of \lv\ members is considered to be at least 95\,\% complete and includes 152 stars, six of them in uDD systems (hence we have 146 systems), and six suspected uDD? systems that we have decided to treat as single stars. 
One of the members, WD\,0211$-$340 = LP\,941-19, cannot be currently well characterised because of its current proximity to a background star. \citet{Holletal18} commented on its high proper motion and Gaia G and JHK magnitudes agree with a degenerate star of $\teff=5270$\,K and $\log g = 8.0$, but its spectral type (and magnetic nature) are unknown. The list is given in Table~\ref{Tab_Stars}, which will be fully described in Sect.~\ref{Sect_Results}. 

The mean mass of all WDs of the local 20\,pc volume is 0.64\,$M_\odot$ with a standard deviation of 0.13\,$M_\odot$, fully consistent with the mean mass of non magnetic WDs of $0.66 \pm 0.14\,M_\odot$ estimated by \citet{Treetal13}. The temperature distribution is highly skewed toward temperatures lower than 10\,000\,K. It appears that the production of WDs was higher in the last few Gyr -- stars younger than 3\,Gyr represent more than a half of the entire sample of WDs, which have age spanning the range between 0.03 and 9.3\,Gyr. Since cooler, hence older, WDs are dimmer than hotter and younger WDs, it is possible already to anticipate that a magnitude limited survey can hardly capture a statistically representative sample of the real population of WDs, as will be more evident in results presented in Sect.~\ref{Sect_Results} and in the analysis that will be carried out in Sect.~\ref{Sect_Analysis}. 

\section{Field detection techniques and measurements from previous literature}\label{Sect_Techniques}
We start with a general overview of the magnetic field measurements of the WDs within 20\,pc from the Sun. 
Magnetic WDs have been discovered with low and high resolution spectroscopy, with narrow- and broad-band polarimetry and with spectropolarimetry. To guide this overview, and to inform its subsequent analysis, it is useful first to recall the sensitivity threshold of the various methods used to detect magnetic fields.

\subsection{Field detection techniques}
\subsubsection{Spectroscopy}
Spectroscopy may reveal Zeeman splitting of spectral lines, which is sensitive to the so-called mean magnetic field modulus \bs, i.e., the surface field strength averaged over the visible stellar disk. Low-resolution spectroscopy of WDs may detect a magnetic field when its strength is at least 1--2\,MG, while high-resolution spectroscopy (say $R \ga 50\,000 $) may detect fields as weak as 50\,kG in DA stars with sharp Balmer line cores \citep[e.g.,][]{Koeetal98}. 

\subsubsection{Circular spectropolarimetry of spectral lines}
Circular spectropolarimetry is sensitive to the longitudinal component of the magnetic field averaged over the stellar disk, or mean longitudinal field \bz, and may probe fields of much lower strength than spectroscopy. The actual sensitivity of spectropolarimetric measurements varies with spectral type. In the brightest DA stars with deep H lines, \bz\ may be measured with an uncertainty of a few hundred Gauss \citep[see e.g.][]{Lanetal15,BagLan18}. Circular polarisation integrated over a spectral line is zero, therefore broadband polarimetry is not useful to detect magnetic fields in WDs, unless the field is sufficiently strong to polarise the continuum.

\subsubsection{Circular polarimetry of the continuum}
  In WDs with featureless spectra (DC spectral type), fields with a longitudinal field of at least $\sim 0.5$\,MG (in absolute value) may be detected via the measurement of circular polarisation in the continuum. Polarisation in the continuum may be measured by spectropolarimetry or by narrowband or broadband circular polarimetry. Since polarisation of the continuum (or of very broad spectral lines) may change its sign with wavelength, spectropolarimetry has the potential to reveal signals that might cancel out in broadband polarimetric measurements. However, spectropolarimetry is often more affected by instrumental effects than simple imaging polarimetry, and our experience with FORS2 and ISIS shows that only fields stronger than a few MG may be confidently detected by spectropolarimetry of the continuum in featureless stars \citep{BagLan20}. In practice, multi-colour narrowband circular polarimetry may be the most suitable detection method when the polarisation signal is $\la 0.3$\,\% in featureless spectra. We finally note that polarimetry of the continuum is the detection method of choice not only for featureless cool WDs, but also for hotter DA and DB WDs in the regime of a very strong field (say $\ga 100$\,MG), in which spectral lines may become very difficult to recognise, as well as for DQ stars, since we still do not know how intensity profiles of C$_2$ bands are modified by the presence of a very strong magnetic field.

\subsection{The Oblique Rotator Model and its implications for detection and characterisation of a magnetic field}
As the star rotates, the magnetic field, if not symmetric about the rotation axis, changes its configuration with respect to the observer. In general, \bs\ does not vary dramatically as the star rotates, and it is unlikely that a spectral line may split at certain rotation phases and not at others. Therefore, no repeated spectroscopic sampling is needed to assess whether a star possesses a magnetic field strong enough to split its spectral lines. In contrast, even a strong magnetic field may have a small average component along the line of sight, and in some cases may be detectable only at certain rotation phases. Ideally, two or three spectropolarimetric observations per star are needed, either to rule out the presence of a detectable field, or, in case of field detection, to confirm that a field exists, and check if it is variable. If the magnetic field is strong enough to polarise the continuum, it is unlikely that the entire $V/I$ spectrum is zero at all wavelengths at a given rotational phase (or at least we have never encountered a such a situation). Therefore we expect that even a single polarimetric measurement of the continuum may be used to assess the presence of a very strong magnetic field. Needless to say that the characterisation of strength and morphology of a stellar magnetic field requires good sampling of the rotational cycle, for instance, ten measurements taken about 0.1 cycles apart from each other.

\subsection{Measurements of magnetic fields of WDs from previous literature}\label{Sect_Old_Obs}
Magnetic WDs have been discovered with low and high resolution spectroscopy, with narrow- and broad-band polarimetry and with spectropolarimetry. Low resolution (classification) surveys have allowed the discovery of hundreds of MWDs in the 2--80\,\,MG regime \citep{Kepetal13}, while the numbers of MWDs discovered via high-resolution spectroscopy and polarimetry is an order of magnitude smaller. 
Observations of MWDs may be broadly divided into four groups.

\noindent
\textit{i)} Narrow band and broadband circular polarisation measurements made possible the discovery of the first MWD by \citet{Kemetal70}, and were systematically employed in small-scale surveys during the 1970s, leading to the discovery of a dozen MWDs \citep[][and references therein]{Angetal81}. In the local 20\, pc volume, these surveys have discovered six MWDs (one DC, two DQ and three DA WDs), and set useful upper limits to the magnetic fields of another 13 DC stars in which no circular polarisation was detected. These early surveys reached a degree of accuracy for $V/I$ often of the order of 0.05\,\% or better. This accuracy could not be improved by modern spectropolarimetry, where instrumental effects make it hard to reach an accuracy in the continuum better than $\sim 0.2$\,\% \citep[see Sect.~6 of][]{BagLan20}, although we must take into account that modern telescopes and detectors are able to reach much fainter objects. Aperture polarimetry is ideal to detect WDs with a magnetic field strong enough to polarise the continuum. However, we are not aware of any systematic broadband or narrow band circular polarisation surveys of WDs after the work of \citet{LieSto80} and \citet{Angetal81}. The early surveys have also targetted numerous DA WDs, but for them, much higher precision would be obtained later with spectropolarimetric techniques. 

\noindent
\textit{ii)} MWDs were discovered with spectroscopic techniques from the 1970s on, but mainly (because of the introduction of highly-efficient CCDs) since the late 1990s. Low-resolution spectroscopic surveys have made it possible to discover hundreds of MWDs \citep{Kepetal13}, mostly in the range 1--80\,MG. Six MWDs have been discovered in the 20\,pc volume with spectroscopic techniques. 

\noindent
\textit{iii)} In the 1990s, spectropolarimetry started to be applied more systematically to observation of WDs, mainly thanks to the the survey of \citet{SchSmi95}, \citet{Putney97}, and later \citet{Kawetal07}. These surveys allowed the measurement of \bz\ with a typical uncertainty of $\sim 10$\,kG, and led to the discovery of another five MWDs in the local 20\,pc volume \citep[including an earlier spectropolarimetric discovery by][]{Lieetal78}, while another 32 WDs were observed with no field detection. 

\noindent
\textit{iv)} With the use of the FORS instrument at the ESO VLT, the sensitivity of the \bz\ measurements was brought down to the 1\,kG region, as shown by \citet{Aznetal04}, \citet{Joretal07}, \citet{Lanetal12} and \citet{Lanetal17}. Later, \citet{Lanetal15} demonstrated that the ESPaDOnS instrument of the CHFT could also be used to measure \bz\ in DA and DB WDs with high accuracy, and \citet{BagLan18} showed that the ISIS instrument of the WHT could compete with the FORS instrument in terms of accuracy. (We note that ESPaDOnS does not have capabilities in the continuum, while FORS and ISIS have.) All these projects led to the discovery of another four MWDs in the local 20\,pc volume \citep[one of which, WD\,2047$+$372, had already been observed by][but with precision not sufficiently high to detect its weak magnetic field]{SchSmi95}, and set meaningful upper limits to \abz\ for 17 WDs in which no magnetic field had been detected. Ten of these non magnetic WDs had been in fact already observed with spectropolarimetric techniques but with a much lower precision. 

\subsection{Motivations for additional observations}
None of these past surveys specifically targetted a volume-limited sample of WDs, although \citet{Kawetal07} reported some statistics of the local neighborhood, finding that nine out of the 43 WDs within 13\,pc from the Sun were magnetic, for a global frequency of $21 \pm 8$\,\%. At that time, the exploration of the 20\,pc volume was still incomplete, with only 116 WDs known members, 15 of which were known to be magnetic. For the \lv, \citet{Kawetal07} estimated an incidence of the magnetic field of $13 \pm 4$\,\%. In fact, apart from the incompleteness of this sample, it is important to note that the works by \citet{Aznetal04} and \citet{Joretal07}, and later \citet{Lanetal15} and \citet{BagLan18} made it clear that all WDs with spectral lines in which \citet{Putney97} and \citet{SchSmi95} failed to detect a magnetic field should be revisited with larger telescopes and more efficient instruments, on the expectation that modern measurements could reach a sensitivity up to a factor of 10 times higher than in the past. 

We finally decided to use the ESPaDOnS, FORS2 and ISIS instruments to perform a spectropolarimetric survey of all WDs of the 20\,pc volume that were not known to have magnetic fields, but for which no highly-sensitive measurements had been performed previously. At the beginning of our project, the target list consisted of $\sim 70$ WDs that were never observed polarimetrically before, and another $\sim 20$ that should be re-observed with much higher accuracy. During the last few years we have presented the discoveries of 12 new MWDs in the 20\,pc volume
\citep{LanBag19a,LanBag19b,BagLan19b,BagLan20}, which is more than 1/3 of all MWDs now known in that sample. In the next section we will present our still unpublished spectropolarimetric observations of WDs, which include mainly non detections, and a few marginal detections.

\section{New spectropolarimetry with ESPaDOnS, FORS and ISIS}\label{Sect_New_Obs}
\begin{figure}
\centering
\includegraphics[width=8.7cm,trim={1.4cm 6.3cm 0.7cm 3.0cm},clip]{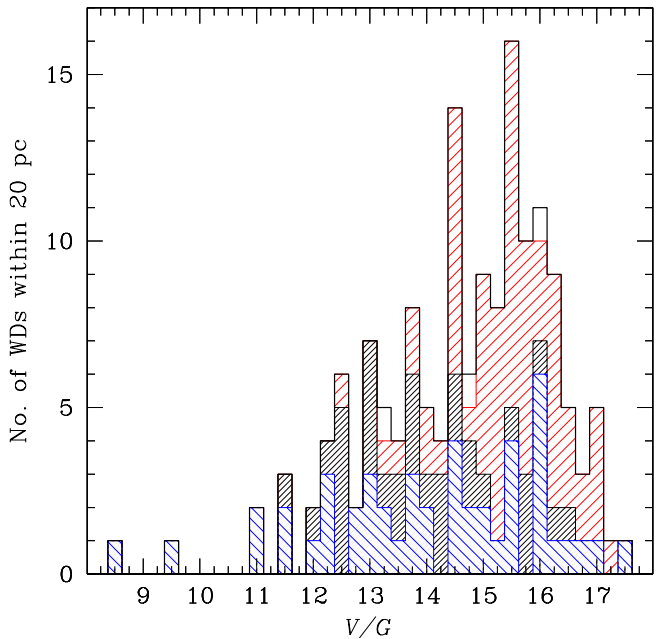}
      \caption{\label{Fig_Magni} Distribution in apparent magnitude (either $V$ or Gaia $G$ magnitude, see Sect.~\ref{Sect_Table}) of the WDs of the \lv. Red lines refer to the new spectropolarimetric observations obtained in the course of this survey, black lines to the WDs that were observed in spectropolarimetric mode prior to our survey and that we have re-observed with higher precision; blue lines refer to WDs that were observed in previous work and that we have not re-observed. The three small unshaded regions correspond to the three WDs that were not checked for magnetic field.
}
\end{figure}
We present 122 new spectropolarimetric observations of 85 WDs. These new observations add to the about 20 spectropolarimetric observations of 12 MWDs that we have already presented, and bring the compilation of a database of highly-sensitive magnetic field measurements of the local population of WDs nearly to completion. As anticipated above, all our new measurements were obtained with the FOcal Reducer and low dispersion Spectrograph (FORS2) instrument \citep{Appetal98} of the ESO Very Large Telescope (VLT), with the Intermediate-dispersion Spectrograph and Imaging System (ISIS) instrument of the William Herschel Telescope (WHT), and with the Echelle SpectroPolarimetric Device for the Observation of Stars (ESPaDOnS) instrument \citep{Donaetal06} of the Canada France Hawaii Telescope (CFHT). Eleven WDs were observed with two instruments, and another two (WD\,0738$-$172 and WD\,2336$-$079) with all three instruments. The relevance of this new survey can be appreciated in Fig.~\ref{Fig_Magni}, which shows the distribution in magnitude of the observations of the WDs of the \lv. Red lines show the distribution in magnitude of the WDs that were observed for the first time in spectropolarimetric mode in the course of this survey (including the 12 MWDs presented in our previous four papers); black lines refer to the distribution in magnitude of the WDs that were observed in spectropolarimetric mode prior to our survey but that we have re-observed with higher precision; blue lines refer to WDs that were checked for a magnetic field in previous works, and that we have not re-observed because we considered that literature data provided sufficient constraints to their magnetic field. Indeed, among the WDs of the latter group, about 20 were already known as magnetic ones, hence did not need to be re-observed to confirm their magnetic nature, a dozen had been found non magnetic by high-sensitivity measurements either with the ESPaDOnS, the FORS or the ISIS instrument in various works from 2004 to 2018; another dozen are DC WDs that were observed with high-precision broadband polarimetric techniques in the seventies. Only three stars could not be checked for magnetic field for the reasons explained in Sect.~\ref{Sect_Accuracy}. Their positions in the histogram are shown by unshaded areas. We note that in Fig.~\ref{Fig_Magni}, each of the six uDD systems of the \lv\ were counted as two individual stars, even if they were observed with just one pointing (but of course it is not trivial to estimate field values or upper limits to the individual system members).

 \subsection{ESPaDOnS data}
With ESPaDOnS we obtained 27 new observations of 24 WDs (although one star, WD\,1334$+$039, turned to be featureless, and no useful field measurements could be made, because ESPaDOnS lacks capabilities in the continuum). ESPaDOnS data cover a useful spectral range of $\sim 3800-8900$\,\AA, with a spectral resolving power of 65\,000 \citep[see][ for details of the use of this instrument for WD observations]{Lanetal15}. ESPaDOnS is a high-resolution spectropolarimeter on a relatively modest telescope of 3.6-m aperture, and lacks the ability to measure continuum polarisation, and because of this it was used to observe only WDs brighter than $V=15$, and mainly WDs with strong spectral features. 
 
\subsection{FORS2 data}
We present 58 new observations of 43 WDs obtained with the FORS2 instrument using different grisms. To observe DC stars we mainly used grism 300\,V with no order separating filter, covering the spectral range $\sim 3700-9300$\,\AA; we set the slit width to 1.2\arcsec\ for a spectral resolving power of 360 \citep[for a justification of the choice of not using the order separating filter see App.~A of ][]{Patetal10}. For DQ stars we used grism 600\,B, covering the spectral range $\sim 3600-6200$\,\AA; slit width was set to 1.0\arcsec\ for a spectral resolution $R\sim 780$. For DA stars we mainly used grism 1200\,R $+$ filter GG435, covering the spectral range $\sim 5800-7300$\,\AA; with a slit width of 1\,\AA\ we obtained a spectral resolving power of 2140. For DZ stars and some DA stars we used the grism 1200B, which covers the spectral range $3700-5100$\,\AA. Slit width was set to 1.0\arcsec, for a spectral resolving power of 1400. We also report two \bz\ measurements of one star, WD\,0233$-$242, that were obtained with FORS2 in 2013, but from which previous literature had reported only the \bs\ value.

\subsection{ISIS data}
With the ISIS instrument we obtained 33 new observations of 30 WDs, using (simultaneously) the grating R600B in the blue arm and grating R1200R in the red arm, covering the spectral ranges of $\sim 3700-5200$\,\AA\ with a spectral resolving power of 2600, and $\sim 6100-6900$\,\AA\ with a spectral resolving power of 8600, respectively (for a 1\arcsec\ slit width). We note that for some of the observations we erroneously adopted a 2x2 binning readout, which led to a slight undersampling, effectively reducing the spectral resolution in the red arm by $\sim 30$\,\%. 

\subsection{Data reduction and field measurements}
Data reduction for FORS and ISIS data is fully described in \citet{BagLan18} and references therein. The methods used to measure the longitudinal field from polarised spectral lines are summarised in  \citet{BagLan18} and \citet{LanBag19a}. The method used to carry out mean longitudinal field measurements with ESPaDOnS is explained by \citet{Lanetal15,Lanetal17}. Because of its high resolving power, ESPaDOnS spectra of adequate S/N can be used to measure \bs\ in MWDs with values above about 50\,kG \citep{Lanetal17}. In case of stars where the magnetic field has to be estimated from a weak level of polarisation in the continuum, we use the relationship
\begin{equation}
\frac{V}{I} = \gamma \bz
\end{equation}
where we adopt $\gamma = 15$\,MG per percent of polarisation, as discussed by \citet{BagLan20}. The polarisation in the continuum was estimated from inspection of heavily rebinned spectra (up to $\sim 400$\,\AA\ spectral bins). 

\citet{BagLan20} have shown that in FORS2 data, the background illuminated by the moon may appear circularly polarised because of cross-talk from linear to circular polarisation. This cross-talk varies rapidly with the position in the field of view, hence background subtraction may produce artefacts in the form of a spurious signal of linear polarisation, stronger in the blue and decreasing with wavelength, being generally little noticeable at $\lambda \ge 7000$\,\AA. The effect depends on the degree of lunar illumination, and is more prominent in fainter than in brighter stars, but normally will be more noticeable on stars fainter than $V \simeq 15.5$ when not observed in dark time. We have found that spurious polarisation is minimised when background is estimated from the same Wollaston strip where the source spectrum is recorded, which is possible to do when observations are taken under seeing conditions better than $\sim 2\arcsec$. 

The observing log and derived magnetic field values are given in Table~\ref{Table_Log}. In App.~\ref{Sect_Log} we discuss individual observations of WDs whenever results may be ambiguous, which happens in particular for the WDs of spectral class DC for the reasons explained above.

\section{Merging earlier with new field measurements: the magnetic fields of the local population of WDs}\label{Sect_Results}
\begin{figure*}
\centering
\includegraphics[angle=270,width=18cm,trim={0.8cm 1.5cm 1.9cm 1.0cm},clip]{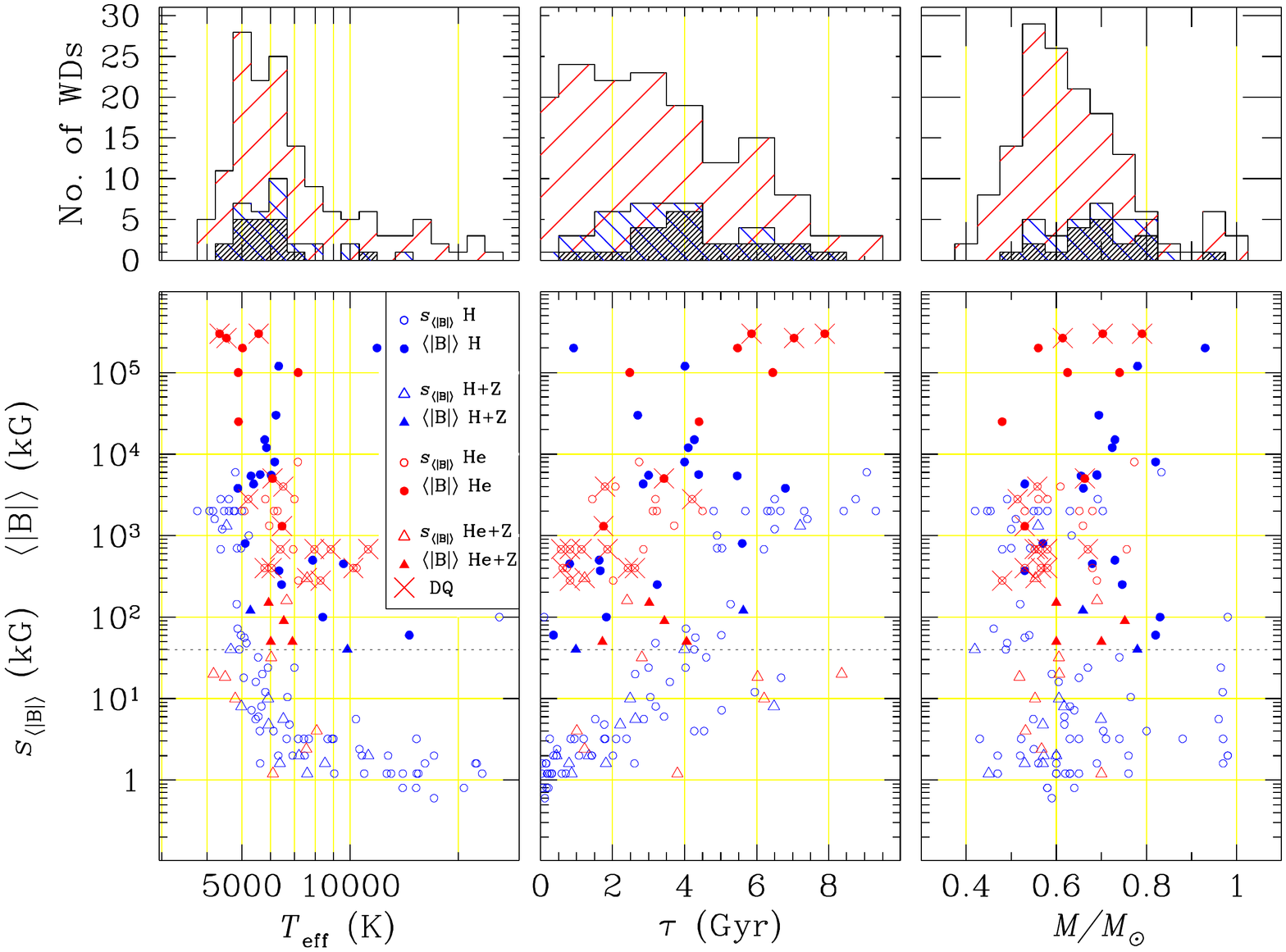}
      \caption{\label{Fig_Histos}
      {\it Top panels:}    Histogram of WDs of the local 20\,pc volume. The red stripes refers to all WDs, the blue stripes to all MWDs, black stripes to all MWDs with $\bs \ge 2$\,MG. 
      {\it Bottom panels:} Full symbols: average \bs\ values versus temperature, cooling age $\tau$, and mass of the MWDs. Empty symbols: sensitivity \sbs\ of the field measurements on stars in which a magnetic field was not detected, calculated as explained in the text (Sect.~\ref{Sect_Table}). Circles refers to non metal-polluted WDs: blue circles to WDs with an H-rich atmosphere (DA and some DC stars), and red circles to WDs with a He-rich atmosphere (DQ and some DC stars). Triangles refers to WDs with metal lines: blue triangles refer to stars with H-rich atmospheres (DAZ stars) and red triangles to stars with He-rich atmosphere (DZ and DZA stars). Crosses over symbols refer to DQ and DQpec WDs.}  
      \end{figure*}
In the 20\,pc volume there are 146 systems: 140 are presumably single WDs, or in visually resolved systems, and six are uDD, for a total number of 152 of currently known WDs. 
Only five WDs have not been observed in polarimetric mode:

{\it i)} WD\,0121$-$429B, a presumed DC member of a uDD system that, being featureless, may be checked for magnetic field only with polarimetric techniques; its companion WD\,0121$-$429A was discovered to be a MWD via spectroscopy.

{\it ii)} WD\,0208$-$510 (= HD\,13445B, spectral class DQB), which is too close to a bright companion to be observed in polarimetric mode with ground-based facilities. Its spectrum \citep[observed with HST by][]{Faretal13} shows only C$_2$ absorption bands that are too broad to give any useful constraint on \bs.

{\it iii)} WD\,0211$-$340, which cannot be currently observed because it is too close to a background star (see Sect.~\ref{Sect_Sample}). In fact, even its spectral type is unknown.

{\it iv)} WD\,0642$-$166 (= Sirius B), a WD of spectral class DA, is too close to its companion to be observed with polarimetric techniques from ground-based facilities. HST spectroscopy allows us to set a constraint on its magnetic field ($\bs \la 100$\,kG), but it should be recalled that some DAs have fields so weak that they may be detected only via spectropolarimetry.

{\it v)} WD\,0736$+$053 (= Procyon B, spectral type DQZ) is in a binary system somehow similar to that of  WD\,0208$-$510. Mg\,{\sc i} and Mg\,{\sc ii} lines around 2800\,\AA\ \citep[observed with HST by][]{Proetal02}
are not split by Zeeman effect, setting a limit of 300\,kG for \bs.\\

The first three stars of this list should be considered as non observed and will be ignored for statistical purposes, while we will keep in mind for that for the magnetic field of Sirius\,B and Procyon\,B there are at least some constraints from spectroscopy. 

Altogether, combining data from the literature as described above with our new observations, we have useful field measurements or upper limits for 149 WDs of the approximately 152 WDs of the 20\,pc volume. The presence of a magnetic field is firmly established in 33 of these stars. 
The remaining non magnetic (or presumed non magnetic) 116 WDs of the \lv\  were observed with spectropolarimetric techniques, with a sensitivity that varies greatly with the object spectral type and temperature. 

\subsection{The accuracy of the observations of WDs in which magnetic field was not detected}\label{Sect_Accuracy}
For the purpose of this work, together with the identification of the MWDs, it is equally important to assess how reliably we can declare that among the remaining 116 WDs, no magnetic fields are present, or at least, no magnetic field detectable through the currently available techniques. Among these 116 WDs, 114 have been checked for the presence of a magnetic field with polarimetric techniques, with a sensitivity that varies from a one or few kG for the brightest DA WDs, up to $0.5 - 2.0$\,MG for the featureless DC WDs. Obviously, observations of uDDs gives looser constraints to the magnetic field of individual members. DQ stars should be practically considered as featureless stars, in that only continuum polarisation may firmly reveal the presence of a magnetic field \citep{Berdetal07}. 

The presence of metal lines allows us to measure the magnetic field in many of the cool, and otherwise featureless stars, with a precision that depends on the number and strength of metal lines. For the DZ targets of our survey, the precision varies between $\simeq 0.4$ and 8\,kG, except for one DZ star (WD\,0552$-$106) for which the uncertainty of its non detection was $\sim 40$\,kG, and another star (WD\,1743$-$545) that, because of the weakness of its metal lines, can be checked for magnetic field only through the measurement of their intensity profiles ($\bs \la 1$\,MG).

There are four DA stars that are so cool that spectropolarimetry of H$\alpha$ does not provide any useful constraint (at least at the {\it S/N} levels that were obtained), therefore the constraint on their field is still given either by the lack of observed Zeeman splitting in their extremely weak H$\alpha$ and/or by the lack of polarisation in the continuum, for a precision that varies between 0.5 and 1.5\,MG. 
This is the case of WD\,1345$+$238, for which we can set $\abz \la 1.5$\,MG, 
WD\,0727$+$482A for which at best we can say that $\bs \la 1$\,MG, and
WD 0751$-$252 and WD\,1823$+$116, for which we can set an upper limit of 0.5\,MG to \bs. We also note that the fainter component of the uDD system  WD\,0727$+$482B could also be a DA star, but its characterisation is uncertain and it will not be included as member of the DA or the DC class in the next sections.
WD\,1820$+$609 is a fifth case of a DA star with a very weak H$\alpha$ in which spectropolarimetry is not very useful. However, the presence of a very narrow core allows us to set an upper limit for \bs\ at about 50\,kG. One DA was observed with $\sigma_z \simeq 35$\,kG, and five DA/DAZ WDs with $\sigma_z = 10-15$\,kG (all these DAs have $\teff \la 5000$\,K except WD\,0728$+$642). About 20 DA/DAZs have $\sigma_z$ between 1 and 10\,KG (but mostly $\simeq 2$\,kG), and all the remaining non-magnetic DA and DAZ WDs, about 40 stars, were observed with sub-kG precision.

In App.~\ref{Sect_Stars} we comment on individual stars, grouped by spectral class, and we justify why each star may or may not be considered magnetic, or suspected magnetic. Whenever necessary, we make additional comments, for instance, regarding the criteria that we have adopted to estimate of stellar parameters of individual components of the uDD systems. Appendix~\ref{Sect_Stars} contains a detailed compilation of the magnetic observations of WDs which may be used to guide further spectropolarimetric observations of the local 20\,pc population of WDs, but is not needed to follow the rest of this work.

\subsection{Four WDs that were erroneously considered magnetic in previous literature}
It is important to report that in the \lv, four WDs were erroneously identified as MWDs in the previous literature. Rectifying this situation is essential in order to correctly estimate the true incidence of magnetic fields as a function of stellar age, especially because three of them are among the youngest WDs of our sample.

{\it i)} Based on a spectropolarimetric measurement obtained at the Mount Stromlo Observatory ($\bz= -6.1 \pm 2.2$\,kG), star WD\,0310$-$688 was identified as a suspected magnetic by \citet{Kawetal07}.  However, the same star was observed with much higher accuracy with FORS1 by \citet{Aznetal04}, who measured $\bz=-0.10 \pm 0.44$\,kG,  and with FORS2 by \citet{BagLan18}, who found $\bz = -0.20 \pm 0.23$\,kG. We consider this star to be non magnetic.

{\it ii)} 40\,Eri\,B = WD\,0413$-$077 was long considered a weakly magnetic WD \citep{Fabetal03,Valetal03,Feretal15}. However, \citet{Lanetal15} made a number of extremely precise measurements of the longitudinal field using both the ESPaDOnS instrument of the CHFT and the ISIS instrument at the WHT. Although most of their measurements had uncertainty as low as 85--90\,G, no magnetic field was detected. The conclusion of \citet{Lanetal15} was that field detections previously reported in the literature were spurious, and the star is actually non-magnetic. In this paper we have presented an additional measurement obtained with ISIS, that has led again to a null detection.

{\it iii)} \citet{Holbetal16} erroneously list the 0.4\,Gyr old WD\,2326$+$049 as magnetic, citing \citet{Aznetal04} as the source of the measurement. In fact, this star was not observed by \citet{Aznetal04}, while the FORS2 measurements by \citet{Faretal18} demonstrated that the star is not magnetic. 

{\it iv)} \citet{LieSto80} observed the cool DA star WD\,1820+609 (at that time believed to be DC), and found a signal of broadband polarisation consistent with zero. \citet{Putney97} re-observed the same star in spectropolarimetric mode and found non-zero continuum polarisation in the instrument red arm ($\simeq -0.5$\,\%, judging from her Fig.~2i, and reporting this value of the polarisation with our sign convention). \citet{Putney97} declared this signal as possibly spurious, but stated that H$\alpha$ revealed a weak field, and that the star needed to be re-observed.  As pointed out by \citet{Lanetal16}, in her Table~1, \citet{Putney97} erroneously reported the measurement by \citet{LieSto80} with a ten times smaller error bars, making that measurement appear as a $6\,\sigma$ detection; this error propagated in the review paper by \citet{Feretal15}, and the star was considered as a magnetic one. Our ISIS measurements have a low \snr, and does not improve on the upper limit of \abz\ set by \citet{LieSto80}, but H$\alpha$ spectroscopy set for \bs\ the upper limit of 50\,kG. We conclude that the star should be considered as non magnetic.

\subsection{The WDs of the \lv: stellar parameters and magnetic field}\label{Sect_Table}
The overall situation is fully summarised in Table~\ref{Tab_Stars}, a list of physical parameters of WDs in the \lv. This list is organised as follows. Column~1 gives the name of the star (if the star is magnetic, its name is printed in boldface);  col.~2 the $V$ magnitude generally taken from the SIMBAD database; when this was not available (for instance, for the newly discovered WDs) we reported the Gaia $G$ magnitude. We note that in the context of this work, we do not need an accurate estimate of the star's apparent magnitude. Colums~3 reports the distance, generally as derived from the parallax of Gaia DR2. Column 4 shows the spectral type -- for MWDs we have added the symbol "H". Traditionally, a MWD would be designated with "H" if the field detection was made with spectroscopic techniques, and with "P" if the detection was made with polarimetric techniques. As already discussed by \citet{BagLan20}, this distinction simply refers to the detection method, and does not point to a physical stellar feature. In fact, many MWDs can be identified as such both because they show line splitting and because line components are polarised. Therefore we prefer to adopt the suffix H for all MWDs, regardless the actual observing technique that was employed for their discovery. A small number of WDs should be considered as suspected MWDs and their cases are discussed in detail in App.~\ref{Sect_Stars}. Suspected MWDs are labelled in col.~4 with a question mark ('H?'). Column~5 says whether the star is in a binary system, either as a visual binary (VB) or visual multiple system (VM), as a resolved double degenerate system (DD) or unresolved double degenerate system (uDD). The symbol s designate stars for which there is no significant evidence of binarity. This classification is largely based on the Tables compiled by \citet{Tooetal17}, with additional inputs as detailed in Sect.~\ref{Sect_Unresolved}.

For the determination of the stellar parameters such as atmospheric composition (col. 6), temperature (col.~7), log g (col.~8), mass (col.~9) and age (col.~10) we have used the references given in col.~11 (see also Sect.~\ref{Sect_Sample}). Column~12 gives an estimate of the average \bs\ of the MWDs. For stars in which \bs\ was estimated through the analysis of the Zeeman splitting observed in intensity, it is possible to measure the average field strength at the surface of the star. For rotating WDs, \bs\ changes as the star rotates, but usually not by very much, so that even a single \bs\ measurement allows us a good estimate of the average field strength of a star. For weaker field MWDs, however, \bs\ cannot be measured directly, and what we know with good accuracy is the average longitudinal component of the magnetic field \bz, at one or several rotational phases. In these cases, the estimate of the average field modulus at the surface of the star relies on modelling results, or educated guesses. Again, we refer to App.~\ref{Sect_Stars} for the discussion of the details of individual stars. 

For stars in which a magnetic field was not detected, in the second last column we have given an indication of the sensitivity of the field measurements, noting either the uncertainty of the best \bz\ measurement and the number of measurements available, or the upper limit for \abz\ as deduced from polarimetry of the continuum (for featureless stars), or the upper limit for \bs\ from spectroscopy, for the few cases in which polarimetric measurements are not available. From these information, it is useful to estimate an upper limit for the surface field \bs. When only spectroscopic data are available, as in the case of fours WDs only, this estimate is straightforward and provided directly in col.~13. For the large majority of stars, a number of \bz\ measurements are available; these measurements set a stronger constraint than spectroscopy alone, but estimating an upper limit for \bs\ is modelling-dependant. Bayesian statistics may allow one to estimate upper limits for individual stars based on the available measurements. However, as crude estimate we decided to define as field sensitivity $\sbs$ the lowest value of $\sigma_z$ multiplied by four, that is, $\sbs = 4\,{\rm MIN}(\sigma_z)$. For example, for star WD\,1202$-$232, the upper limit for \bs\ from direct spectroscopic measurement of its H$\alpha$ is 50\,kG, but there are also four \bz\ measurements available, the most precise of which has $\sigma_z \simeq 0.4$\,kG. We say that this star has $\sbs = 1.6$\,kG. Under the approximation that the field has a dipolar structure, \sbs\ may be interpreted as an approximate estimate of the upper limit for \bs\ (this interpretation will be discussed again in Sect.~\ref{Sect_Dipolar_Approx}). In the majority of cases of Table~\ref{Tab_Stars}, this estimate is between 0.5 and 2 times the upper limit of \bs\ found via a more rigorous Bayesian approach, but there exist outlier cases, as the suspected but not confirmed MWDs, for which our approach definitely underestimates the upper limit for \bs. Finally, for featureless stars, for which only polarisation of the continuum is available, \sbs, expressed in MG, has been set equal to 15 times the upper limit of $\vert V/I\vert$, expressed in percent units.
\begin{figure}
\centering
\includegraphics[angle=270,width=8.7cm,trim={5.1cm 1.4cm 1.9cm 0.9cm},clip]{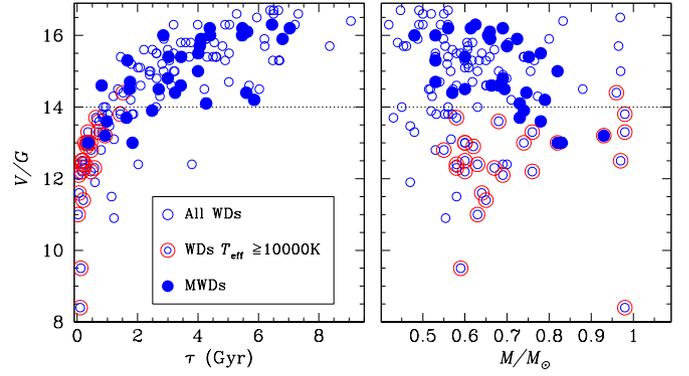}
      \caption{\label{Fig_MagLimited} {\it Left panel:} age-apparent magnitude scatter plot of the \lv, showing all WDs cooler than 10\,000\,K (empty blue circles), all WDs hotter than 10\,000\,K (larger empty red circles) and all MWDs (filled blue circles). The apparent $V$ or $G$ magnitude of the $y$-axis is the same as reported in Table~\ref{Tab_Stars} (see Sect.~\ref{Sect_Table} for more details).
      {\it Right panel:} scatter plot mass-magnitude for the same sample and with the same meaning of the symbols as in the upper panel. The horizontal dotted lines help to visualise what the impact of a target selection based on a magnitude limit could have on the sampling of stars with different parameters.}
\end{figure}

The last column of Table~\ref{Tab_Stars} contains a reference to the literature reporting field measurements for the star. The references to older broadband polarimetric measurements of non-featureless WDs, superseded by modern spectropolarimetric data, are given between parenthesis. More details about individual stars are given in App.~\ref{Sect_Stars}.

A general overview of the statistics that can be deduced from Table~\ref{Tab_Stars} is shown in Fig.~\ref{Fig_Histos}. The top panels show the distributions of magnetic and non magnetic WDs with temperature, age and mass. The bottom panels help to visualise the field strength distribution, but also show how stellar parameters affect the sensitivity of the field measurements, in particular that the detection threshold increases with age. In young WDs, fields may be detected with a sensitivity of the order of a few kG, while only fields with MG strength can be detected in most older stars. In contrast, there is no obvious bias against detecting fields as a function of stellar masses. Regarding temperature, field measurement sensitivity varies quite non-linearly with \teff. For DA stars, for example, sensitivity is affected by the depth and width of the Balmer lines, and becomes poor both at high temperature (above 20–30000 K) as H ionises and the lines weaken, and below about 7000 K as the lines weaken because the population of the $n=2$ levels of neutral H diminish.

It is particularly interesting to relate the physical parameters of the local population of WDs with the scope of a magnitude limited survey. The top panel of Fig.~\ref{Fig_MagLimited} shows that if we were to observe in the \lv\ only WDs brighter than apparent magnitude $V=14$, apart from sampling a smaller number stars, we would miss virtually all WDs older than 3\,Gyr. If in addition we were also setting colour limits to our survey, focusing our attention for instance to stars hotter than 10\,000\,K, we would look only at the very youngest stars of the local population. No matter how deep a survey could go, if the sample is magnitude limited, we will look at a population of WDs that is not representative of the local situation, and all but the youngest WDs will be vastly underrepresented. However, Fig.~\ref{Fig_MagLimited} offers no evidence that a magnitude-limited survey of the \lv\ would be biased against higher-mass WDs. This will be discussed again in Sect.~\ref{Sect_Cool}.

\section{Analysis}\label{Sect_Analysis}
Having established that a magnetic field is present in 33 out of 149 observed WDs, we conclude that the overall frequency of MWDs in our volume-limited, nearly complete sample is 22\,\%. Extrapolated to the Galactic sample, the fraction of magnetism is $22 \pm 4$\,\%, although we should bear in mind that we know that weaker fields of stars with featureless spectra, if present, have escaped detection.

In this section we analyse whether there are hints that magnetic fields are more frequent or stronger in WDs of specific spectral classes  (Sect.~\ref{Sect_Class}) or chemical composition of the atmosphere (Sect.~\ref{Sect_Comp}).  We also study the distribution of the field strength (Sect.~\ref{Sect_Strength_Distribution}), in particular whether the fields are found preferably in a certain range of strength. We finally explore  whether there are correlations of field frequency with  age (Sect.~\ref{Sect_Cool}) or stellar mass (Sect.~\ref{Sect_Mass}).

\subsection{The fraction of MWDs among WDs of different spectral classes}\label{Sect_Class}
\begin{figure}
\centering
\includegraphics[width=8.7cm,trim={0.3cm 5.9cm 0.7cm 3.0cm},clip]{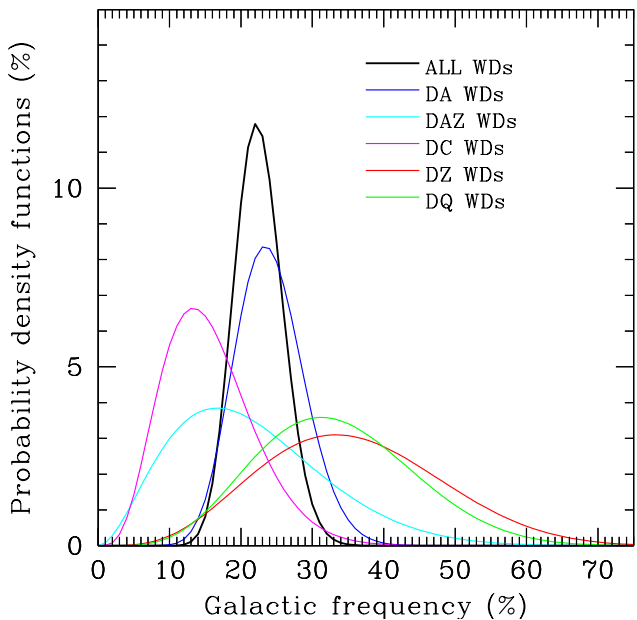}
      \caption{\label{Fig_Stats} Probability density function of the Galactic frequency of the magnetic field in WDs of difference spectral classes.}
\end{figure}
\begin{table}
\begin{center}
\caption{\label{Tab_Stats} Summary of basic data about the incidence of magnetic fields in the local population of WDs.
The frequency $f=M/N$ of col. 4 is the ratio between confirmed MWDs (col. 2) and all WDs (col. 3) for the classes of stars of col. 1; column~5 is the corresponding expected value of $r$, $(M+1)/(N+2)$.}
\tabcolsep=0.12cm
\begin{tabular}{lrrr@{$\pm$}rcc}
\hline\hline
Spectral type        & $M$ & $N$  &\multicolumn{2}{c}{$f$}& $\langle r P_r \rangle$&Notes\\
\hline 
DA                   & 18  & 77  & 23.4 & 4.8\,\% & 24.1\,\%& {\rm 1}\\         
DAZ                  &  2  & 12  & 16.7 &10.8\,\% & 21.4\,\%&        \\	       
DA+DAZ               & 20  & 89  & 22.5 & 4.4\,\% & 23.1\,\%&        \\	       
DC                   &  4  & 30  & 13.3 & 6.2\,\% & 15.6\,\% &      \\	       
DZ+DZA               &  4  & 12  & 33.3 &13.6\,\% & 35.7\,\% &{\rm 2}\\ 
DQ                   &  5  & 16  & 31.2 &11.6\,\% & 33.3\,\% &        \\ [1mm]
Observed  WDs        & 33  &149  & 22.1 & 3.4\,\% & 22.5\,\% &{\rm 3} \\      
\hline
\end{tabular}
\end{center}
\begin{small}
\noindent
{\rm 1} If we discard from the statistics four extremely cool DA stars for which field measurements have very low sensitivity ($\simeq 0.5$\,MG), $f \simeq 25$\,\%.\\
{\rm 2} Non magnetic star WD\,1743$-$545 = PM\,J17476$-$5436 has extremely weak H$\alpha$ and metal lines, and the sensitivity of our field measurement is not better what could be obtained in a DC star; therefore this star is not considered in the statistics of DZ (nor DC) WDs.\\
{\rm 3} Star WD\,0727+482B = G\,107-70B belongs to a uDD system and we were not able to assess its spectral classification, hence it is considered only in the global statistics. \\
\end{small}
\end{table}

We first discuss magnetism in WDs of different spectral classes. The general statistical results are summarised in Table~\ref{Tab_Stats}, while Fig.~\ref{Fig_Stats} shows the probability density functions $P_r$ such that $P_r\,{\rm d}r$ is the probability that the frequency of galactic MWDs is comprised between $r$ and $r+{\rm d}r$.

\subsubsection{The fraction of MWDs among DA and DAZ stars}\label{Sect_DA_DAZ}
A slight majority of WDs have spectra that show only H lines and that are classified as DAs. There are 77 DA stars in Table~\ref{Tab_Stars}. DAZ stars, which are DA stars with a spectrum polluted by the presence of metal lines, are treated separately below. Eighteen DAs have a confirmed magnetic field, for a frequency of  $23 \pm 5$\,\%. We note, however, that four DA stars in which no magnetic field was detected have in fact such weak H$\alpha$ lines that constraints on their field measurement come only from continuum polarimetry ($\sbs = 0.5 - 1.5$\,MG, see Sect.~\ref{Sect_Accuracy}). If we discard these stars from our statistics, on the ground that their field sensitivity is much lower than typical for this class of stars, the frequency of the occurrence of magnetic fields in DA WDs would be $\simeq 25$\,\%. Field strength in magnetic DA WDs spans the entire observed range in WDs, from kG level \citep[for instance, WD\,2150$-$820,][]{LanBag19a} to the $\sim 200$\,MG \citep{Jor03} level of Grw\,+70$^\circ$\,8247. 

About 25\,\% of DA WDs studied by \citet{Zucetal03} were found to have a metal-polluted atmosphere, where the metals are believed to be accreted from the debris of disintegrating members of a former planetary system.  In the \lv, the ratio between DAZ and DA+DAZ stars is about 12\,\% (to discuss the incidence of metal pollution in WDs it may be more interesting to consider the ratio between all metal polluted WDs, including therefore not only DAZ but also DZ and DZA WDs, and all WDs; in the \lv\ this ratio is about 10\,\%). Two of the 12 DAZ WDs of the \lv\ are magnetic, for a frequency of $17 \pm 11$\,\%.  The incidence of the magnetic fields in DA and DAZ taken together is $22 \pm 4$\,\%. 

\citet{Kawetal19} analysed a sample of 15 DAZ stars with \teff\ between $\simeq 5200$ and 7000\,K,  and noted that five of them are magnetic (see their Table~5 -- only one star is in common with our sample), for a frequency of $33 \pm 12$\,\%. However, when they restrict their analysis to DAZ WDS cooler than \teff = 6000\,K, \citet{Kawetal19} find that four out of seven DAZ WDs are magnetic, and suggest that about 50\,\% of DAZ WDs cooler than \teff = 6000\,K are magnetic (the frequency is $50 \pm 18$\,\%). 
If we consider only the cooler DAZ WDs of the \lv\ and merge them into the sample studied by \citet{Kawetal19}, we find that out of 21 DAZ cooler than 7000\,K, five are magnetic, for a frequency of $24 \pm 9$\,\%, fully consistent with the frequency estimated for normal DA WDs of all ages,  while four out of 12 DAZ cooler than 6000\,K are magnetic, for a frequency of $33 \pm 14$\,\%. The analysis of this extended sample does not offer statistical support to the claim that magnetic fields are more frequent in cool DAZ than in normal DA WDs:  11 out of 31 DA WDs with $5200 \le \teff \le 7000$\,K are magnetic, and five out of 18 DA WDs with $5200 \le \teff \le 6000$\,K are magnetic, for a magnetic frequency of $35\,\pm 9$\,\% and $28\,\pm 11$\,\%, respectively. 

\subsubsection{DB stars}
There are no DB stars (WDs with He-lines in the optical spectrum) in the \lv, the closest to this class is a DBQA star (WD\,1917$-$077) which will be listed among the DQ stars.

\subsubsection{The fraction of MWDs among DC stars}
There are 31 known DC WDs in the \lv; their magnetic fields may be revealed only by broadband- or spectro-polarimetry. Before our survey, only 10 DC WDs of the \lv\ volume had been observed in polarimetric mode, and only one of them had been found magnetic. In the course of this survey we have observed for the first time in polarimetric mode 20 DC WDs of the \lv, and re-observed one that had been already observed in broadband imaging mode in a previous study.  Only one of the DC stars of the \lv\ (WD\,0121$-$429B, a member of a uDD system) remains not observed in polarimetric mode. Our survey discovered three new DC MWDs \citep{BagLan20}. Among the observations newly published in this paper, we report that WD\,1116$-$470 is a suspected DC MWD, that could host a field of the order of a few MG, and should be re-observed. The magnetic frequency among DC WDs is $\simeq 13$\,\%. We note that the sensitivity of most of the new measurements is $\simeq 5$\,MG (in fact, the weakest field firmly detected has a strength estimated to be about 20\,MG); this situation  could certainly be improved even with current instrumentation, potentially leading to the discovery of further magnetic fields in the DC WDs of the \lv.

More than a half of the magnetic DA and DAZ have a magnetic field that is not strong enough to be detected with polarimetric measurements of the continuum, at least with the sensitivity typical of the current measurements.  If the distribution of field strength of DC stars is similar to that  observed in DA WDs, one could hypothesize that the actual frequency of magnetic fields in the DC WDs of the \lv\ is between 27 and 33\,\%, for an estimated Galactic frequency of, say, $30 \pm 10$\,\%.

\subsubsection{The fraction of MWDs among DZ and DZA stars}
The origin of the metal lines of DZ and DZA WDs is explained in the same terms as for DAZ WDs. All 13 known DZ and DZA WDs of the local 20\,pc volume have been observed for the first time in polarimetric mode in the course of our survey of the \lv, except for WD\,2138$-$332 observed by \citet{BagLan18}, and reported as suspected magnetic, and the DZA star WD~0738$-$172, reported as suspected magnetic by \citet{Frietal04}. 

\citet{BagLan19b} reported the preliminary results of our survey,  confirming the magnetic nature of WD\,2138$-$332, and reporting the discovery of three additional magnetic DZs. At that time, we were also led to suggest that the occurrence of magnetic fields in DZ stars could be particularly frequent, having reached the conclusion that at least 40\,\% of DZ WDs of the \lv\ are (weakly) magnetic. However our subsequent field measurements (presented in this paper) failed to discover any new magnetic DZ WDs, and did not confirm the field detection by \citet{Frietal04} on WD~0738$-$172, a star which we now consider as an only (marginally?) suspected MWD. Since the work by \citet{BagLan19b}, two new DZ stars have been identified in the \lv,  WD\,0552$-$106 and WD\,1743$-$545 (that were previously classified DC), and we found that neither of them is magnetic. We note, however, the metal lines of WD\,1743$-$545 are so weak that the sensitivity of our field measurement is only at the MG level, hence this star should not be included in the statistics. In conclusion, our new estimate of the frequency of the occurrence of magnetic fields in the DZ stars of the \lv\ is 4/12, a fraction that is admittedly lower than the lower limit that we had previously estimated, for a Galactic frequency of $33 \pm 14$\,\%. This range is formally consistent with that of DA WDs, and fully consistent with that of DA WDs within a similar age range as that of DZ and DZAs.

Metal lines allow us to detect weak magnetic fields that would be undetected in stars that without pollution from disk-debris would have featureless spectra. The high frequency of field detection in DZ stars is consistent with the hypothesis that DC stars are actually magnetic at least as frequently as DA WDs, except that the magnetic fields of DC stars are revealed only when their strength is of the order of several MG.

\subsubsection{The fraction of MWDs among DQ stars}
DQ stars are WDs with spectroscopic traces of carbon (neutral carbon lines or molecular C$_2$ Swan bands). Most of them, at least below effective temperatures of about 10000\,K, have helium-dominated atmospheres \citep{Koeetal20}, and the presence of carbon is a result of dredge-up by the extending convection zone in the upper helium layer \citep{Koeetal82,Peletal86}. The 17 DQ stars of Table~\ref{Tab_Stars} represent more than 10\,\% of the local 20\,pc WD population, a proportion much bigger than the fraction of 1.6\,\% of DQs among all WDs classified by the SDSS \citep{KoeKep19}. This large discrepancy may be explained at least in part as a selection effect of the SDSS. Cool WDs are probably underrepresented in the SDSS, and it is also probable that many DQs with weak bands are mistaken for DC stars, because many SDSS spectra have low {\it S/N}. 

Five of the DQ WDs of the local 20\,pc volume are confirmed magnetic, although only 15 out of 17 could be observed in polarimetric mode. The presence of a K0 companion at 1.9$"$ prevented us from observing WD\,0208$-$510 = HD\,13445B with FORS2. WD\,0736$+$053 = Procyon~B has a very bright F0 companion at 5.3$"$, and could only be observed spectroscopically by \citet{Proetal02} with the HST. However,  from these spectra we can still derive an useful upper limit for \bs\ ($\simeq 300$\,kG). All DQs in the local 20\,pc volume are cooler than 10,000\,K,  except for the DBQA WD\,1917$-$077. 

Two out of three peculiar DQ WDs \citep[DQ stars with Swan band-like molecular band depressions that are not at the expected wavelengths for C$_2$ molecules, see ][]{Schetal95} are magnetic. The ratio of five MWDs out of 16 DQ WDs results in a magnetic frequency of $31 \pm 12$\,\%, which is higher than, but still consistent with that of DA WDs, but with an important difference: the typical field strength that can be detected in DQ stars is much higher than in DA WDs. Similarly to the case of DC stars, the magnetic frequency could well be underestimated by a factor of two because of the insensitivity to fields below MG level, in which case we could hypothesise that 2/3 of DQ stars are magnetic. An alternative possibility is that the frequency of the occurrence of magnetic fields in DQ stars is similar to that of DA WDs, but DQ MWDs host much stronger fields than average (in fact, DQ stars are the WDs of the local population that host the strongest fields). In either case, a magnetic field and the presence of C$_2$ in the stellar atmosphere seem correlated, and one could speculate that a magnetic field is responsible for C$_2$ buoyancy.

No WDs belonging to the class of rare hot DQs, WDs with a C-rich atmosphere and little or no trace of H or He \citep{Dufetal07}, and with $\teff = 18000-24000$\,K \citep{Dufetal08}, are present in the local 20\,pc volume, and we are not able to discuss previous claims that magnetic fields may be nearly ubiquitous in this class of stars \citep{Dufetal13,Dunlap14}. 

\subsection{The frequency of the occurrence of magnetic field versus stellar atmospheric composition}\label{Sect_Comp}
Twenty out of 103 WDs with an H-rich atmosphere are magnetic, for a frequency of $19 \pm 4$\,\%; and 13 out of 43 WDs with a He-rich atmosphere are magnetic, for a frequency of $30 \pm 7$\,\%. Except for the four magnetic DZ stars discussed by \citet{BagLan19b}, all He-rich MWDs have a field strength from 1 to hundreds MG. This field strength distribution is biased by the fact that the \lv\ does not include DB WDs, the only He-rich stars, apart from DZ stars, in which fields weaker than 1\,MG may be detected. There are no He-rich WDs younger than 0.5\,Gyr in the \lv, which may represents another bias for the statistics, if the frequency of field occurrence changes with cooling age (see Sect.~\ref{Sect_Cool}).  Five out of eight He-rich WDs older than 5\,Gyr are magnetic, for a frequency of $62 \pm 17$\,\%. Such a high frequency among the oldest stars is not mirrored by H-rich WDs, as only four out of 25 H-rich WDs older than 5\,Gyr are magnetic, for a frequency of $16 \pm 7$\,\%. However, a proper comparison between the magnetic properties of H-rich and He-rich WDs requires to go through a careful revision of the chemical composition of the atmosphere of the coolest stars of the sample, which is not so secure as for the hottest ones. 

\subsection{Distribution of magnetic field strength over the sample}\label{Sect_Strength_Distribution}
\begin{figure}
\centering
\includegraphics[width=8.7cm,trim={0.8cm 6.0cm 0.7cm 3.0cm},clip]{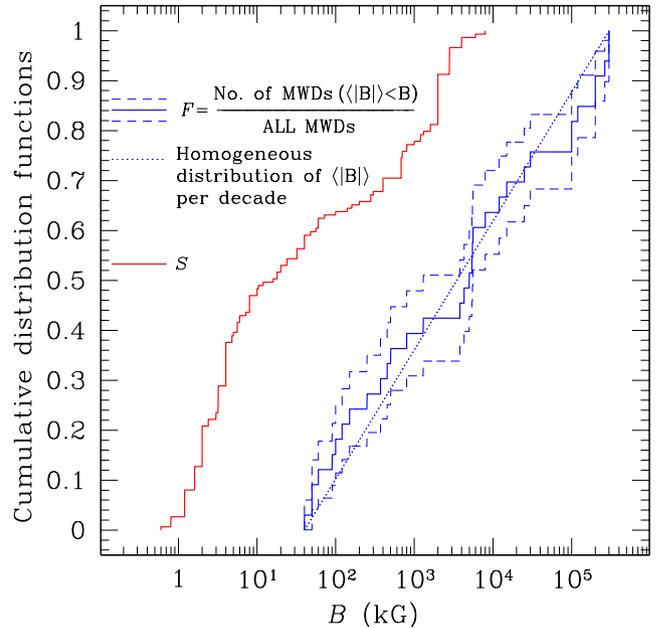}
      \caption{\label{Fig_CumB} Blue solid line: the ratio $F$ between the number of MWDs with ${\bs}$ smaller than the abscissa value $B$ and the number of all MWDs; the dashed lines show the $\pm 1\,\sigma$ uncertainty.
      Blue dotted line: the expected behaviour of a field distribution constant per decades.
      The red solid line shows the ratio $S$ between the number of field measurements with sensitivity \sbs\ smaller than the abscissa value $B$ and the total number of observed WDs.} 
\end{figure}
\begin{figure}
\centering
\includegraphics[width=8.7cm,trim={0.8cm 5.7cm 0.7cm 3.0cm},clip]{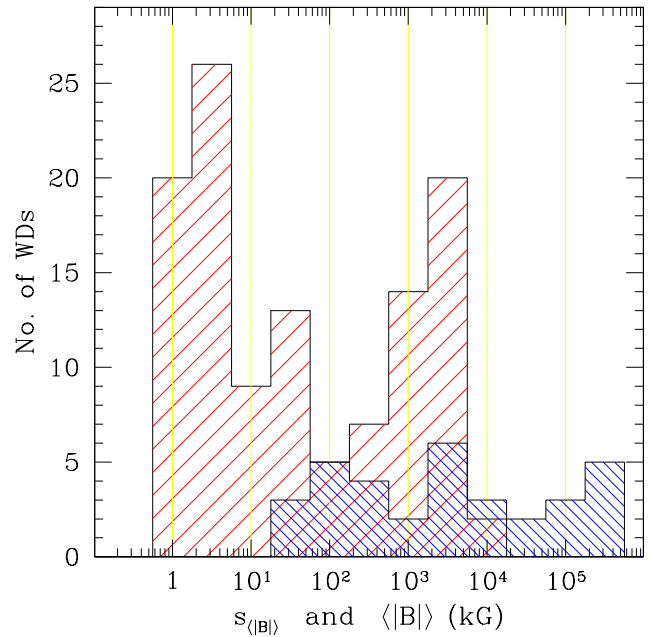}
      \caption{\label{Fig_Histo_Sigmas} The red strips shows the distribution of the sensitivity of the measurements of the magnetic field (practically four times the uncertainty of the best \bz\ measurement of each star, see Sect.~\ref{Sect_Table}) and the blue strips shows the distribution of the observed magnetic fields strengths.} 
\end{figure}
\citet{Feretal15} have proposed that the distribution of field strength of MWDs is peaked between 2 and 80\,MG, a picture that is in conflict with the finding by \citet{Kawetal07}, who carried out a Kolmogorov-Smirnov (K-S) test for the WDs within 13\,pc, and showed that the field strength is constant for each decade interval with a probability of 70\,\%. In this section we present the results obtained from the analysis of the \lv. 

\subsubsection{The distribution of the field strength per decades is rather homogeneous}
Compared to previous studies, we are now able to extend the analysis to a volume more than twice as large, in which we have probed the magnetic field of many DA WDs with sensitivity up to 1-dex better than in the past, and have also searched for the presence of MG magnetic fields in the entire population of DC WDs. Figure~\ref{Fig_CumB} shows the cumulative distribution function for the magnetic field strength. This figure clearly demonstrates that within the range of field strength found in the 20\,pc volume, which extends between about 40\,kG and 300\,MG, the probability of fields occurring is roughly constant per dex of field strength. This same behaviour is observed in Fig.~\ref{Fig_Histo_Sigmas} (which is effectively the marginal distribution of field strengths plotted in the lower panels of Fig~\ref{Fig_Histos}) as well as the distribution of best measurement uncertainties for undetected stellar fields. 

It is important to emphasise that this distribution is very different from the apparent distribution of field strengths as assembled from a census of all known MWDs, however discovered \citep[e.g][Fig.\,8]{Feretal15}. The ensemble of currently known MWDs is dominated by discoveries made from low resolution, low S/N optical and near IR flux spectra obtained for the SDSS project \citep[e.g.][]{Schetal03,Kepetal13,Kepetal16}. Such spectra are most sensitive to magnetic fields in the range of about 2--80\,MG, and the observed distribution has an excess frequency in this range about five times higher than weaker and stronger fields outside of this range. By contrast, Figures~\ref{Fig_CumB} and \ref{Fig_Histo_Sigmas} show that $\sim 40$\,\% of MWDs have fields weaker than 2\,MG, and that MWDs with fields in the regime between 2 and 80\,MG account for another 40\,\% of the total. 
Figures~\ref{Fig_CumB} shows also that the total of fields weaker than 500\,kG contribute roughly to one third of the total observed fields in the 20\,pc volume. These weaker fields are often detected only through highly-sensitive spectropolarimetric techniques, and strongly magnetic DC and DQs may be detected only through polarimetry, no matter how high is the S/N and the resolution of traditional spectroscopy. These considerations clearly highlight how spectroscopy alone cannot provide the correct picture of magnetism in degenerate stars.  

\subsubsection{There are no MWDs with field strength $\la 40$\,kG}\label{Sect_40kG}
A still more notable, and quite surprising, feature of Fig.~\ref{Fig_CumB} is the complete lack of MWDs with detected fields below about 40\,kG. Extrapolating the frequency distribution of Fig.~\ref{Fig_Histo_Sigmas} down about two more bins, a constant field frequency suggests that there should be of the order of five or six MWDs in our survey volume with fields between 4 and 40\,kG. Instead, the distribution comes to an abrupt halt. We have already noted that many of the available \bz\ measurements are obtained with extremely high precision. This is quantified by the red solid line in Fig.~\ref{Fig_CumB}, that shows, for instance, that about half of the WDs were observed with a sensitivity of 10\,kG or better, and 20\,\% of the WDs (that is, 33 WDs) were observed with a sensitivity of 3\,kG or better. The same situation may be appreciated also by looking at the histograms of Fig.~\ref{Fig_Histo_Sigmas}.  While it is remarkable that no field weaker than 40\,kG has been detected in the 20\,pc volume, we should bear in mind that only a fraction of the WDs were checked for the presence of a magnetic field with a sensitivity sufficient to detect the rare and ultra-weak fields. Assuming for instance that the real frequency of MWDs with field strength up to 40\,kG is 4\,\%, the probability to find no magnetic field in that range, out of 33 observed WDs, is still a non-negligible 26\,\%. Observations do offer a stronger constraint, because all measurements with a sensitivity better than 40\,kG which fail to detect a field lower than 40\,kG contribute to reduce the probability that field weaker than 40\,kG, if existing, would pass undetected. Assuming that the frequency of the occurrence of field with strength $\le B_{\rm max}$ is constant and equal to $f_B$, the probability $p$ to find no star with field weaker than $B_{\rm max}$ in a set of observations with sensitivity $\sbs$ may be approximated by
\begin{equation}
p= \prod_{i=1}^N  \left( 1 - f_B \frac{B_{\rm max} - \sbs^{i}}{B_{\rm max}} \right)
\end{equation}
where the index $i$ runs over all stars for which $\sbs^{i} \le B_{\rm max}$. For $B_{\rm max} = 40$\,kG and $f_B=0.04$, we find $p=5.5$\,\%.

Is it possible that we have found the effective lower limit of the overall distribution of MWD field strengths? In that case we would conclude that MWD fields are generally limited to the range between a few tens of kG and a few hundreds of MG, a span of about four dex.

A firm answer to this question will come after the systematic analysis of a larger volume-limited sample of MWDs, further observations of the suspected MWDs (such as WD\,0738$-$172), and a more accurate estimate of the \bs\ upper limits (see our considerations about \sbs\ of Sect.~\ref{Sect_Table}). So far, outside the \lv\ we are aware of only three MWDs with \bs\ possibly smaller than 30-40\,kG, namely WD\,1105$-$048 and WD\,0446$-$789, which were already discovered to be magnetic or suspected magnetic WDs by \citet{Aznetal04}, and a suspected weakly magnetic star, WD\,0232$+$525, newly discovered by us. Some or all of these stars may well have a stronger field that is seen approximately  equator-on, or for which, as in WD\,2359-434, (unusually) \abz\ never rises above about 10\,\% of the value of \bs. However their existence calls into question the general validity of the result found in the \lv, and suggests at the very least that the field strength distribution may have a soft lower field strength edge rather than a very sharp lower limit, if such an edge exists at all.

In conclusion, although the details of the field strength distribution of the MWDs are actually not strongly constrained by observations, the overall shape of the cumulative distribution function of Fig.~\ref{Fig_CumB} appears consistent with a distribution of the field strength constant per decades from approximately 40\,kG to 300\,MG, and substantially lower outside these limits. Further data are needed to firmly extend the validity of this result outside of the \lv.

\subsubsection{In the weak-field regime we are sensitive only to the dipolar field component}\label{Sect_Dipolar_Approx}
We finally note that both the lower and the upper limits of the field strength are probed mainly by polarimetry. A 30\,kG dipolar field would polarise spectral lines, but would not be strong enough to produce a noticeable broadening of their intensity profiles, even with high resolution; at the other extreme, magnetic fields of hundreds of MG strength make it very difficult to recognise spectral features (if they are present at all), so that detection relies on measurements of circular polarisation of the continuum. 

At the lowest field strengths, it should be kept in mind that the Stokes $V$ profiles of spectral lines are sensitive to \bz, which is the component of the magnetic field along the line of sight, averaged over the stellar disk. This average is dominated by the dipolar component of the magnetic field, while higher-order multipolar components tend to contribute very little to \bz, since their line-of-sight component averaged over the stellar disk, for a similar field strength at the magnetic poles, is much smaller than the dipolar contribution. In particular, if we compare a quadrupolar component and a dipolar component of similar strength, the quadrupolar contribution to \bz\ will be about 1/10 the dipolar contribution \citep{Schwarz50,Lanetal98}.  To make this as clear as possible, even if \bz\ is consistent with zero with a sub-kG sensitivity, we cannot prove that a magnetic field is totally absent from the surface of a WD. For instance, a WD could have a non-linear quadrupolar magnetic field \citep{Bagetal96} that produces a surface field \bs\ up to 30--40\,kG, and still be undetectable either via spectroscopic and spectropolarimetric techniques. In this respect, polarimetric weak field detection techniques for WDs are less efficient than in the case of main-sequence stars, in which often the projected surface velocity resolution of Stokes $V$ introduced by rotation may allow one to detect a field even when $\bz \simeq 0$ \citep[the so-called "crossover effect", see][]{Babcock51}.

\subsection{The frequency of the occurrence of magnetic field as a function of cooling age}\label{Sect_Cool}

It is particularly important to understand whether the frequency of the occurrence of a magnetic field is correlated to the star's cooling age, as this information may provide valuable constraints on field evolution both into and during the WD stage. However, it is important to keep in mind that the sensitivity of the field measurements used in this work decreases with stellar age, except for the case of stars with metal lines. 

\subsubsection{Previous results from the literature, and their interpretations in terms of a bias against high-mass WDs.}
The question of the occurrence of fields as a function of cooling age has been investigated in the past. \citet{LieSio79}, \citet{ValFab99} and \citet{Lieetal03} suggested that the magnetic field could be more frequent in older and cooler than in hotter and younger WDs (in fact,
\citeauthor{Greetal71} \citeyear{Greetal71} had already practically assumed that magnetic fields were a characteristics of the cooler WDs only, because no hot MWDs had been found in Greenstein's very large collection of WD spectra at that time). In particular, \citet{ValFab99} suggested a threshold age around 1\,Gyr for the occurrence of magnetic fields with strength $\ga 1$\,MG, finding that the frequency of MWDs with MG field strength is $3.5 \pm 0.5$\,\% among WDs hotter than 10000\,K, and $20 \pm 5$\,\% among WDs cooler than  10000\,K.

\citet{Lieetal03} brought in statistical considerations, by examining the frequency of fields in three qualitatively different samples. They examined blue spectra of the WDs found in the Palomar-Green (hereafter P-G) survey of faint blue objects, and found that among the observed WDs, which generally have \teff\ near or above 10\,000\,K, only $2 \pm 0.8$\,\% have a magnetic field, which, owing to the selection effect introduced by the limited sensitivity of low-resolution spectroscopy, have a strength of about 2\,MG or more. They also examined two large surveys that contain a much higher fraction of older WDs: the sample of 110 cool WDs (mostly with $4000 \la \teff \la 10000$\,K) modelled by \citet{Beretal97}, and an earlier version of the 13 and 20\,pc volume-limited samples, collected by \citet{Holbetal02}, which are dominated by cool older WDs. In these samples the frequency of MWDs is much higher: in the sample by \citet{Beretal97}, the frequency of MG fields is $7 \pm 2$\,\%, and in the volume limited samples the frequency of MG fields is $11 \pm 5$\,\% in the complete 13\,pc volume, and $8 \pm 3$\,\% in the less complete 20\,pc sample. Taken at face value, this result clearly suggests that strong magnetic fields are less common in young WDs than in old samples, and indeed, \citet{Lieetal03} did acknowledged the possibility that some mechanism existed that would cause the frequency of MWDs increase with decreasing \teff. However, \citet{Lieetal03} also argued that this apparent difference could probably due to a very strong selection effect caused by the combination of two facts: (1) that typical masses of MWDs are significantly higher than those of non-magnetic WDs, and (2) that higher mass WDs are significantly smaller in radius than low-mass WDs, hence fainter. They argued that the small radii of relatively massive MWDs led, in the Johnson B magnitude-limited P-G survey, to sample a volume only about 1/4 as large as the one surveyed for lower-mass, presumably less frequently magnetic WDs. Making a correction for this effect, they concluded that the fractional occurrence of MWDs in the P-G survey is actually very close to that of the surveys of cooler stars.

Later, \citet{Kawetal07} analysed an incomplete sample of WDs of the \lv, and again found no evidence of a higher incidence of magnetism among older WDs, within uncertainties.

\begin{figure}
\centering
\includegraphics[width=8.6cm,trim={0.7cm 2.1cm 1.0cm 1.0cm},clip]{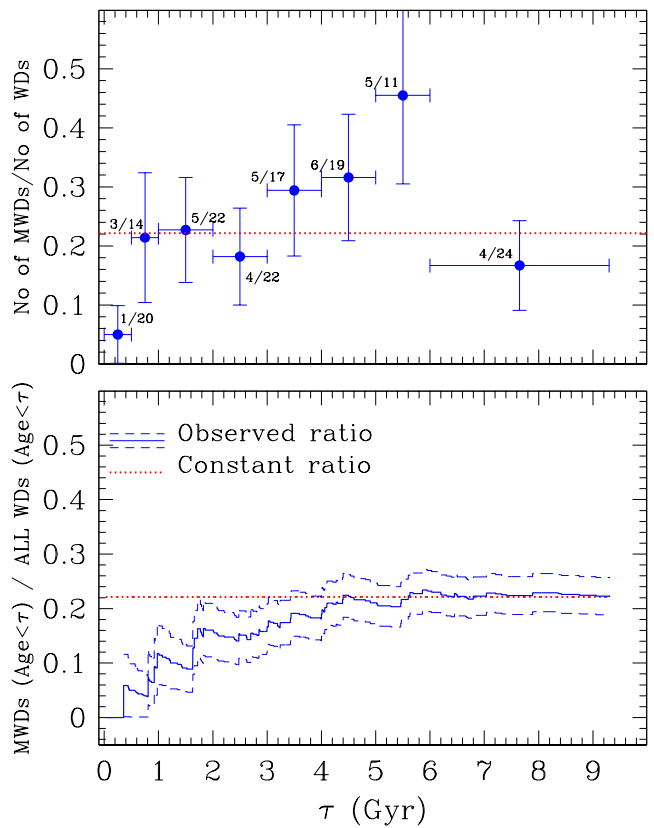}
      \caption{\label{Fig_CumAge} 
      {\it Top panel:} The ratio between the number of MWDs and the number of all WDs that are produced during the interval of time represented by the horizontal errorbars, as a function of cooling age $\tau$. The fraction $M/N$ printed close to the symbols represents the number of MWDs ($M$) and all WDs ($N$) in the interval of time. The vertical errorbars represent the uncertainties associated to the frequency of the occurrence of MWDs extrapolated to the Galactic sample.
      {\it Bottom panel:} The blue solid line represents the ratio between the observed MWDs younger than the abscissa value $\tau$ and all WDs younger than $\tau$; the blue dashed lines show the $\pm 1\,\sigma$ uncertainties of that frequency. The red dotted line shows a constant ratio between MWDs and all WDs.} 
\end{figure}
\subsubsection{In the \lv\ there is a shortage of young magnetic DA WDs}
What does the analysis of our greatly enlarged data set for the  \lv\ say? From Table~\ref{Tab_Stars} we can see that only four WDs out of 34 DA WDs younger than 1\,Gyr are magnetic, and that there is only a single MWD among 20 DA WDs younger than 0.5\,Gyr, the very weakly magnetic star WD\,2047$+$372, discovered by \citet{Lanetal16}, with age $\simeq 0.36$\,Gyr. The frequency of magnetism of one out of 20 (or 19, if we consider Sirius B as not observed) is remarkably smaller than the overall frequency of magnetic fields of about 20--25\,\% derived for the full sample of DA WDs.\footnote{The dearth of younger, hence hotter, hence more luminous magnetic WDs is reflected also in the fact that there are no MWDs brighter than $G \sim 13$ in our sample (see Fig.~\ref{Fig_MagLimited}).} These data predict the galactic frequency of young magnetic DA WDs to be $5.0 \pm 4.9$\,\%. The top panel of Fig.~\ref{Fig_CumAge} shows the ratio between MWDs and all WDs produced in various bins of interval of cooling age, and suggests that the occurrence of magnetic fields increases with age. This behaviour may also be appreciated by looking at the cumulative distribution functions. The bottom panel of Figure~\ref{Fig_CumAge} shows the ratio $r(t)$ between the number of all MWDs younger than age $t$, $M(t)$, and the number of all WDs younger than $t$, $N(t)$. This curve may be extrapolated to the galactic frequency of WDs with an uncertainty $ ( r(t) (1-r(t)) / N(t) )^{1/2}$, and compared with the constant production rate of 33/149. We can see that in the local 20\, WD population, fields are relatively rare in young stars, and more common in WDs older than 1\,Gyr, in spite of the fact that detection thresholds for older stars are definitely higher than for younger stars. The lack of young MWDs in the \lv\ strongly suggests that the onset of higher magnetic frequency occurs later than 0.5\,Gyr.  We note that in the \lv\ there are no WDs younger than 0.5\,Gyr other than of spectral type DA. Therefore, based on the analysis of this sample, we are not able to extrapolate our finding for instance to DB and DQ WDs.

\subsubsection{In fact, magnitude-limited surveys have little bias against high-mass WDs}\label{Sect_MagLim_Bias}
We now face two conflicting interpretations of the observational data. On the one hand, the argument proposed by \citet{Lieetal03} suggests that magnitude-limited surveys are biased against high-mass (hence more frequently magnetic) WDs, and that this is the reason that such surveys sometimes find a low frequency for the occurrence of MWDs. On the other hand, a volume limited survey, which should be exempt from any bias against stellar mass, nevertheless strongly suggests that magnetic fields are more frequent in older than in younger WDs.  How can we reconcile these views?
\begin{figure}
\centering
\includegraphics[width=8.6cm,trim={0.7cm 1.6cm 1.0cm 1.0cm},clip]{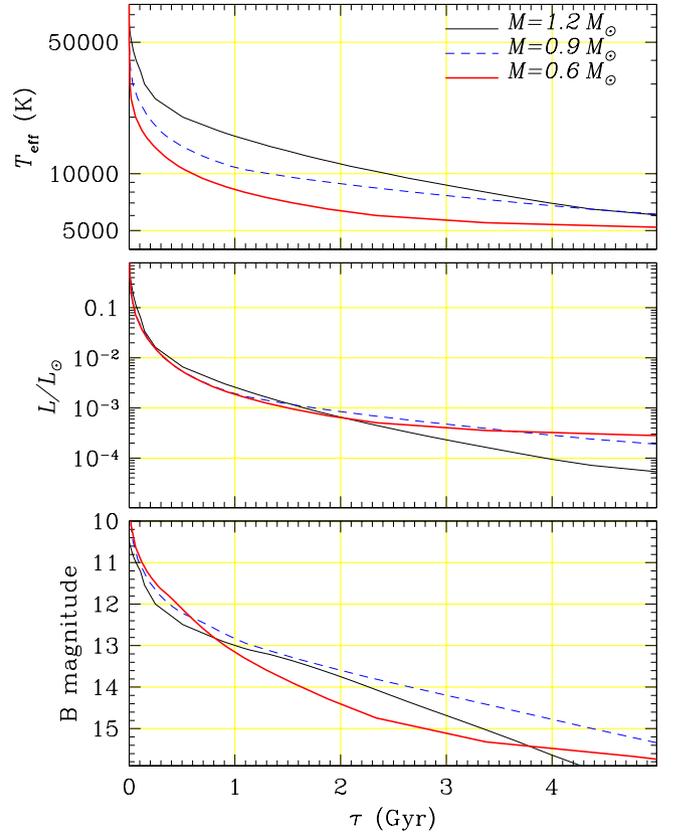}
   \caption{\label{Fig_Cool_Curve}  Cooling history of  $1.2\,M_\odot$ (black solid lines), $0.9 M_\odot$ (blue dotted lines) and $0.6 M_\odot$ (tick red solid lines) WDs. {\it Top panel}: \teff\ vs. cooling age $\tau$; {\it mid panel}: luminosity vs. cooling age; {\it bottom panel}: Johnson B absolute magnitude vs. cooling age. Data from Montreal cooling curve database. 
} 
\end{figure}

The key concept is that for a given \teff, because of its smaller surface, a higher mass WD will certainly have a lower luminosity ($L \propto M^{-2/3}$) than a lower mass WD. However, the higher mass WD, exactly because of its smaller surface and larger internal mass and thermal energy, will also cool more slowly than the lower mass WD,  so luminosity will drop more slowly for a higher mass WD than for a lower mass WD. While luminosity will be always higher for a lower mass than a higher mass WD at the same \teff, the same quantities as a function of cooling age may compare in a quite different way. That is, the comparison we must make is not of massive and less massive WDs of the same \teff, but of pairs of WDs of the same age.

To explore this issue in a quantitative way, we have used the Montreal cooling curve database, which reports \teff, $\log g$ and absolute magnitude in numerous photometric bands as functions of cooling time for a grid of WD masses. From these data, we have plotted the values of \teff\ (upper panel), luminosity (mid panel) and the absolute B magnitudes (lower panel) of WDs of $0.6 M_\odot$, $0.9 M_\odot$ and $1.2 M_\odot$ as functions of cooling age from formation, covering the typical range of non-magnetic WDs, fairly massive MWDs, and extremely massive MWDs.
If we compare an $0.6 M_\odot$ (a typical non-magnetic WD) and an $0.9 M_\odot$ WD (a typical MWD) that formed at the same time and at the same distance from us, the more massive WD is more than a magnitude fainter than the less massive WD during a short period of a few Myr after initial creation. However, after about 20\,Myr, when the  \teff\ values have declined to about 25\,000\,K ($0.6 M_\odot$) and 35\,000\,K ($0.9 M_\odot$), the visible brightness difference has dropped to roughly 0.1 or 0.2\,mag, and after about 600\,Myr the more massive star is actually significantly brighter than the less massive one (see the bottom panel of Fig.\,\ref{Fig_Cool_Curve}). Within the age range 0.02--0.5\,Gyr, the magnitude difference between WDs of 0.6 and $0.9\,M_\odot$ of the same age is only about 0.2\,mag, and the brighter star would be visible in a magnitude limited survey to a distance $\sim 10$\,\% greater than the fainter, which amounts to a difference in survey volume of $\sim 30$\,\%. Since it is the lower mass WD that is slightly brighter during this cooling age interval, the volume of the higher mass MWD should be increased by about 30\,\% to correct for the different volumes surveyed. 
Note however that the extremely massive $1.2\,M_\odot$ WD is as much as 0.5\,mag fainter in B relative to the WD of $0.6\,M_\odot$ for much of the first 800\,Myr of cooling, then its magnitude drops rapidly below the lower mass comparison after about 4\,Gyr.  Overall, it appears that the bias that exists in magnitude-limited samples of WDs with ages between about 1 and 4\,Gyr will actually lead to overestimates of numbers of the most massive WDs compared to WDs of normal mass. 

We conclude that the view that magnitude-limited surveys of WDs are strongly biased against high-mass stars is not generally correct, although various biases do exist for WDs of different masses at some ages. The relationships between luminosity and cooling age change with mass in a complex way, underscoring the fact that statistically studies of WDs must be based on volume-limited rather than magnitude limited surveys.

\subsubsection{Magnitude-limited surveys support the idea that there is a shortage of young MWDs} 
We re-examine the results of magnitude-limited surveys in light of the considerations of Sect.~\ref{Sect_MagLim_Bias}. We first note that even if weak (few kG to 2\,MG) fields were common among young WDs, these would be still missed in the P-G survey (and even more so by the SDSS), therefore the frequency of the occurrence of young MWDs in the P-G survey should be compared with the frequency of the occurrence of MWDs with field strength between 2 and 100\,MG that is deduced from the analysis of the \lv, which is $13 \pm 4$\,\%.

From the stellar parameters of the 347 WDs identified in the P-G survey \citep{Lieetal05}, we find that about 90\,\% of these stars have cooling ages $\la 0.5$\,Gyr. This young group includes about 312 WDs, six of which are magnetic, for a frequency of $1.9 \pm 0.8$\% MWDs. In their 2003 study, \citet{Lieetal03} corrected this frequency for their estimate of the volume bias by a factor of 3.96, bringing their estimate of actual MWD frequency in the P-G survey to $7.9 \pm 3$\%, in very good agreement with their estimates for cooler WDs, and also consistent with our estimate of $13 \pm 4$\,\%.  However, we have argued in the preceding section that the surveyed volume should only be corrected by a factor of roughly 1.3, leading to an estimated frequency of $2.5 \pm 1$\%, substantially smaller than predicted from a volume-limited survey. This result supports, with better statistics, our contention that WDs with ages less than 0.5\,Gyr are significantly less likely to show magnetic fields than older WDs. 

Our interpretation of the fact that magnitude-limited surveys of hot stars find fewer MWDs than volume-limited surveys is that this is not the result of a significant selection effect against higher-mass MWDs, but of a selection effect in favour of younger WDs, among which we believe that frequency of MWDs is substantially depressed from the general average. The striking difference between the frequency of MWDs among younger and older WDs probably reflects the action of the mechanisms that produce magnetic fields in WDs.

\subsubsection{Future work: investigating a larger volume-limited sample}\label{Sect_Outside_Age}
The clearest way to confirm or reject the idea that a deficiency of magnetic fields  in young WDs is a general property of WDs and not limited to the \lv, is to extend our investigation to a larger volume limited sample.
There is no complete spectropolarimetric dataset for any volume larger than the \lv, but it appears that the 40\,pc volume, which has about 8 times the number of WDs as the \lv, is likely to become the next volume of wide interest \citep{Treetal20,McCetal20}. We look ahead to the possibilities offered by this volume for testing our proposal. 

The northern hemisphere half of the 40\,pc volume (including the \lv) has about 81 WDs with ages of less than 500\,Myr \citep{McCetal20}, corresponding to a total population of approximately 162 such stars. These young stars have been fairly intensely studied because of their brightness (the faintest WDs of this young age group in the 40\,pc volume generally $V$ or $G$ brighter than 15\,mag). At least 90\% are DA stars, in which fields of a MG or more are readily identified in good classification spectra, and fields ten times smaller may be identified with high resolution spectroscopy. In addition, our own surveys have explored this sample rather extensively. Our estimate is that we and others have obtained spectropolarimetric observations on roughly half of the  MWDs of ages less than 0.5\,Gyr in this volume, and a very large fraction have at least been examined with low- to mid-resolution spectroscopy \citep[see, e.g.][]{Treetal20,McCetal20,Napetal20}. 

Within the 40\,pc\ volume we know of nine MWDs younger than 0.5\,Gyr (see Table~\ref{Tab_Young}, which includes WD\,2047$+$372, the only young MWD in the \lv, and WD\,0232$+$525, a new MWDs recently detected by us). It would be outside of the scope of this paper to report on a partial analysis of this sample, but as a very preliminary result we note that at present only about 6\,\% of the young sample are known to have magnetic fields. Even if further searches double the number of young MWDs in the volume, which because of previous searches seems unlikely to us, the magnetic fraction will still be only about half of the average frequency of MWDs that we have clearly shown to apply to the \lv. Thus we predict that the result of this larger future survey will still support the conclusion that there is a dearth of MWDs among the youngest WDs. 

\begin{table}
\caption{\label{Tab_Young} List of known MWDs younger than 0.5\,Gyr within 40\,pc from the Sun.}
\begin{center}
\tabcolsep=0.14cm
\begin{tabular}{lcrlc}
\hline\hline
     & \bs\ & $M$          &            &         \\
Star & (MG) & ($M_\odot$)  & References & Notes   \\
\hline
WD\,0041$-$102 & 20            & 1.1 & \citet{Liebetal77}&  \\
WD\,0232$+$525 & 0.005:        & 0.82& unpublished & \\
WD\,0301$+$059 & 200           & 1.12& \citet{LanBag20}& \\
WD\,0316$-$849 & 300           & 0.86& \citet{Baretal95} & 1\\
WD\,0945$+$245 & 670           & 0.80& \citet{Lieetal93}&\\
WD\,1105$-$048 & 0.010:        & 0.58& \citet{Aznetal04}&\\
WD\,1105$-$340 & 0.150         & 0.67& \citet{LanBag19a}&\\
WD\,1658$+$440 & 2.3           & 1.32& \citet{Schetal92WD}&\\
WD\,2047$+$372 & 0.06          & 0.82& \citet{Lanetal17}&\\
\hline
\end{tabular}
\end{center}
\begin{small}
\noindent
1. Star WD\,0316$-$849 = RE~J0317$-$853 =  EUVE~J0317$-$853 = V*~CL~Oct is often confused with star WD\,0325$-$857 = EQ~J0317-855 = LB~9802, a hot and young non magnetic WD \citep[][Table~1]{Kawetal07}. 
\bigskip

\end{small}
\end{table}

\subsubsection{The higher incidence of magnetic fields in certain class of stars may just reflect a higher incidence of magnetic fields in older/cooler WDs}
That DZ stars might exhibit a frequency of the occurrence of magnetic fields higher than average would naturally be explained in the scenario of magnetic fields that become more frequent as cooling age increases; weaker fields would be present in many cool (hence old) H and He-rich WDs, but would be detected only if the stellar spectrum exhibit metal lines, while the field would pass unnoticed in featureless or nearly featureless WDs. This interpretation is consistent also with the considerations about DAZ stars of Sect.~\ref{Sect_DA_DAZ}, where we noted that the frequency of magnetic DAZ WDs is actually similar to the frequency of magnetic DA WDs, once we compare the situation in the same temperature ranges.

Admittedly, it is possible to speculate that the magnetic field may be originated during disk debris accretion, and its presence is longer lasting than that of metal lines in the stellar spectrum. Some of the cooler magnetic DA WDs could be former magnetic DAZ WDs. However, that temperature and age are important discriminating factors is supported by the fact magnetic fields are much more frequent in DZ and DAZ WDs cooler than $7500-8000$\,K than in hotter WDs of the same spectral class \citep{Holletal15,Kawetal19}.

We have also seen in Sect.~\ref{Sect_Comp} that the frequency of the occurrence of magnetic fields in He-rich WDs ($30 \pm 7$\,\%) is higher than in H-rich WDs ($19 \pm 4$\,\%), and we have noted that in the \lv\ there are no He-rich WDs younger than 0.5\,Gyr. The higher frequency of He-rich MWDs may be explained again as an effect of temperature/age, although we must keep in mind that in the \lv\ we have observed a specially high frequency of MWDs among the oldest He-rich stars. In the cooling age range 0.5 to 5\,Gyr, the frequency of the occurrence of magnetic field in H-rich and He-rich WDs are very similar: $25.9 \pm 5.7$ and $22.9 \pm 7.1$\,\%, respectively.

\subsubsection{Do older MWDs have stronger fields than younger MWDs?}\label{Sect_Young_Strength}
\begin{figure}
\centering
\includegraphics[width=8.6cm,trim={0.7cm 2.1cm 1.0cm 1.0cm},clip]{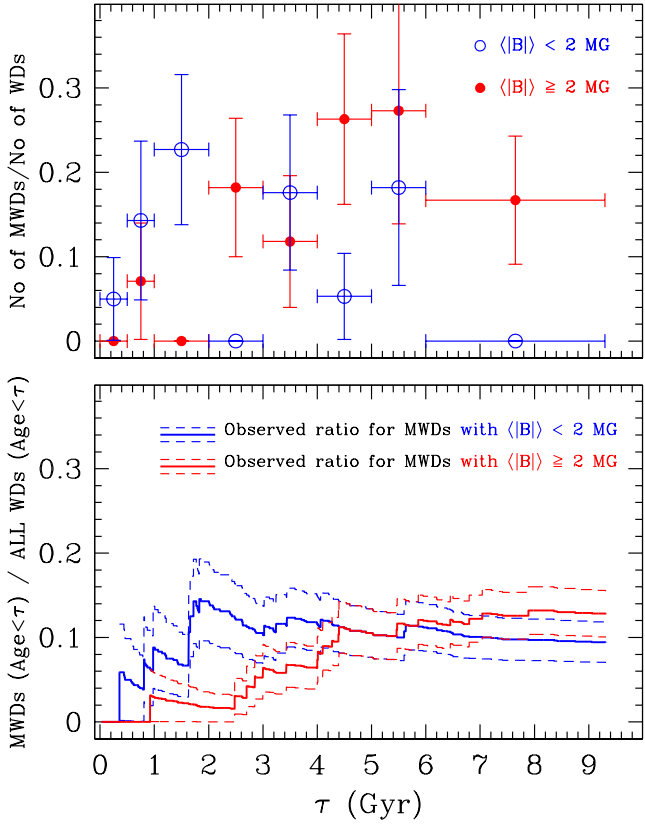}
      \caption{\label{Fig_CumAgeB} 
Same as Fig.~\ref{Fig_CumAge}, but for MWDs with fields weaker than 2\,MG (blue symbols) and stronger than 2\,MG (red symbols). Because fields weaker than $1-2$\,MG cannot be detected in stars older than a few Gyr (unless there are metal lines in the stellar spectra), the zero frequency of weakly magnetic WDs at $\tau=7.5$\,Gyr may well be due to an observational bias.
} 
\end{figure}
It is very important to check next whether young MWDs have different features than older MWDs, apart from being more rare. We note that in the \lv, fields with strength higher than 2-5\,MG are more common in older than in younger MWDs.  The most notable exception, of a relatively young (0.92\,Gyr) WD with a very strong field (and with a higher than average mass, $0.93\,M_\odot$) is Grw\,$+70^\circ\,8247$. The next youngest high-field stars  in the \lv\ all have cooling ages older than 2.5\,Gyr. Figure~\ref{Fig_CumAgeB} shows the same quantities as in Fig.~\ref{Fig_CumAge}, but separately for MWDs with fields weaker than 2\,MG and for MWDs with fields stronger than 2\,MG, and demonstrates in a convincing way that in the \lv\ there is a pronounced dearth of higher-field MWDs younger than 2--3\,Gyr.

This issue can be further investigated only by mean of a complete survey of a larger volume sample of WDs. With the data available we are not able to investigate this finding beyond the 20\,pc volume with unbiased statistics, therefore we will not consider this characteristic as an observational constraint. We also note that \citet{WicFer00} had concluded that there is no evidence that field strength depends on temperature. 

\subsection{The frequency of the field occurrence versus stellar mass}\label{Sect_Mass}
\begin{figure}
\centering
\includegraphics[width=8.6cm,trim={0.7cm 2.1cm 1.0cm 1.0cm},clip]{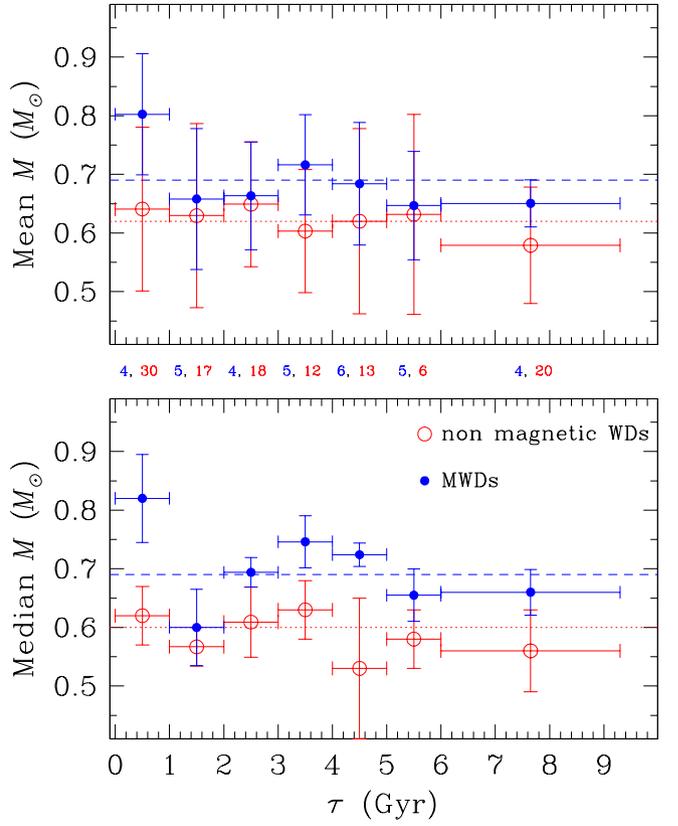}
   \caption{\label{Fig_Masses} Mean (top panel) and median values (bottom panel) of the mass of non magnetic (red empty circles) and magnetic (blue filled circles) WDs as a function of cooling age, binned in the intervals of time represented by the horizontal errorbars. Vertical errorbars represent the standard deviation of the mean values, and the semi interquartile range for the median values, calculated for each time bin. The horizontal blue dashed (red dotted) lines represent the mean (top panel) and median (bottom panel) for the magnetic (non magnetic) WDs. The numbers $M$, $N$ between the two panels show the number $M$ of MWDs and the number $N$ of  non magnetic WDs in each time interval bin.
} 
\end{figure}
Figure~\ref{Fig_Histos} shows no obvious correlation between field strength and stellar mass, but it is clearly apparent that magnetic and non magnetic WDs have different mass distributions. This is not a surprise, as \citet{Liebert88} had already suggested that MWDs are more massive than average, and this conjecture has been confirmed by several subsequent studies, the most recent ones being those of \citet{Kepetal13} and \citet{McCetal20}. We note that there are no WDs with $M > 1\,M_\odot$ in the \lv, therefore we cannot explore the full parameter space.

\subsubsection{The average mass of MWDs is higher than the average mass of non MWDs}
We recall that the mean mass of all WDs of the local 20\,pc volume is $0.64 \pm 0.13\,M_\odot$. The mean mass of non magnetic WDs is $0.62\,M_\odot$, with a standard deviation of $0.13\,M_\odot$, and the mean mass of MWDs is $0.69\,M_\odot$, with a standard deviation of $0.10\,M_\odot$. Marginal evidence for correlation with age will be discussed in Sect.~\ref{Sect_Young_Mass}.

\subsubsection{Mean mass of MWDs and field strength}
Formally, the field strength of the MWDs of the \lv\ seems independent of stellar mass: the mean mass is $0.69 \pm 0.10\,M_\odot$ (standard deviation) whether we consider fields weaker than 2\,MG, or fields stronger than 2\,MG. However, things might be different in certain age bins (see Sect.~\ref{Sect_Massage}).

Our results may be compared with the statistics gathered by \citet{Feretal15}, mainly on the basis of MWDs discovered via low-resolution spectroscopy  (hence highly biased in favour of MWDs with field strength between 2 and 80\,MG).  \citet{Feretal15} found that the mean mass of high-field MWDs is $0.78 \pm 0.05\,M_\odot$, but they found also that the average mass of more weakly magnetic WDs is consistent with that of non-magnetic WDs, a feature that we do not see in the \lv. Based on a 40\,pc volume-limited sample of northern WDs (but probably still biased in favour of strongly magnetic WDs, because field detections in this sample have relied heavily on spectroscopic data), \citet{McCetal20} found that the mean mass of MWDs is $0.75 \pm 0.05\,M_\odot$.

\subsubsection{Does the average mass of WDs and MWDs depend on age?}\label{Sect_Young_Mass}
In the \lv, the average mass of the MWDs older than 2\,Gyr is $0.67 \pm 0.08\,M_\odot$, while that of MWDs younger than 2\,Gyr is $0.72 \pm 0.13\,M_\odot$. The mean mass of MWDs younger than 1\,Gyr is $0.80 \pm 0.10\,M_\odot$ (all standard deviations, not standard errors).  Figure~\ref{Fig_Masses} shows the mean and median masses of magnetic and non magnetic WDs as a function of cooling age.  There is marginal, not fully convincing evidence that younger MWDs are more massive than older MWDs.

\subsubsection{What do the observations outside the \lv\ tell us about the mass of MWDs versus their age?}\label{Sect_Massage}
Although the result that younger MWDs would appear not only rarer, but also more massive than the average MWD, is very suggestive, it cannot be extended with very strong statistical support, because the observational sample in the \lv\ is too small. There are only nine MWDs younger than 2\,Gyr, only four younger than 1\,Gyr, and only one younger than 0.5\,Gyr, the star WD\,2047$+$372, which has a relatively weak field, and $M \sim 0.82\,M_\odot$. Is it possible to gather information outside of the \lv? We have already noted that the targets of the P-G survey were mostly young WDs, and Table~1 of \citet{Lieetal03} shows that the MWDs found in that survey have mass $\ga 0.75\,M_\odot$ and field strength $\ga 1$\,MG, with the exception of the 2.3\,MG star PG\,2329+267, which is listed with $M \simeq 0.61\,M_\odot$. However, a revision  based on the catalogue of \citet{Genetal19} of the parameters of the same list shows that all MWDs have $M > 0.75\,M_\odot$, and PG\,2329+267 has $M = 0.91\,M_\odot$. If we turn our attention to the part of the local 40\,pc volume of Sect.~\ref{Sect_Outside_Age} (see in particular Table~\ref{Tab_Young}), we find that younger MWDs are more massive than the average MWD. However, we note that MWDs younger than 0.5\,Gyr with low mass do exist. Table~\ref{Tab_Young} includes WD\,1105$-$048 ($M \simeq 0.58\,M_\odot$, $\bs \simeq 10$\,kG) and WD\,1105$-$340 ($M=0.67\,M_\odot$, $\bs \simeq 150$\,kG). Outside the 40\,pc volume we know of WD\,0446$-$789 ($M \simeq 0.55\,M_\odot$), a very young MWDs already mentioned in Sect.~\ref{Sect_40kG}, with an extremely weak field. Unfortunately, there is not sufficient information, especially on older WDs, to reach any firm conclusion. We suggest, as a working hypothesis for further investigation, that among the (relatively rare) young MWDs, strong fields are found mainly or only in WDs more massive than the average MWD. Lower-mass young MWDs generally exhibit weak, possibly extremely weak fields.

\subsubsection{Summary}
While all these results are consistent with the conjecture that MWDs are more massive than non-magnetic WDs, first proposed by \citet{Liebert88}, it seems that in the \lv\ we have found comparatively more low-mass MWDs than in other investigations, possibly because we have given a more appropriate statistical weight to WDs of various ages. In contrast to what found in magnitude-limited surveys, from which stronger magnetic fields seem to be associated to higher mass WDs, in the \lv, MWDs with fields stronger than 2\,MG have average mass similar to weaker field MWDs. However, MWDs younger than 2\,Gyr are marginally more massive than older MWDs. After also looking at data available for stars outside of the \lv, we suggest that young MWDs are rare, and among these young MWDs, stronger fields (of the order of 1\,MG or more) are found mainly in higher-mass MWDs; lower-mass young MWDs do exist, but seem to have weaker fields.

\subsection{The frequency of MWDs in binary systems}
There are 35 binary systems, including wide binary systems that comprise a WD and a main-sequence (MS) star, DD and uDD systems. Five of these systems (three CPM and two uDD systems) include a MWD, for a frequency of $14 \pm 6$\,\%, marginally lower than in single WDs. We note, however, that some of these systems include a DC WD and that the polarisation signal of a MWD in an unresolved system would be diluted by the radiation of the companion, so that this frequency may actually be underestimated.

\section{Discussion}\label{Sect_Discussion}
In this section we begin by summarising the observational constraints that we have gathered from our volume-limited survey of WDs (Sect.~\ref{Sect_Constraints}). Then (Sect.~\ref{Sect_Origin}) we consider the relevance of our conclusions to two kinds of broader questions. First, we will examine the extent to which our findings appear to support or contradict various theories of how the magnetic fields of MWDs are created in an previous evolutionary phase and retained in the cooling phase, or possibly generated in close binary systems. Secondly, we address the question of possible further field generation that may occur while WDs are in the cooling phase. We then consider the apparent lack of evidence of Ohmic decay (Sect.~\ref{Sect_Evolution}).

\subsection{Observational constraints}\label{Sect_Constraints}
The main characteristics of the WDs of the \lv\ can be summarised as follows.

\subsubsection{Overall incidence of magnetic fields}
About 20-25\,\% of WDs have a magnetic field. The number of WDs in the \lv\ is large enough to provide strong statistical support to this result, and we may assume it is approximately valid in the Galaxy. The presence of a magnetic field does not seem correlated with the spectral class, except maybe for DQ WDs, in which the magnetic field seems either stronger or more frequent than average. Furthermore, there are hints that DZ WDs may be more frequently magnetic than other WDs. However this result could simply be due to the fact that magnetic fields seem more frequent in older than in younger WDs (see Sect.~\ref{Sect_Age}).

\subsubsection{Behaviour with age}\label{Sect_Age} 
There is a dearth of MWDs younger than 0.5\,Gyr. This result seems to be supported by 
the outcome of previous spectroscopic surveys, outside of the \lv. In fact, there is some evidence that the frequency of magnetism in WDs rises with time until cooling times of the order of 5\,Gyr (upper panel of Fig.\,\ref{Fig_CumAge}).  Beyond that age, the frequency drops, possibly because we are not able to detect weak fields in older stars. There is no strong evidence that field strength decays with time. 

\subsubsection{Field strength distribution}
The field strength is uniformly distributed (with roughly equal probability of finding fields per dex of field strength) over four decades, from 40\,kG to 300\,MG. It appears that these field values limit approximately the range of field strengths frequently found in WDs. Outside of the \lv, we are aware of only three known MWDs with  a field strength marginally lower than the lower threshold, and of a few MWDs above the general upper limit, with field strengths of several hundreds MG. 

\subsubsection{Correlations between magnetic field and mass}
MWDs are more massive than non-magnetic WDs. This result is fully consistent with previous studies not confined to the \lv, which are probably biased in favour of younger and stronger field MWDs. In fact, there is some marginal evidence that younger MWDs are more massive than older MWDs. There are no WDs with $M > 1\,M_\odot$ in the \lv, which represents a limitation of our analysis of the parameter space. Data from outside of the \lv\ show marginal evidence that young MWDs with strong fields are more massive than young MWDs with weak fields, but data from the \lv\ show that there exist several old MWDs with low mass and strong fields.

\subsubsection{Magnetic fields and binarity}
There are no strong indications that the overall frequency of MWDs in binary systems is different than in single WDs. However, it should be noted that in the \lv\ there are no known close (spectroscopic) binary systems including a dM and a WD, so this result is not inconsistent with the finding by \citet{Lieetal05} that MWDs in close (but not interacting) systems with a non-degenerate companion are very rare.

\subsection{Field origin and observational constraints}\label{Sect_Origin}
In the following we discuss how various major hypotheses concerning the origin of MWD fields stand up against the observational constraints of Sect.~\ref{Sect_Constraints}.

\subsubsection{The fossil field hypothesis}
The possibility of some level of magnetic flux conservation during evolution from the main sequence to a final compact state was first discussed by \citet{Wolt64}, applied to WDs by \citet{Lan67}, and discussed more fully by \citet{Angetal81}. Flux conservation suggests that magnetic flux might be roughly conserved as a star evolves from a main sequence structure to a WD structure two orders of magnitude smaller in radius (the fossil field hypothesis). Such a transformation could convert the main sequence fields of the order of $10^4$\,G found in Ap and Bp main sequence stars \citep[e.g.][]{DonLan09} to fields of roughly $10^8$\,G = 100\,MG. 

It is now very clear that the strongly magnetic Ap and Bp main sequence stars are not sufficiently frequent in space to account for the high frequency of MWDs, as already pointed out by \citet{Kawetal07}. Strong fields are found in about 8\% of A and B main sequence stars \citep{Powetal08}. During the first one or two Gyr of Milky Way evolution, these stars, because of their relatively rapid evolution,  provided most of the WDs. If flux conservation from Ap and Bp stars provided all the magnetic fields of the oldest MWDs, we would expect no more than about 8\% of oldest WDs to be magnetic (assuming that the fraction of Ap and Bp stars has been constant with time). More recently, many WDs have also formed from main sequence F stars, which require a few Gyr to end their nuclear-burning lives. These lower mass (ca. $1.2 - 1.6 M_\odot$)  progenitors have almost no fossil magnetic fields, so WD fields formed by flux conservation would be expected to make up a substantially smaller fraction than 8\% of the younger WDs, well below the observed frequency of more than 20\%.  Although this might be a minor MWD formation channel, it is clearly not the most important process. 

\subsubsection{Origin of the fields as a result of deep-seated convective dynamo action}\label{Sect_Seated}
Another potential source of magnetic fields inherited from pre-WD evolution arises from the possibility of dynamo action in the interior of a star either in the main sequence or in the AGB. This dynamo would leave behind a remnant magnetic field that is retained in the stellar interior as a fossil field, possibly compressed and amplified during later evolution, and finally revealed after much mass loss in the newly formed WD. 

It appears that dynamo-driven magnetic fields may originate in the convective cores of main sequence A or B stars (which are unrelated to the surface magnetic fields of the Ap and Bp stars). A series of papers summarised by \citet{Steetal16} discusses how a dynamo-driven field produced in the strongly convective core during main sequence evolution of an intermediate-mass star \citep[e.g.][]{BraSpr04,Bruetal05} can leave a strong, stable field \citep{Dueetal10} which can survive into the red giant phase, where it could be detected by its effect on the oscillation spectrum. Such fields would then be presumably amplified by flux compression during the final collapse of the star into WD. It is thought that this kind of mechanism can produce MG fields, that slowly diffuse outward to the surface of the WD while decaying by Ohmic losses, on a timescale of order 1\,Gyr. 

Field generation based on a dynamo operating in the region outside the core in a star powered by nuclear fusion in shells, possibly stimulated by the angular momentum added by ingestion of a planet, is discussed by \citet{KisTho15}. The resulting field can reach 10\,MG. 

Dynamos operating in single WD progenitors of various masses and rotation rates (or in various kinds of interacting binary systems, see Sect.~\ref{Sect_Merger}) and leaving behind internal fields in the evolving stars, could occur over a wide range of initial stellar parameters. Such dynamos may be capable of generating both the range of field strengths and the high frequency of fields found in white dwarfs, although further numerical explorations are needed to more fully understand the capabilities of these models. The observed dearth of young MWDs compared to the frequency of fields in older WDs may be explained by slow emergence of such internal fields to the stellar surface.  

It is worth noting the recent detection claimed by \citet{Caietal20} of magnetic fields in three young WDs that are members of open clusters, in which the mass and age of the progenitor of each can be determined fairly accurately. All three WDs have quite high masses (around $1\,M_\odot$) and progenitor masses of around 5 or $6\,M_\odot$. \citet{Caietal20} argue that these new MWDs are so young that it is very unlikely that they formed from binary merger processes discussed in Sect.~\ref{Sect_Merger} below; instead they are almost certainly descended from single upper main sequence stars. These discoveries, if fully confirmed, appear to require that at least some MWDs descend from single main sequence stars. 

\subsubsection{Field generation during merger in a close binary}\label{Sect_Merger}
Cataclysmic variables (CVs) are close binary systems in which a non-degenerate star is transferring mass onto a very close WD companion. In about 25\% of such systems the WD companion has a MG magnetic field. CVs form from binary systems containing one star which has already become a WD, in which the separation is small enough that when the non-degenerate star starts to evolve into the giant, the WD is engulfed in the expanding envelope. If the WD survives this common envelope event without merging with the core of the red giant, the system becomes a CV.  However, an extensive search  by \citet{Lieetal05}, failed to discover any MWDs among the pre-CV systems, which are close binaries with a main sequence star and a WD. 

Following this result, \citet{Touetal08} proposed that the available orbital energy and angular momentum could lead to the creation of a strong magnetic field via dynamo action during the common envelope phase. The predicted outcome of the process is that if merging occurs, the result would be a single, high-mass high-field MWD, while those systems that finally do not quite merge would become magnetic CVs. Later, \citet{Brietal15,Brietal18Isolated,Brietal18CV} carried out synthetic evolution calculations of binary star populations, using simple parameterised models of the common envelope phase and of dynamo action during the common envelope event, to show that this mechanism might well lead to generation of strong fields, up to the upper field limit observed in WDs. In particular, they found parameter choices that enabled the population synthesis field strength predictions to reproduce the distribution of magnetic field strength depicted by \citet{Feretal15}, according to which most of MWDs have a magnetic fields strength between 2 and 80\,MG. 

A probable example of MWD creation from a binary pathway may be provided by the magnetic WD\,0316--849 = RE\,J0317--853 \citep{Baretal95}, which is not within the \lv. This extremely hot, "young" and massive DAH WD, with a field of about 500\,MG and a rotation period of 725\,s  has a common proper motion WD companion that has a much older total age. The two WDs were presumably created together in a wide binary system. The very discordant ages may be understood if WD\,0316--849 is the result of a recent merger, which presumably also led to the creation of the huge magnetic field \citep{Feretal97}.

If we consider the possibility that some version of common envelope evolution followed by merger is the dominant mechanism for formation of MWDs, then we would expect a high frequency in space of initially close binary systems that will evolve into single MWDs, a frequency close to the upper limit of systems that will finally merge into single WDs \citep{Tooetal17}.  However, it is not yet clear that such systems are common enough to produce more than 20\,\% of all WDs. Furthermore, if the mechanism operates as described by \citet{Brietal15,Brietal18Isolated}, it is predicted to produce mainly fields with a rather limited range of field strengths; by contrast, the predicted distribution is not at all similar to the observed roughly constant frequency per dex of field strength in the \lv. 
We also note that  \citet{BelSch20} have recently discussed the consequences of these models for a more comprehensive range of CV and CV-like binary situations. \citet{BelSch20} argue that the basic model predictions have important conflicts with observations of CVs and pre-CVs, and that the original mechanism needs to be substantially revised. However, they offer several suggestions for changes that would bring this model of field generation into better agreement with observations. In conclusion, it appears to us that the merging mechanism requires further development to be shown to be a major contributor to the observed population of MWDs.

\subsubsection{Dynamo action during WD core crystallisation}\label{Sect_Isern}
\begin{figure*}
\centering
\includegraphics[angle=270,width=18.0cm,trim={5.1cm 2.0cm 1.9cm 0.9cm},clip]{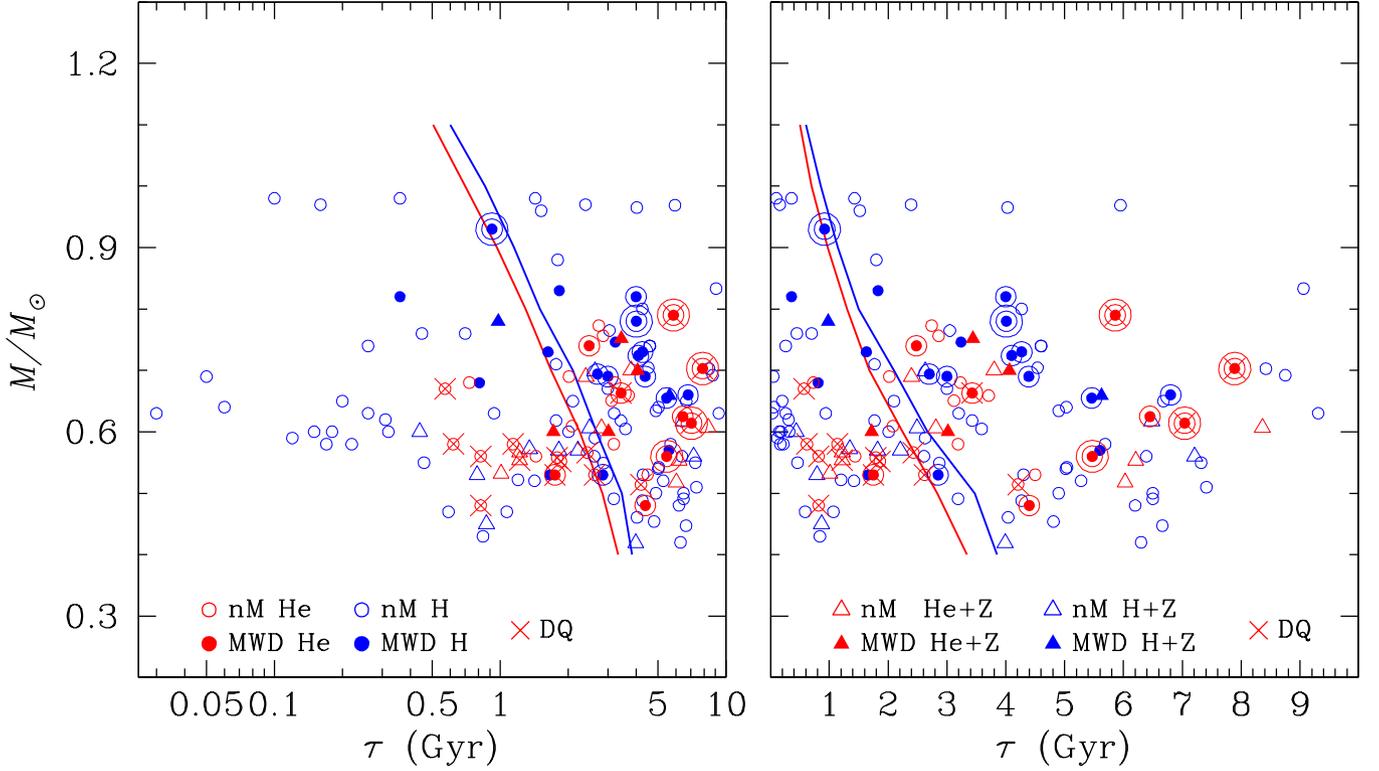}
      \caption{\label{Fig_Crystal} 
      Cooling age -- mass diagram for magnetic (solid symbols) and non magnetic (empty symbols) WDs, showing the boundary at which crystallisation convection begins as a WD cools with age, as described by \citet{Iseetal17}, for stars with "thick" (blue solid lines) and "thin" (red solid lines) hydrogen layers \citep[from][]{Badetal20}. Stars for which we have not determined atmospheric composition have been arbitrarily considered as H-rich. The same plot is shown with cooling age in logarithmic scale (left panel) and linear scale (right panel). Circles represent DA and and DC stars, while triangles represents stars with metal lines (DAZ, DZ, DZA). DQ stars are represented with crosses over the symbols. Solid symbols surrounded by one circle represent MWDs with field strength between 1 and 100\,MG, and solid symbols surrounded by two circles represent MWDs with field strength $>100$\,MG.
      } 
\end{figure*}
The magnetic fields of some WDs could be slowly generated during a time of the order of 1\,Gyr by some internal physical mechanism during the cooling of the white dwarf itself. \citet{ValFab99} suggested that electric conductivity in WDs could increase with time, or that magnetic field diffuses from the inner region of the star.  \citet{Iseetal17} have shown that during crystallisation in the C-O core of a typical WD, separation and sinking of the solidifying O component leads to strong convection in the core. Provided that the WD is rotating rapidly, a dynamo of the same type that produces the magnetic fields of Earth, Jupiter, and M dwarfs can operate. As in those objects, this dynamo requires rapid rotation to function in a saturated state (i.e. to produce the largest fields of which the mechanism is capable). In WDs, a rotation period of less than about $10^3$\,s is sufficient to ensure saturation. A straight-forward extrapolation of the achievable field strength versus convective energy density suggests that this mechanism could produce global fields only up to about 1\,MG in strength. However, \citet{Schetal21} have suggested that taking account of the probable dependence of the dynamo efficiency on the magnetic Prandtl number leads to the expectation that much larger fields than 1\,MG could be generated. 

Figure~\ref{Fig_Crystal} shows the cooling age -- mass diagram for magnetic and non magnetic WDs, compared to the lower limit of the age for different masses at which crystallisation convection begins, as calculated by \citet{Badetal20} for WDs with thick ($q_{\rm He}=10^{-2}$, $q_{\rm H}=10^{-4}$) and thin  ($q_{\rm He}=10^{-2}$, $q_{\rm H}=10^{-4}$) hydrogen layer, where $q_X = M_X/M_*$. It appears that the frequency of the occurrence of magnetic fields after crystallisation has begun, $29 \pm 5$\,\%, is twice the frequency in WDs with a fully liquid core, $14 \pm 4$\,\% (these frequencies are estimated assuming that the three MWDs between red and blue lines have not reached yet in the crystallisation sequence). Figure~\ref{Fig_Crystal} suggests that a dynamo linked to the crystallisation could be a primary mechanism that leads to the generation of magnetic fields in degenerate stars. However, we recall that the operation of this mechanism requires (1) a large increase by an unknown “scaling law” factor of the fields estimated by \citet{Iseetal17} to reach the observed field strengths, and (2) rapid rotation (periods of minutes or hours) to allow the crystallisation dynamo to operate at peak output. It is not known at present whether either of these conditions is satisfied.  Thus, although the crystallisation dynamo may explain the general rise in frequency of magnetic fields already noted in Figure~\ref{Fig_CumAge}, this possibility needs further study before being accepted as an important mechanism. 

\subsubsection{Mass differences between magnetic and non-magnetic WDs}
A clear feature of the MWDs of the \lv\ is that their average mass is about $0.07\,M_\odot$ larger than the mean for the non-magnetic WDs, although there is substantial overlap between the two distributions. 

A straight-forward interpretation of the overall excess mass of MWDs is that they may be formed preferentially from massive progenitors. This would be consistent with the hypothesis that MWD fields descend from single (magnetic and non-magnetic) A and B main sequence stars, or anyway from high-mass progenitors. This observation would also be consistent with the possibility that MWDs are produced in binary mergers, which will tend to lead to more massive than average WDs \citep{Brietal15}. If most MWD fields are generated earlier in the evolution of the progenitor star by dynamo action, it is not unreasonable to suppose that the field creation process should depend fairly strongly on main sequence progenitor mass, and thus lead to rather different outcomes in WDs of different masses.

A particularly interesting aspect of the mass distribution of the MWDs of the \lv\ is that MWDs with cooling ages below about 1--2\,Gyr appear to be still significantly more massive than the average MWD mass. This constraint fits into the scenario in which at least some MWDs develop their magnetic field after crystallisation: higher-mass WDs crystallise sooner (even much sooner) than lower mass WDs, so there must be more MWDs among younger higher-mass WDs than among younger lower-mass WDs. After a certain amount of time, all WDs have gone through the process of crystallisation and perhaps the residual difference in mass between MWD and non-magnetic WDs is due to the fact that WDs with higher mass progenitors are most likely to be magnetic than WDs with lower mass progenitors and/or the result of binary mergers.

Finally, the observation that among the youngest WDs, the strongest fields are found in WDs with mass substantially higher than average, while in older stars, strong magnetic fields are found also among WD with average mass, needs to be verified with additional data. If confirmed, it would suggest again that the more massive MWDs are visibly magnetic almost immediately after formation, while the less massive MWDs require 1--2\,Gyr to reveal their fields.

\subsection{Evolution of the magnetic field during the WD cooling phase}\label{Sect_Evolution}
The apparent deficiency of magnetic fields in the youngest WDs is consistent with different scenarios: a fossil field that emerges slowly from the interior to the surface by a diffusion process, and which can have a characteristic time scale as long as 1--2\,Gyr; or a dynamo generated field that forms when crystallisation begins. At some stage, though, all field generation mechanisms discussed above, including dynamo acting during crystallisation, must cease operation. The fossil theory suggests that any fields produced by earlier active mechanisms should evolve simply by diffusion and Ohmic decay. In this case the time scale for surface field strength growth (by field diffusing to the surface) or field decline (because of resistive losses to the supporting currents) could be as slow as the global decay time, which has been estimated to range from 0.6 to 4\,Gyr for WD masses between $0.4 M_\odot$ to $1 M_\odot$ \citep{Fonetal73}. 

Because we now have a record of field occurrence and strength in WDs as old as 5--8\,Gyr, at which ages we find, in fact, the strongest magnetic fields in the \lv\ sample, with only a modest decline in the frequency of field occurrence and no significant change in the field strength distribution (see Figs\,\ref{Fig_Histos} and \ref{Fig_CumAge}), it does not appear that simple Ohmic decay dominates evolution during the WD cooling phase. It is quite possible that internal field reorganisation of large, invisible internal field structures (initially possibly of several thousand MG in strength?) replace surface magnetic flux that is lost by Ohmic decay, or that an internal dynamo like the one discussed in Sect.~\ref{Sect_Isern} is acting until late stages of the cooling phase. We are generally not sensitive to weak fields in older stars, but a highly-sensitive survey of broadband circular polarisation aimed at probing Ohmic decay in cooler WDs could help to better understand the evolution of the magnetic fields in WDs.

\section{Conclusions}\label{Sect_Conclusions}
We have carried out a spectropolarimetric survey of WDs that, complemented with literature data, has made it possible for us to compile an almost complete database of highly-sensitive magnetic field measurements of the local 20\,pc WD population.  This volume-limited sample contains approximately 152 known WDs (including secondaries from a somewhat uncertain number of unresolved double degenerate stars). In the course of our survey, we have observed 70 stars that have never been observed in spectropolarimetric mode before, and another 20 that had been already observed with polarimetric techniques, but with typically 1 dex lower precision than we have achieved with our new measurements. In the course of our survey of the \lv\ we have discovered 12 new MWDs, that is, more than a third of all known MWDs  of the \lv. Combining our new data with those from previous literature, field strength estimates and upper limits are now available for 149 WDs of the \lv. This sample is the best current approximation of a statistically unbiased observational database of magnetic white dwarfs, and some of its characteristics may well be representative of the entire Galactic population of WDs, or at least of WDs formed at several kpc from the Galactic Centre. 

We have found that at least 33 of the observed 149 WDs have magnetic fields. 
We have also used the results of this survey to explore and test proposals in the literature concerning abnormally high or low frequencies of magnetic fields in various spectral classes of WDs, and to check if the frequency of the occurrence of MWDs correlates with cooling age and mass.

We found that among the commonest class, the DA WDs, $23 \pm 5$\% of the stars have detected magnetic fields. This figure serves as a reference frequency. 

The incidence of magnetic fields in DC stars is $13 \pm 6$\%, approximately half that of WDs of spectral class DA, but this is likely an artefact due to the fact that in featureless WDs we can detect only magnetic fields with a strength $\ga 1$\,MG. Since we know that about half of the magnetic DAs have fields that do not produce a significant polarisation in the continuum, it is quite possible that the incidence of magnetism in DCs is actually underestimated by a factor of order two, while being effectively similar to that found for DA WDs. 

There are claims in the literature that magnetic fields are more frequent in WDs with metal lines, especially in the cooler ones. Indeed, combining our data with those of \citet{Kawetal19}, we found that the frequency of MWDs among DAZ cooler than 6000\,K is $33 \pm 14$\%, and the same result is obtained if we consider the DZ and DZAs WDs of the \lv. It has been suggested that this could be explained by the formation of a magnetic field during the accretion of rocky debris \citep{Faretal11,KawVen14}. However, we have shown that this statistics is fully consistent with the frequency of the DA MWDs in the similar age/temperature range, therefore we suggest that the high incidence of magnetic fields observed in cooler stars with metal lines simply reflects the fact that magnetic fields are more frequent in cooler/older WDs than in hotter/younger WDs, as it will be best summarised later. In particular, the presence of metal lines in a spectrum otherwise featureless like that of DZ WDs allows us to detect weak magnetic fields that might be pretty common, but undetected, in DC stars.

The \lv\ seems richer in DQ stars than what expected from the estimate of their occurrence (a few percent of the WDs), as more than one out of ten of the local population of WDs belong to this spectral class.  The frequency of fields among the 16 cool DQ and DQpec WDs is about $31 \pm 12$\%.  Taken at a face value, this frequency is not significantly higher than that of magnetic DAs. However, only fields stronger than average may be detected in DQ WDs. If the field strength distribution found in DA WDs applies also to DQ WDs, then DQs stars of the \lv\ could be characterised either by a much higher frequency of magnetic fields than average, or by the presence of magnetic field smuch stronger than average. In either case, we could speculate that the link between magnetic field and presence of C$_2$ bands in a He-rich atmosphere could be due to C$_2$ buoyancy induced by the magnetic field. In some respect, DQ WDs could be the equivalent of chemically peculiar stars of the upper mean sequence. The immediate next step will be to check a much wider sample of DQs for the presence of magnetic field.

An important result of our analysis is that field-strength frequency distribution is approximately constant per decade over a range extending from a few tens kG up to several hundreds MGs. This is in very sharp contrast to the field frequency distribution found in the overall sample of known MWDs, which (because of the impact of SDSS on new discoveries) exhibits a very strong frequency spike between about 2 and 80\,MG. Furthermore, because of the extreme sensitivity of many of our measurements which have nevertheless revealed almost no WD fields below $\bs \approx 40$\,kG, we argue that we may have detected the effective lower limit of the field strength frequency distribution. That is, we suggest that WD fields occur essentially in the range of 40\,kG to several hundred MG. 

We found that in the local 20\,pc volume, the frequency of the occurrence of the magnetic field in young stars is significantly lower than average (only one out of 20 WDs younger than 0.5\,Gyr is magnetic). A qualitatively similar result had been obtained in the past, but called into question as possible effect of an observational bias against high-mass WDs. With the help of theoretical models, we have argued that magnitude-limited surveys are not strongly biased against hot high-mass WDs, except when the stellar mass starts to exceed $1\,M_\odot$. Therefore, the low number of MWDs found by magnitude-limited surveys reflects mainly the fact that magnetic fields are rare in younger WDs. This result is tentatively supported by a superficial analysis of the situation in the 40\,pc volume limited sample of WDs. The shortage of young MWDs suggests that some surface fields may be produced during  WD evolution, or at least revealed during this evolution, rather than all being inherited from earlier evolution and being revealed as soon as the WD is born. We also note that in the \lv, strong magnetic fields are more common in older than in younger WDs, but this result is statistically too weak to be extended outside of the \lv, where we have no data to confirm its validity.

We have found that the mean mass of the MWDs of the \lv\ is higher than average, that is, $0.69 \pm 0.10\,M_\odot$ (where the uncertainty is the variance of the distribution). This result is qualitatively in agreement with what has been previously suggested in the literature, except that the mean mass value of the MWDs of the \lv\ is lower than what had been estimated from magnitude-limited spectroscopic surveys. Considering that magnitude-limited spectroscopic surveys were probably biased in favour of young MWDs, and considering that in the \lv, MWDs younger than 2\,Gyr are on average more massive than older MWDs (but at a low-significance level), it is possible that in general, younger MWDs are more massive than older MWDs. In the \lv\ we did not find a correlation between the mass and the field strength, except perhaps for the younger MWDs. When we consider MWDs of all ages, the mean mass of MWDs with field weaker than 2\,MG is the same as the mean mass of the MWDs with field stronger than 2\,MG. Observations outside of the \lv\  suggest that in young MWDs, strong fields, when found, occur only or mainly in WDs with mass higher than the average ($\ga 0.75 - 0.80\,M_\odot $). Young lower-mass MWDs seem to host weaker fields (which as such tend to escape detection in spectroscopic surveys). However, further observations are needed to better clarify the relationship between mass, age and field strength.

We have explored possible constraints on the mechanism(s) leading to the occurrence and evolution of magnetic fields in WDs. We have shown that the frequency of the occurrence of magnetic fields in WDs appears to rise steadily with increasing cooling age. Interestingly, this rise appears to coincide roughly with the ages at which cooling WDs start to experience core crystallisation, and using the onset of this process as a divider line in a age-mass diagram, we find that the frequency of MWDs is roughly twice as large on the cool side of the crystallisation boundary as on the hot side.  This suggests that the presence of a dynamo as predicted by \citet{Iseetal17} is a possible mechanism for the production of magnetic fields. This dynamo mechanism, which is similar to the one that produces the magnetic field of the Earth, is due to a combination of rapid rotation and core convection that is driven by sinking of solid O. While the mechanism as originally presented by \citet{Iseetal17} could only explain fields up to $\sim 100$\,kG strength, \citet{Schetal21} have argued that it could actually produce much stronger fields. We note that the operation of this mechanism would be qualitatively consistent with the observed dearth of young MWDs, and with the higher average mass of the MWDs compared to the non-magnetic WDs among younger stars (because higher mass WDs crystallise earlier than older WDs). However, both the suggested strong increase of the dynamo efficiency, and the necessary rapid stellar rotation are still to be demonstrated, therefore the connection of the increased frequency with the core crystallisation must remain speculative at present. That fields appear more frequent in older than in younger WDs could be also explained by slow emergence of fossil fields at the stellar surface, and the fact that MWDs have a mass higher than average is consistent with other possible channels of field formation, including that MWDs are the results of merging. The new data suggest in fact that more than one channel for field formation in WDs exists.

The evolution of fields during WD cooling also provides an important constraint on the magnetic field physics. On the one hand, after formation, the field evolution of MWDs should be governed mainly by diffusion and Ohmic dissipation on time scales of the order of 1--2\,Gyr. This naively suggests that typical field strength should decline gradually with cooling age. In contrast, the distribution of field strengths is observed to remain roughly constant with age out to several Gyr. Possible solutions to this dilemma are that the declining complexity of very large internal fields tends to replace surface flux lost from the surface by Ohmic decay, and/or the presence of a dynamo acting until late in the cooling phase.  Apart from improving the statistics of the occurrence of magnetic fields, the modelling of individual stars of different masses and ages may further help to identify the mechanism(s) that generate and possibly sustain the magnetic field of white dwarfs.

Extending our conclusions beyond the local 20\,pc volume, we infer that in this part of the Galaxy, and potentially on a Galactic scale, $22 \pm 4$\,\% of WDs have a magnetic field. This is likely an underestimate of the real frequency of the occurrence of magnetic fields, because of two reasons: in DC WDs, weaker fields may well exist but are non-detectable with the current techniques; and the frequency of the occurrence of magnetic fields in WDs increases with age, hence the fraction magnetic WDs is higher if we consider WDs older than a certain age. It is reasonable to conclude that in our region of the Milky Way, at least one star out of four will end its life as a MWD. This high frequency demonstrates clearly that magnetism can not be considered a rare and exotic phenomenon among WDs. It appears to be even more common than metal pollution. Magnetism is a major aspect of WD physics that must be integrated with efforts to fully understand the structure of degenerate stars, and to explore the multiple evolution pathways leading to the creation of WDs and evolution through this state. 

\section*{Acknowledgements}
The new observations presented in this work were made 
with the FORS2 instrument at the ESO Telescopes at the La Silla Paranal Observatory under programmes ID 0101.D-0103, 0103.D-0029 and 0104.D-0298;
with ESPaDOnS on the Canada-France-Hawaii Telescope (CFHT) (operated by the National Research Council of Canada, the Institut National des Sciences de l’Univers of the Centre National de la Recherche Scientifique of France, and the University of Hawaii), under programmes 15BC05, 16AC05, 16BC01, 17AC01, and 18AC06; 
and with the ISIS instrument at the William Herschel Telescope (operated on the island of La Palma by the Isaac Newton Group), under programmes P15 in 18B, P10 in 19A and P8 in 19B. This research as made use also of additional FORS2, UVES and X-Shooter data obtained from the ESO Science Archive Facility.
JDL acknowledges the financial support of the Natural Sciences and Engineering Research Council of Canada (NSERC), funding reference number 6377-2016. 
The authors would like to acknowledge the great help consistently offered by the support astronomers and instrument and telescope operators at the three observatories during the entire observing campaign.
This work has made use of data from the European Space Agency (ESA) mission
{\it Gaia} (\url{https://www.cosmos.esa.int/gaia}), processed by the {\it Gaia}
Data Processing and Analysis Consortium (DPAC,
\url{https://www.cosmos.esa.int/web/gaia/dpac/consortium}). Funding for the DPAC
has been provided by national institutions, in particular the institutions
participating in the {\it Gaia} Multilateral Agreement.
\section*{Data Availability}
All raw data and calibrations of FORS2, ISIS and ESPaDOnS data are available at the observatory archives: 
ESO archive at {\tt archive.eso.org}; Astronomical Data Centre at {\tt http://casu.ast.cam.ac.uk/casuadc/}; and the Canadian Astronomical Data Centre at {\tt https://www.cadc-ccda.hia-iha.nrc-cnrc.gc.ca/en/}.

\bibliographystyle{mnras}
\bibliography{sbabib} 

\clearpage

\onecolumn
\begin{small}
\tabcolsep=0.14cm
\begin{longtable}{lrrlllrllllr@{\,}rl}
\caption{\label{Tab_Stars} Physical parameters of the known WDs within 20\,pc from the Sun.
Stars for which a magnetic field has been firmly detected have their name written with boldface fonts. The meaning of the various columns is detailed in Sect.~\ref{Sect_Table}. }\\
\hline\hline
\multicolumn{13}{c}{}\\[-2mm]
         STAR         & $V/G$& $d$    & Sp.   & Bin.      &Atm.     &\teff &$\log g$ &$M$        &Age   &Ref.   &$\overline{\bs}$& Constraints&References for \\[-3mm]  \\
                      &      & (pc)   & type  &           &comp.    & (K)  &  c.g.s. &($M_\odot$)&(Gyrs)&atm.   &    (MG)        &            & field meas.\\            \\
\hline
\endfirsthead \caption{continued.}\\
  \hline\hline
\multicolumn{13}{c}{}\\[-2mm]
         STAR         & $V/G$& $d$    & Sp.   & Bin.      &Atm.     &\teff &$\log g$ &$M$&Age   &Ref.   &$\overline{\bs}$&Constraints&References for  \\ [-3mm]          \\
                      &      & (pc)   & type  &           &comp.    & (K)  &  c.g.s. &($M_\odot$)&(Gyrs)&atm.   &      (MG)      &           &field meas. \\              \\
\hline \endhead
\hline
\endfoot
{\bf WD\,2359$-$434  }&13.1 &  8.34 & DAH   & s         &H        & 8390&  8.37  &0.83  &1.83 &a     &\bbs{0.10}   &\mg &37,27,53,U        \\ 
{\rm WD\,0000$-$345  }&14.9 & 14.81 & DC    & s         &He       & 6280&  8.18  &0.68  &3.23 &a     &\saz{2}{MG}       &32                \\ 
{\bf WD\,0004$+$122  }&16.3 & 17.45 & DCH   & s         &He       & 4885&  8.09  &0.63  &6.45 &b     &\bbs{100\fil}&\mg &62                \\ 
{\bf WD\,0009$+$501  }&14.4 & 10.87 & DAH   & s         &H        & 6445&  8.25  &0.75  &3.24 &b     &\bbs{0.25}   &\mg &21,36,39,(17)     \\ 
{\bf WD\,0011$-$721  }&15.3 & 18.79 & DAH   & s         &H        & 6340&  7.89  &0.53  &1.66 &a     &\bbs{0.37}   &\mg &60                \\ 
{\bf WD\,0011$-$134  }&15.9 & 18.58 & DAH   & s         &H        & 5855&  8.22  &0.72  &4.1  &b     &\bbs{12\fil} &\mg &19,26             \\ 
{\rm WD\,0038$-$226  }&14.5 &  9.1  & DQpec & s         &He(+C)   & 5210&  7.91  &0.51  &4.21 &b     &\saz{3}{MG}       &16,17,22          \\ 
{\rm WD\,0046$+$051  }&12.4 &  4.32 & DZ    & s         &He(+H+Ca)& 6106&  8.2   &0.7   &3.8  &c     &\ssz{0.3}{kG}{2}  &(17),56,Ftw       \\
\hline
\end{longtable}
    \begingroup
    \let\clearpage\relax 
    \onecolumn 

\noindent
{\bf Key to references for the stellar parameters:} \\
 0: no Gaia data available;
 a: \citet{Subetal17};  b: \citet{Bloetal19}; c: \citet{Couetal19}; d: \citet{Giaetal12};  e: \citet{Faretal13};
 f: \citet{Holletal18}; g: \citet{Genetal19}; h: Bergeron web;      i: \citet{Holbetal16}; j: \citet{Limetal15};
 k:\citet{Voretal10};   l: Gaia distance to common proper motion binary companion.\\

 \noindent
 {\bf Key to references of magnetic field measurements:}  \\
 Etw: this work, using the ESPaDOnS instrument; 
 Itw: this work, using the ISIS instrument;
 Ftw: this work, using the FORS instrument;
 U: UVES archive data \citep[see][]{Napetal20}.\\
 1: \citet{AngLan70a};
 2: \citet{Kempetal70}; 
 3: \citet{AngLan70Further};
 4: \citet{AngLan71-Second};
 5: \citet{LanAng71};
 6: \citet{AngLan71-Periodic};
 7: \citet{Angetal72-GRW}
 8: \citet{Angetal72-eph};
 9: \citet{AngLan72};
10: \citet{Angetal74b};
11: \citet{AngLan74};
12: \citet{Angetal75} ;
13: \citet{LanAng75};  
14: \citet{Lieetal75};
15: \citet{Lieetal78};
16: \citet{LieSto80};  
17: \citet{Angetal81};
18: \citet{West89};
19: \citet{Beretal92};
20: \citet{Cohetal93};
21: \citet{SchSmi94};
22: \citet{Schetal95};
23: \citet{SchSmi95};   
24: \citet{PutJor95};
25: \citet{Putney95};
26: \citep{Putney97};  
27: \citet{Koeetal98};
28: \citet{Schetal99};
29: \citet{MaxMar99};
30: \citet{BerPii99};
31: \citet{Maxetal00};
32: \citet{Schetal01};
33: \citet{BeuRei02};
34: field limit deduced from inspection of the HST spectra of \citet{Proetal02}.
35: \citet{Valetal03};
36: \citet{Fabetal03};
37: \citet{Aznetal04} -- note that in this paper, \bz\ was expressed with the opposite sign than usually defined in stellar magnetography. Field values were revised by \citet{Bagetal15} adopting the usual sign;
38: \citet{Frietal04};
39: \citet{Valetal05};
40: \citet{Valetal06};
41: \citet{Kawetal07}; 
42: \citet{Joretal07}; 
43: \citet{Subetal07}; 
44: \citet{Berdetal07};
45: \citet{Valetal08};
46: \citet{Voretal10};
47: \citet{Faretal11};
48: \citet{KawVen12};
49: \citet{Lanetal12}l
50: \citet{Voretal13};
51: \citet{Lanetal15};
52: \citet{Lanetal16};
53: \citet{Lanetal17};
54: \citet{Venetal18};
55: \citet{Faretal18}
56: \citet{BagLan18};
57: field limit deduced from inspection to the HST spectra of \citet{Joyetal18};
58: \citet{LanBag19a}; 
59: \citet{BagLan19a};
60: \citet{LanBag19b};
61: \citet{BagLan19b};
62: \citet{BagLan20};
DC: \citet{Kawetal21}.
\endgroup
\end{small}
    \twocolumn 

\twocolumn
\clearpage

\appendix

\section{New spectropolarimetric observations of WDs: observing log and comments 
on individual stars}\label{Sect_Log}
Our new spectropolarimetric observations are presented in Table~\ref{Table_Log}. In the following we comment on those observations for which interpretation may be ambiguous. Most cases are those of DC stars observed with FORS2 in which we have tried to detect circular polarisation of the continuum. \citet{BagLan20} have argued that, because of cross-talk from linear to circular polarisation, observations of faint stars obtained in grey and bright time may be contaminated by a circularly polarised background that may be difficult to subtract correctly, because it changes rapidly with position in the field of view. We comment also on the observations of WDs of other spectral type in which interpretation or background subtraction is problematic.

\subsection{WD\,0123$-$262} We observed this DC star with FORS2 and grism 300V. The observations were obtained in dark time, so they are not contaminated by polarised background. 
\begin{figure}
\centering
\includegraphics[width=8.7cm,trim={0.0cm 6.3cm 0.5cm 2.0cm},clip]{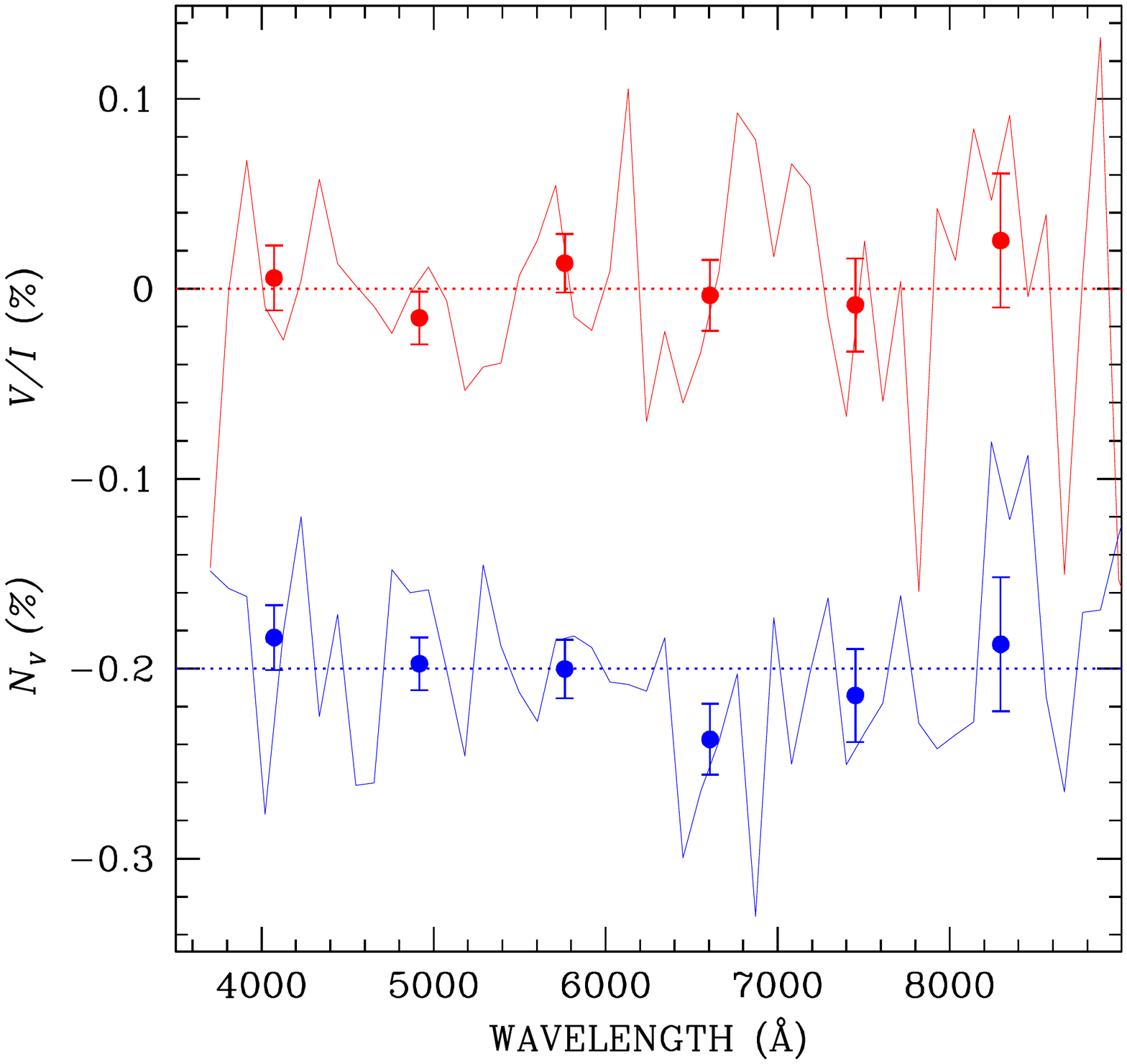}
      \caption{\label{Fig_WD0123} Reduced $V/I$ (red circles and red line) and $N_V$ (blue circles and blue line, offset by -0.2\,\% for display purpose) profiles of star WD\,0123$-$262, rebinned at $\simeq 850$\,\AA\ (circles) and at $\simeq 100$\,\AA\ (solid lines). The horizontal dotted lines represent the zero for $V/I$ (red dotted line) and for $N_V$ (blue dotted line).} 
\end{figure}
The signal of circular polarisation is consistent with zero within the error bars of less than 0.05\,\% in 100\,\AA\ bins between 4000\,\AA\ and 6500\,\AA. At longer wavelengths the uncertainty increases up to 0.1\,\% per 100\,\AA, but no systematic deviations from zero are noted. If we consider 400\,\AA\ spectral bins, it is safe to assume that the observed polarisation is consistent with zero within $0.05$\,\%, for a $\abz \la 0.75$\,MG. For a 800\,\AA\ rebinning, the upper limit is 0.02\,\%, for  $\abz \la 0.3$\,MG. Note though that observations were obtained with detector E2V, and therefore are affected by fringing at wavelengths longer than H$\alpha$. However, with heavy rebinning, fringing does not seem to leave spurious signatures in circular polarisation, and this dataset provides us with what is probably our most precise measurement of circular polarisation in the continuum obtained with FORS2 (see Fig.~\ref{Fig_WD0123}).

\subsection{WD\,0210$-$083} 
\begin{figure}
\centering
\includegraphics[width=8.7cm,trim={0.0cm 0.0cm 0.0cm 1.0cm},clip]{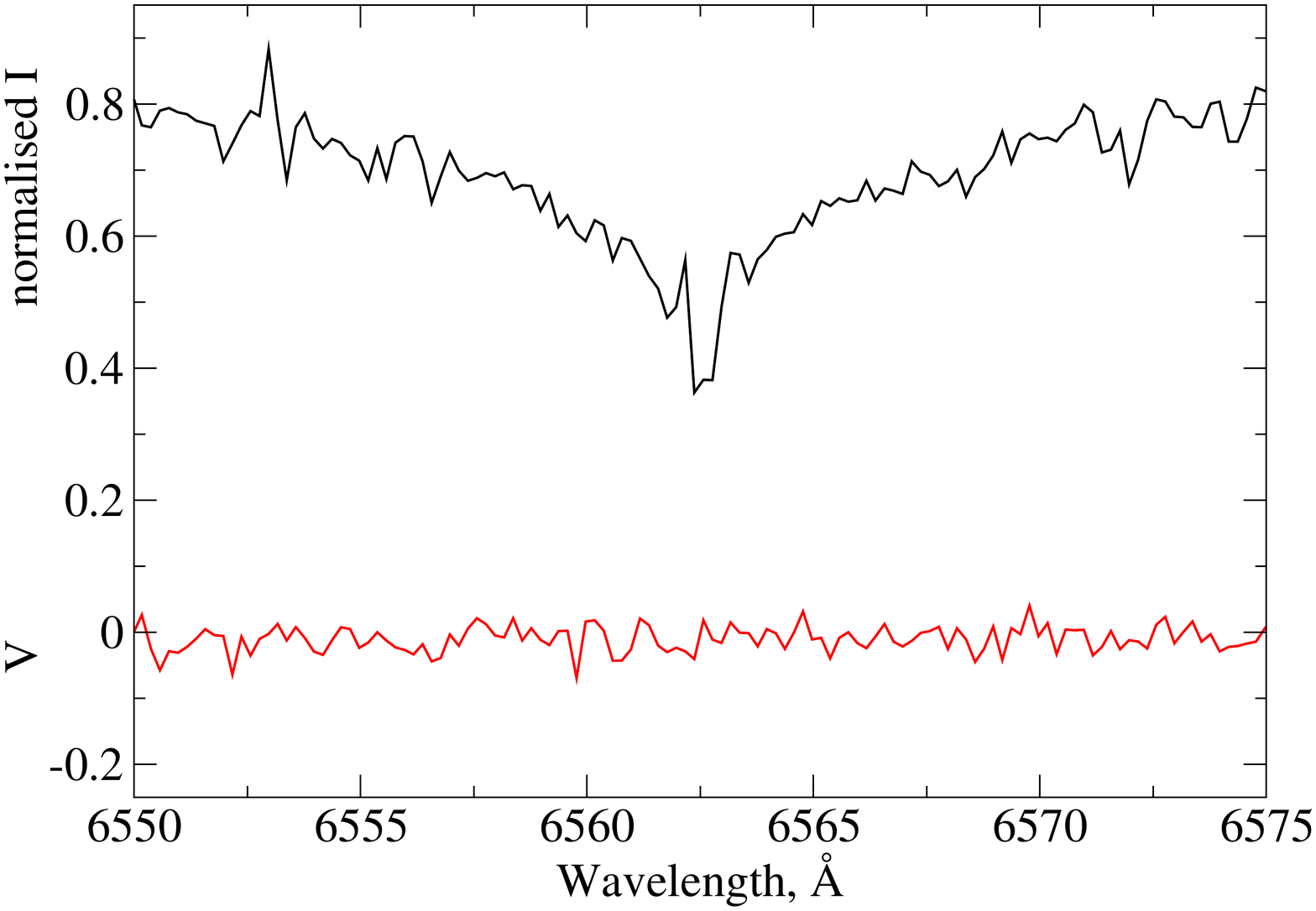}
\includegraphics[width=8.7cm,trim={1.0cm 6.0cm 1.0cm 3.0cm},clip]{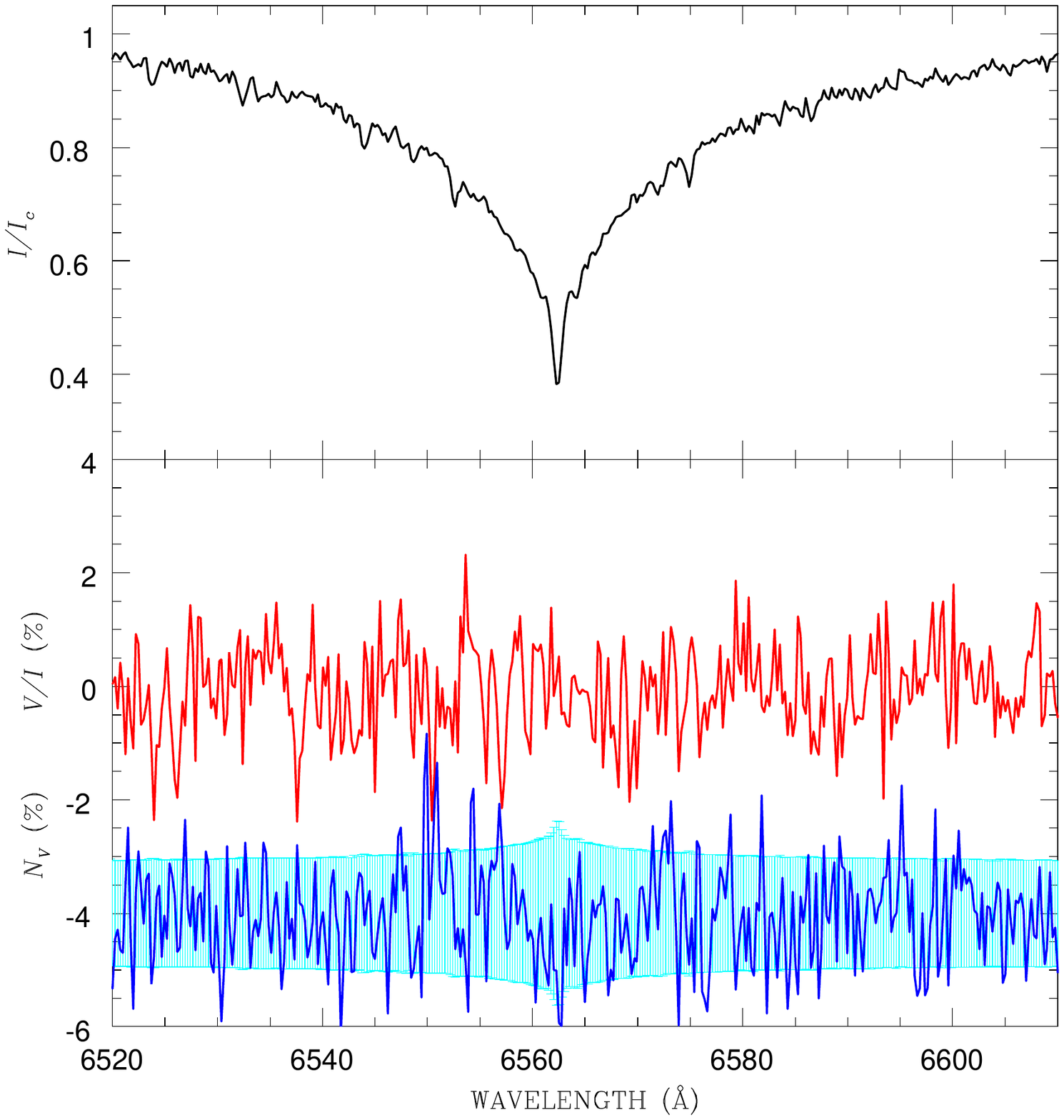}

      \caption{\label{Fig_WD0210} H$\alpha$ $I$ and $V$ spectrum of WD\,0210$-$083 obtained with ESPaDOnS (top panel), and H$\alpha$ $I/I_{\rm c}$ and $V/I$ observed with ISIS (bottom panel). Note absence of any significantly non-zero signal in $V$ and $V/I$ around the line core. These spectra are typical of the survey spectra obtained during our survey of the \lv. 
      } 
\end{figure}

We observed this DA both with ISIS and with ESPaDOnS; the ESPaDOnS intensity profile of the core of H$\alpha$ is slightly shallower than the cores of other WDs of similar temperature in our sample, but about the same width. The Stokes $V$ profile is flat, and the ISIS data in both arms suggest no field (see Fig.~\ref{Fig_WD0210}). We consider this WD to be not magnetic, but the star deserves further observations.
\subsection{WD\,0415$-$594} This DA WD was observed with FORS2. Background subtraction needs a special attention due to the uneven illumination from the bright K2 companion.

\subsection{WD\,0743$-$336} This DC WD was observed with FORS2. We note that the magnitude of the star is $G=15.3$, not $V=16.7$ as found in Simbad and reported by \citet{Kunetal84}. In the $B$ band we measure a signal of $V/I = -0.15 \pm 0.05$\,\%. At longer wavelengths, the polarisation signal is zero within noise. Since the star was observed at 60\degr\ from the Moon, which was 64\,\% illuminated, background is affected by cross-talk from linear to circular polarisation. The star is bright enough that background subtraction is not particularly challenging but the origin of the signal of circular polarisation at shorter wavelengths may well be spurious.  We set $\vert V/I \vert \la 0.15$\,\%.

\subsection{WD\,0806$-$661}
This is a DQ star observed with FORS2. $V/I$ is smaller than $0.02$\,\% except at $\lambda \ga 8000$\,\AA, where both $V/I$ and $N_V$ depart from zero. Ignoring this as a probably spurious signal, we set $\vert V/I \vert \la 0.02$\,\%.

\subsection{WD\,0810$+$489} This is a DC star observed with ISIS. Data were obtained with bad seeing. There is a signal of $\sim -0.2$\,\% in the blue, while in the red the signal is around +0.1\,\%. We consider this as a non detection and we set $\vmax \le 0.2$\,\%.

\subsection{WD\,0856$-$007} This star was observed with FORS2 and grism 300V because it was originally classified as DC, but our spectra revealed the presence of H$\alpha$ and Ca\,{\sc ii} lines. The star is thus a DAZ. From very low resolution spectropolarimetry of H$\alpha$ we obtained a marginal detection ($\bz = 27 \pm 12$\,kG) which prompts for further observations at higher resolution with grism 1200R.

\subsection{WD\,1055$-$172} ISIS observations of this DC star were obtained with bad seeing and strong moon background. The blue part of the spectrum shows a very noisy signal of $V/I \simeq 0.2$\,\% and the red $\simeq -0.4$\,\%, which we do not accept as detection.

\subsection{WD\,1116$-$470} This DC star was observed twice with FORS2 and grism 300V, the first time at 60\degr\ from a half illuminated moon, the
\begin{figure}
\centering
\includegraphics[width=8.7cm,trim={0.0cm 5.6cm 1.3cm 3.5cm},clip]{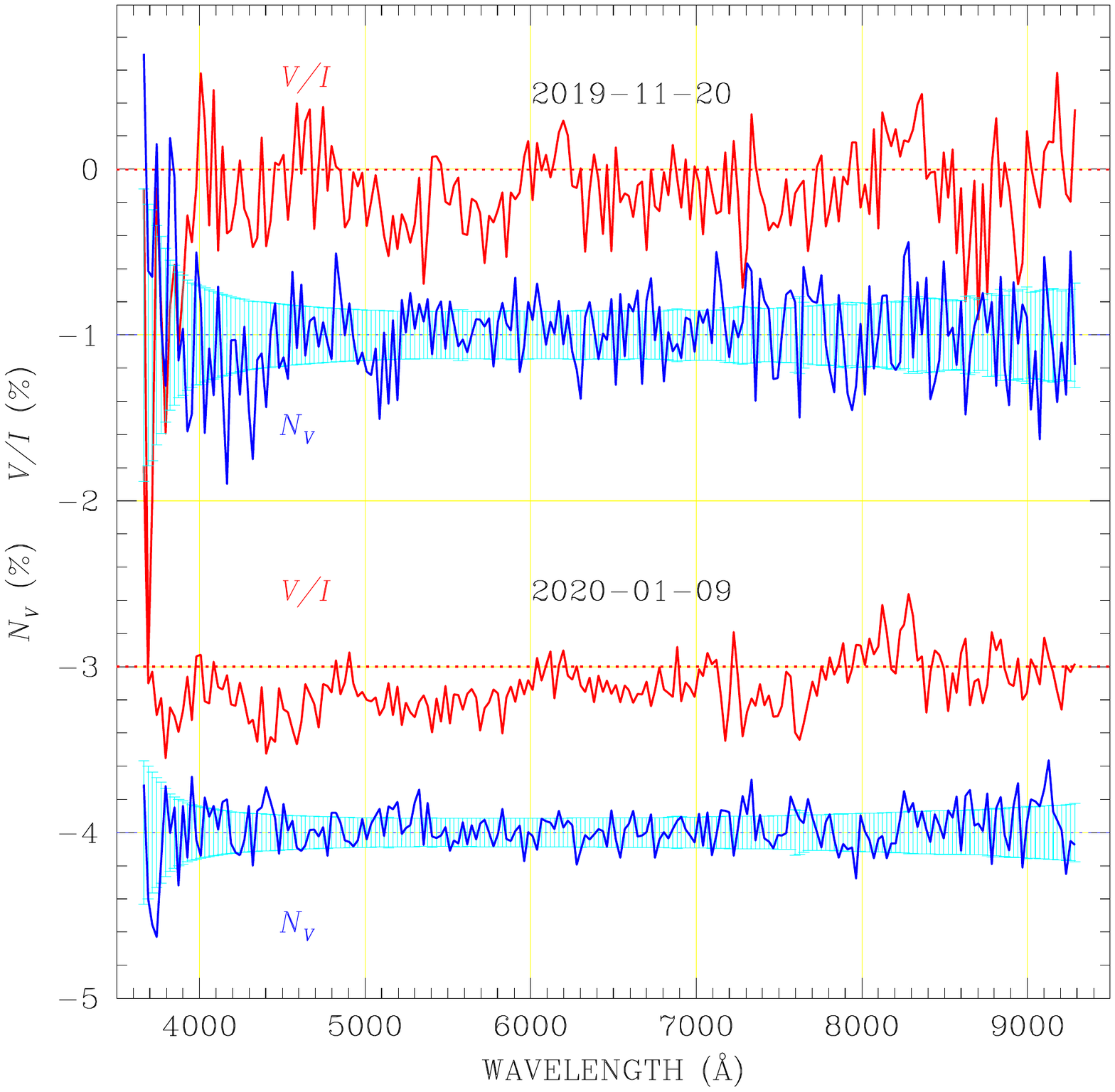}
\caption{\label{Fig_WD1116} Circular polarisation spectra (solid red lines) and null profiles (solid blue lines, overplotted to the light blue error bars) for star WD\,1116$-$470 obtained on 2019-11-20 and 2020-10-09 with the FORS2 instrument and grism 300V. For display purpose, $V/I$ spectrum obtained on 2020-10-09 is offset by $-3$\,\% and null profiles are offset by $-1$ and $-4$\,\%.} 
\end{figure}
second time at 100\degr\ from full moon. In both observations, the background is strongly circularly polarised, and in both observations the star shows a constant signal of $\sim -0.2$\,\%, that goes to zero at $\lambda \ga 8000$\,\AA. The interpretation of this signal is problematic, because on the one hand it is clear that background may contaminate the signal of the star, but on the other hand the signals measured in two different nights are consistent with each others (see Fig.~\ref{Fig_WD1116}), even though the background polarisation had the opposite sign in the two different epochs. Furthermore, contamination from cross-talk should manifest itself more at shorter wavelength than at longer wavelength, while the observed signal of polarisation seems constant with wavelength. In conclusion, while we cannot firmly confirm that the observed signal is real, the star should be considered as a strong candidate magnetic DC WD (with a field strength of the order of $\sim 3$\,MG), and should be re-observed during dark time.

\subsection{WD\,1145$-$747} This DC star was observed during dark time with FORS2, both with grism 600B and with grism 300V. The latter data show no polarisation ($\vert V/I \vert \la 0.1$\,\%, for $\abz \la 1.5$\,MG), but the spectrum obtained with grism 600B has $V/I=-0.2$\,\%. Since the same offset is shown by the null profile, we conclude that the signal seen with grism 600\,B is spurious.

\subsection{WD\,1310$-$472} This DC WD was observed during dark time with FORS2. The background is not polarised and the star shows a nearly constant signal of polarisation $V/I \simeq 0.09$\,\%. This may be explained as cross-talk from $I$ to $V$, but the star certainly should be re-observed with higher accuracy.

\subsection{WD\,1316$-$215} This is one of the faintest DA stars. It was observed with FORS2 twice with grism 1200R, once in bright time, with a strongly polarised background, and once during dark time. This dataset may be used as a benchmark for contamination from a polarised background with grism 1200R. Because H$\alpha$ is present, \bz\ was accurately estimated to be zero on both occasions. Therefore we do not expect circular polarisation. From the observations obtained during bright time we measured $V/I \simeq +0.1$\,\% in the continuum, while during dark time we observed a signal much closer to zero, with peaks of $\vmax \la 0.05$\,\%. 

\subsection{WD\,1334$+$039} This is a DA star observed with ISIS. The spectrum shows a broad H$\alpha$, a bit offset to the red (by 5\,\AA) with respect to the absorption H$\alpha$ of the bright sky background.

\subsection{WD\,1338+052} This DC star was observed during dark time and show a signal of circular polarisation in the continuum ($\simeq -0.1$)\,\%. Although it cannot be explained in terms of contamination from a polarised background, the signal is still too small to be deemed to be real. The star should be re-observed with higher precision.

\subsection{WD\,1345$+$238} This is a very cool DA with an extremely weak H$\alpha$. Because of strong background and bad seeing, the quality of our ISIS observations does not allow us to set a strong constraint, with upper limit $\vert V/I \vert \la 0.3$\,\%.

\subsection{WD\,1821$-$131}
This DA WD was observed both with ISIS and FORS2, and with both instruments, spectropolarimetry of the Balmer lines shows that \bz\ is consistent with zero. However, in the blue arm of ISIS we measure a signal of circular polarisation of +0.3\,\%, and in the red a signal of $-0.2$\,\%. With FORS2+grism 1200R we measured a signal of circular polarisation of  $-0.2$\,\% in the continuum, consistent with what we have observed with ISIS. Because H$\alpha$ is not polarised, and because there is no hint that the star may have a DC companion, we consider the polarisation signal to be spurious.

\onecolumn
\begin{center}
\begin{longtable}{lllcccr r r@{$\pm$}l r@{$\pm$}l}
\caption{\label{Table_Log} New magnetic field measurements of WDs within 20\,pc.}\\
\hline\hline
STAR                             &            
INSTR.                           &            
Grism/                           &            
DATE                             &            
UT                               &            
MJD                              &            
Exp                              &            
S/N                              &            
\multicolumn{2}{c}{\bz}          &            
\multicolumn{2}{c}{\nnz}        \\            
                                 &            
                                 &            
Grating                          &            
yyyy-mm-dd                       &            
hh:mm                            &            
                                 &            
(s)                              &            
\AA$^{-1}$                       &            
\multicolumn{2}{c}{(kG)}         &            
\multicolumn{2}{c}{(kG)}        \\            
\hline
\endfirsthead
  \caption{continued.}\\
  \hline\hline
STAR                             &            
INSTR.                           &            
Grism/                           &            
DATE                             &            
UT                               &            
MJD                              &            
Exp                              &            
S/N                              &            
\multicolumn{2}{c}{\bz}          &            
\multicolumn{2}{c}{\nnz}        \\            
                                 &            
                                 &            
Grating                          &            
yyyy-mm-dd                       &            
hh:mm                            &            
                                 &            
(s)                              &            
\AA$^{-1}$                       &            
\multicolumn{2}{c}{(kG)}         &            
\multicolumn{2}{c}{(kG)}        \\            
\hline
\endhead
\hline
\endfoot
WD\,0046$+$051       &  FORS2      & 1200B & 2019-09-30 & 02:34 & 58756.107 & 2160 &1235&$   0.0  $&0.3 &$  -0.2 $& 0.3 \\  [2mm]
WD\,0123$-$262       &  FORS2      &  300V & 2019-09-29 & 04:26 & 58755.185 & 1440 & 250&\multicolumn{4}{c}{$\vmax \la 0.03$\,\%}\\ [2mm]
WD\,0135$-$052AB     &  ESP        &       & 2015-10-29 & 08:47 & 57324.366 & 3256 & 295&$  -0.2 $&  0.3 &$  0.3 $& 0.3 \\ [2mm]
WD\,0141$-$675       &  FORS2      & 1200R & 2018-06-28 & 08:29 & 58297.354 &  876 & 200&$  -6.3  $&  4.6&\multicolumn{2}{c}{}\\ 
                     &  FORS2      & 1200R & 2018-06-29 & 09:15 & 58298.385 & 5400 & 590&$   0.3  $&  0.4&$   0.5 $& 0.4 \\
                     &  FORS2      & 1200R & 2018-07-02 & 08:28 & 58301.353 & 2700 & 620&$   0.0  $&  0.3&$  -0.4 $& 0.3 \\ [2mm]
WD\,0148$+$641       &  ESP        &       & 2018-08-18 & 12:03 & 58348.502 & 3468 & 125&$   0.9  $& 0.8 &$  -0.5 $& 0.8 \\ [2mm]
WD\,0148$+$467       &  ESP        &       & 2015-10-02 & 09:01 & 57297.376 & 3256 & 205&$  -0.1  $&  0.4&$   0.6 $& 0.4 \\
\hline
\end{longtable}
\end{center}

\twocolumn
\clearpage


\section{Comments on the magnetic field measurements of the WDs of the local 20 pc volume}\label{Sect_Stars}
Here we review the circular broadband and spectro-polarimetric measurements of WDs within 20\,pc from the Sun that help to characterise their magnetic nature. We comment on the magnetic field measurements of individual stars, grouped by spectral types. We recall that while the definition of the sign of the mean longitudinal field has been fairly consistent in the literature, different definitions of the sign of circular polarisation have been adopted -- see \citet{BagLan20} for details. In the following, we define circular polarisation to be positive when the electric field vector, in a fixed plane perpendicular to the direction of propagation of the light, is seen to rotate clockwise by an observer facing the source. This definition conforms to that adopted in textbooks such as \citet{Shurcliff62} and \citet{LanLan04}, in publications with numerical simulations of radiative transfer \citep[e.g.][]{Wadetal01}, and in observational work such as that of  \citet{Kemetal70}, \citet{LanAng71}, \citet{BagLan20}, but opposite to what was probably the more common convention in the literature of WDs \citep[e.g.][]{Angetal81,SchSmi95,Putney97,Voretal10}.  When we report literature values, we adapt them to this convention, hence the $V/I$ values reported in this section may or may not have the same sign as reported in the original paper.

In general, we found that the accuracy of the older measurements in the continuum, obtained through broadband filters, are comparable to or even better than those obtained from polarisation spectra, even when highly rebinned. Therefore, for DC WDs, which have featureless spectra, and for some DQ stars, with very broad molecular lines, null detections obtained with broadband filters still set useful constraints on the upper limit of the field strength. Conversely, for the WDs with spectral lines, low resolution spectropolarimetry permits us to achieve a much higher sensitivity than broadband measurements. In these stars (mainly DA WDs), null detections obtained with broadband or narrowband circular polarimetric measurements are no longer useful, if recent spectroscopic or spectropolarimetric measurements have been obtained. An example is represented by DA WD  40\,Eri\,B, in which \citet{AngLan70a} measured $\bz = 20 \pm 5$\,kG, while \citet{Lanetal15} obtained a series with a typical uncertainty of $80-90$\,G, for an improvement of sensitivity better than 1.5\,dex. \citet{Angetal81} tried twice to measure the magnetic field of WD\,0752$-$676, at that time classified as DC, with an uncertainty of $\sim 350$\,kG; later the star was found to be DA, and with FORS we obtained a field measurement with a 0.4\,kG uncertainty. \citet{LanAng71}, \citet{LieSto80} and \citet{Angetal81} obtained field measurements for 32 of the WDs in the \lv, with a typical uncertainty of hundreds of kG. About half of these stars are DC or DQ, and these measurements are still very useful for this work. The measurements for the remaining half have now been superseded by much more accurate spectropolarimetric spectral line measurements, and references are given in notes referenced in the final column of Table~\ref{Tab_Stars}. 

\subsection{DA stars}\label{Sect_DA}
\subsubsection{{\bf WD\,2359$-$434} = {\bf J\,915} --  well monitored and modelled magnetic variable with 50 to 100\,kG field}%
The presence of a magnetic field in this star was first suggested by \citet{Koeetal98} and confirmed with ESO FORS1 low-resolution spectropolarimetry by \citeauthor{Aznetal04} (\citeyear{Aznetal04}; we note that these data were re-reduced as described by \citealt{Bagetal15}, and the sign of \bz\ changed to conform to our usual convention). \citet{Lanetal17} have modelled the star using 13 ESPaDOnS data, and concluded that it has a complex magnetic configuration. The stars is a photometric variable, and both light and magnetic variations agree on a rotation period of $2.7$\,h. 

\subsubsection{{\bf WD\,0009$+$501} = {\bf GJ\,1004} --  well monitored and modelled magnetic variable with 150 to 250\,kG field}%
WD\,0009$+$50 was discovered to be a magnetic WD by \citet{SchSmi94} through spectropolarimetry. The star's magnetic field was measured by \citet{Fabetal03} and \citet{Valetal05}; the latter work presented a model obtained by adopting a rotational period of $8$\,h. With Valyavin, we have obtained numerous ESPaDOnS measurements that we will present in a future paper.

\subsubsection{{\bf WD\,0011$-$721} = {\bf  NLTT\,681} --  magnetic with $\bs = 365$\,kG, observed only once} %
The star was spectroscopically confirmed as a nearby DA WD by \citet{Subetal08}, and discovered to be magnetic by \citet{LanBag19b} using FORS2 low-resolution spectropolarimetry of H$\alpha$ (which shows Zeeman splitting for $\bz \simeq 75$\,kG and $\bs \simeq 365$\,kG). The star was observed only once, and follow up observations are needed to check if the star is magnetically variable and whether it is a good candidate for monitoring and modelling.

\subsubsection{{\bf WD\,0011$-$134} = {\bf G\,158-45} -- 9\,MG, magnetic variable}%
Discovered as a magnetic WD via spectroscopy by \citet{Beretal92}, who from analysis of H$\alpha$ Zeeman splitting, estimated $\bs \sim 10$\,MG. \citet{Putney97} obtained six measurements and showed that \bs\ varies between 8 and 10\,MG, and that \bz\ reverses sign with a rotation period that lies between some hours and a few days. It should be further monitored for modelling.

\subsubsection{{\bf WD\,0121$-$429A} = {\bf LP 991-16A}  -- 5.5\, MG field in DA component of uDD, flux spectroscopy only}\label{Sect_DA0121}%
The weakness of H$\alpha$ relative to the strength expected for the \teff\ value deduced from the energy distribution ($\sim 6000$\,K) suggests that the system is a DD with a DAH and a He-rich DC \citep{Subetal07,Giaetal12}, with roughly equal contributions to the light around H$\alpha$. Furthermore, the mass deduced by \citet{Giaetal12} for a single star model is only $0.41\,M_\odot$, and both \citet{Holletal18} and \citet{Genetal19} find $\log g = 7.57$, implying $M = 0.39\,M_\odot$. Such a low mass cannot result from single star evolution. From these two lines of evidence, we conclude that this star is a DD, and we list the two components as separate WDs in our tables.  

From the best-fit to the single star H- and He-rich models to the photometry \citep{Giaetal12,Holletal18,Genetal19}, we assume that both the H-rich star and its its He-rich DC companion have \teff\ in the range of 5800--6300\,K. We take as our starting point a single star model of the system with $\teff = 6035$\,K, $\log g = 7.57$, and $M = 0.39 M_\odot$, an average of models assuming either H-- or He--rich atmospheres \citep{Giaetal12,Genetal19}  Assuming that each of the two stars contributes roughly equally to the observed brightness, we take this starting model and exchange the initial model for two stars with radii $R_i$ to $R_i/\sqrt 2$. Keeping the same \teff\ as initially, WDs with the new reduced radius have half the surface area of the initial single-star model, and together they provide the same light as the initial model. However, we need the new mass and $\log g$ parameters of the individual WDs contributing to our simple model. These are estimated from the mass-radius relationship \citep[e.g.][]{Giaetal12,Bedetal17} to be $0.69 M_\odot$ and 8.17 for both stars.  Finally, we obtain new cooling time estimates as discussed in the text.  Of course, the assumption that the two WDs contribute equally to the observed light is only a rough approximation, but it allows us to obtain plausible approximate values of parameters for both stars that are needed for our statistical study. 
 
 \citet{Subetal07} discovered a strong magnetic field through H$\alpha$ flux spectroscopy. The observed splitting corresponds to $\bs \approx 5.5$\,MG. This star has never been observed in broadband polarimetric or spectropolarimetric mode. The clear magnetic splitting of H$\alpha$ and H$\beta$ shown by \citet{Subetal07} indicates that the observed field is found in the DA. Without polarimetry, we have no information about a possible field in the DC star, so it is treated as non-observed (see Sect.~\ref{Sect_DC0121}). This system should be monitored with high-resolution spectroscopy to search for radial velocity variations of the DA, and spectropolarimetrically, to check for magnetism in the DC star and to test for possible magnetic field variability of either component.

\subsubsection{WD\,0135$-$052A = GJ\,64A (uDD-SB2)}%
The system WD\,0135$-$052 is a clear SB2, with radial velocity semi-amplitudes of about 70 and 75\,\kms, and an orbital period of 1.56\,d. The SB2 system was analysed in detail by \citet{Safetal88}, and important data are summarised by \citet{Safetal98}. It is composed of two nearly identical DA WDs, the masses of which have been estimated for both components from the radial velocity curves. \citet{Beretal89} have used mass data, distance and photometry to estimate the values of \teff\ and $\log g$ for the individual components. Ages have been estimated by \citet{Holbetal16}. The resulting parameters are listed in Table\,\ref{Tab_Stars}. 
Three spectra obtained by \citet{Koeetal98} show clearly split H$\alpha$ varying on a time scale of days (Fig.~5 of their paper). 

\citet{SchSmi95} measured $\bz = -1.6 \pm 6.1$\,kG. Later, \citet{Aznetal04} and \citet{BagLan18} obtained much more sensitive measurements with FORS1 and FORS2 ($\bz = 0.2 \pm 0.4$\,kG and $0.0 \pm 0.2$\,kG respectively), without detecting any field. The single ESPaDOnS spectrum presented in this paper was fortuitously obtained at an orbital phase at which no splitting is evident ($\bz = -0.2 \pm 0.3$\,kG). Since H$\alpha$ does not show any sign of Zeeman splitting in our ESPaDOnS spectrum, we can set the upper limit for \bs to 50\,kG for each star. 
However, it should be noted that this upper limit refers to the combined signal of the two stars, therefore the upper limit for each star should be considered to be twice as large. 

\subsubsection{WD\,0135$-$052B = GJ\,64B (uDD-SB2)}%
See section above.

\subsubsection{WD\,0148$+$641 = EGGR\,268 (VB: M2 at 12")}%
Visual binary with a M2 companion at 12\arcsec. \citet{SchSmi95} measured $\bz= 0.4 \pm 4.5$\,kG. We observed it with ESPaDOnS (this work) and found $\bz = 0.9 \pm 0.8$\,kG. 

\subsubsection{WD\,0148$+$467 = GD\,279}%
\citet{SchSmi95} measured $\bz = -5.8\pm 7.8$\,kG. \citet{BagLan18} observed the same star with 20 times higher precision with three ISIS measurements, but did not detect a field ($\bz = -0.2 \pm 0.5$\,kG, $0.0 \pm 0.2$\,kG, $-01 \pm 0.4$\,kG). No detection was obtained in two new ESPaDOnS measurements presented in this paper ($\bz = -0.1 \pm 0.4$\,kG, $+0.1 \pm 0.3$\,kG). 

\subsubsection{WD\,0210$-$083 = LP\,649$-$67 (VB: dM at 3.6\arcsec\ and uDD?)}%
Recently discovered as a WD within 20\,pc by \citet{Holletal18}, the star belongs to a CPM visual binary system with a 3.6\arcsec\ separation \citep{Heintz93,Luyten97} to the primary of the common proper motion VB, LP\,649--66, a dM star. One new ISIS ($\bz = +0.4 \pm 0.8$\,kG, averaging the measurements in the blue and red arm) and one new ESPaDOnS measurement ($\bz = -3.1 \pm 1.2$\,kG), all presented in this paper, show that the star is a DA, but did not reveal the presence of a magnetic field (although ESPaDOnS data give a $2.5\,\sigma$ detection). 

The WD is possibly an unresolved double degenerate, based on the photometrically inferred mass of 0.43\,$M_\odot$ \citep{Genetal19}, which is too low to have resulted from single star evolution. One peculiarity is the profile of the core of H$\alpha$, which is clearly slightly shallower than the cores of other WDs of similar temperature in our sample, but about the same width (see Fig.~\ref{Fig_WD0210}). We treat this system as a non magnetic single WD, and mark it as uDD? until more data are obtained.

\subsubsection{WD\,0230$-$144 = LHS 1415}%
Observed in broadband circular polarimetric mode by \citet{LieSto80}, who measured $V=-0.002 \pm 0.10$\,\%. This star was then observed three times by \citet{Putney97} in spectropolarimetric mode; one of these measurement is a 2.5\,$\sigma$ detection ($\sigma \simeq 14$\,kG), and one is a 8\,$\sigma$ detection, but described as "faulty" in the text. We observed the star once with ISIS (this work), but we did not detect any field (with $\sigma \simeq 4$\,kG). In our ISIS spectra, H$\alpha$ does not show any hint of Zeeman splitting, nor does it in the $R\sim 18000$ UVES spectra available in the ESO Archive, so we can set 50\,kG as the upper limit for \bs. This star should be probably observed again, but at present we consider it as non magnetic.

\subsubsection{{\bf WD\,0233$-$242A} = {\bf NLTT 8435A} -- 3.8\,MG, variable (uDD)}\label{Sect_DA0233}%
This star was first observed by \citet{VenKaw03}, who obtained a low resolution, low S/N spectrum at Mount Stromlo and classified it as a very cool DC WD, with an estimated \teff\, around 5300\,K. It was modelled by \citet{Giaetal12} (using the Mount Stromlo spectrum) as a DA on the basis of a communication from Kawka indicating that a magnetically split H$\alpha$ line had been detected in new data, again assuming $\teff \approx 5300$\,K. A new FORS red spectrum was reported by \citet{Venetal18}, showing very clear magnetic splitting, which was modelled with a slightly decentred dipole of polar field strength about 6.1\,MG. \citet{Venetal18} also noted a radial velocity shift of at least 60\,\kms\ between the two spectra, which they attributed to binary motion, and reported a photometric time series which suggested a stellar rotation period of about 95\,min. 

We retrieved the two circularly polarised FORS spectra \citep[taken during 2013 January for ][]{Venetal18} from the ESO Archive. The first spectrum, from 2013-01-02, shows very clean Stokes $I$ Zeeman components, with a rather boxy but narrow $\pi$ component, and quite broad $\sigma$ components. The second $I$ spectrum, from 2013-01-09, shows a sharp $\pi$ component slightly deeper than in the first spectrum, and $\sigma$ components very similar to those of the first spectrum, except for rather severe terrestrial line contamination of the blue $\sigma$ component. Comparing the second spectrum of WD\,0233--242 to the terrestrial lines observed in a rapidly rotating main sequence B star indicates that one contaminating line appears in the $\pi$ component and is responsible for the morphological differences between the $\pi$ line core shapes observed in the two spectra. 

As mentioned by \citet{Venetal18}, the sharp $\pi$ components of the two spectra differ in radial velocity. We measure a velocity difference of $90 \pm 10$\,\kms\ (a shift of $1.97 \pm 0.2$\,\AA) between the two H$\alpha$ $\pi$ components. We considered the possibility that this difference might be due to a problem with FORS wavelength calibration, but this is rejected because telluric emission lines and the O$_2$ atmospheric absorption bands that bracket H$\alpha$ appear at exactly the same apparent wavelengths in both spectra. We also considered the possibility of a wavelength shift of the $\pi$ component as a function of mean magnetic field strength. However, the value of \bs, as measured by the position of the centroid of the red $\sigma$ line component (the blue component is badly distorted in the second spectrum by water vapour lines) changes by less than 0.1\,MG between the two spectra, which in turn would shift the $\pi$ component by only about 0.03\,\AA\ \citep{SchWun14}. 

We conclude that the velocity shift found by \citet{Venetal18} is real and we agree with their conclusion that that this WD is a member of an SB1 binary star system. As there is no sign whatever in the observed spectra of the absorption bands characteristic of main sequence red dwarfs, the companion is most probably a DC WD. Because the observed DA is found to have a \teff\ value close to 5000\,K, the expected strength of the H$\alpha$ line is fairly sensitive to the exact value of \teff, and cannot be used to estimate the fractional contribution to the light of the DC without detailed modelling. Thus the only useful constraint we have on the DC companion is that its value of \teff\ is probably also close to 5000\,K because of the consistency of the photometric data with both H and He models having this temperature \citep{Giaetal12}. Nevertheless, we have included the companion DC in the census of the 20\,pc volume (see Sect.~\ref{Sect_DC0233}), although we cannot provide much useful information about the DC component of the system. The \teff\ value found by \citet{Bloetal19}, 4555\,K, seems too cool to account for the clear H$\alpha$ line in the A component, so we will assume that both stars have the same  $\teff = 4840$\,K, as derived by \citet{Holletal18}. We then make the assumption that both stars contribute roughly the same amount to the observed luminosity, so we reduce the radius $R$ of the initial single star model to $R/\sqrt 2$ in order to divide the surface area of the star by 2. Then we recompute $M/M_\odot$ and $\log g$ from the mass-radius relation, as discussed in the notes for WD\,0121--429 above. The resulting individual  parameter values, shown in Table\,\ref{Tab_Stars}, are of course quite uncertain. 

We have measured the value of \bs\ by computing the separation of the centroid of the red $\sigma$ Zeeman component from the $\pi$ component. We find value of 3.9 and $3.8 \pm 0.1$\,MG for the first and second spectra respectively \citep{SchWun14}. However, in both spectra the large spread of the sigma components indicates a wide dispersion in local field strength $|B|$ over the visible hemisphere, ranging between about 2.8 and 5.2\,MG. (Note that the 6.1\,MG field strength mentioned by \citet{Venetal18} is the deduced polar field strength of a model decentred dipole field, not to a direct measurement of \bs.) We have also used the circular polarisation $V/I$ data of the two available FORS spectra to measure the mean longitudinal field strength \bz\ revealed by each spectrum, using the mean separation between the centroids of the spectral line as viewed in right and left circular polarisation \citep[e.g.][]{Lanetal15}. The measured values of the field moment \bz\ for the two spectra are $+428 \pm 24$\,kG and $-360 \pm 20$\,kG. The measured values thus clearly indicate that the field \bz\ is variable, and the MWD is rotating as well as orbiting around a companion. Note that these \bz\ values are both rather small compared to the \bs\ values revealed by the $I$ spectra, and if taken at face value suggest a large inclination of the global field axis to the line of sight. However, these polarisation measurements may well not provide simple snapshots of the longitudinal field of the MWD for two reasons. First, the unseen companion may significantly dilute the observed polarisation, without affecting the measured value of \bs. Secondly, the duration of each FORS observation, 2600s, is almost 50\,\% of the rotation period of 95\,min = 5700\,sec suggested by time series photometry of the system \citep{Venetal18}. That is, each measurements of $V/I$ may be smeared over nearly 1/2 rotation, probably leading to much polarisation cancellation and a small residual computed longitudinal field. This smearing may also lead to the strong similarity of the Zeeman splitting as observed in the two $I$ spectra. However, neither dilution nor rotational smearing act to reduce the measured value of \bs\ below its true value, but simply average \bs\ over about 1/2 rotation of the WD.

\subsubsection{WD\,0310$-$688 = GJ\,127.1}%
Based on a spectropolarimetric measurement obtained at the Mount Stromlo Observatory, this star was identified as a suspected magnetic WD by \citet{Kawetal07}, who measured $\bz -6.1 \pm 2.2$\,kG. However, the same star was observed with much higher accuracy with FORS1 by \citet{Aznetal04}, who measured $\bz=-0.10 \pm 0.44$\,kG, and with FORS2 by \citet{BagLan18}, who found $\bz = -0.20 \pm 0.23$\,kG. We do not consider the star to be magnetic. 

The shape of H$\alpha$ as seen in spectra from a ESO UVES Archive with $R\sim 20000$ allow us to set an upper limit of $\bs \la 50$\,kG.

\subsubsection{WD\,0357$+$081 = G\,7-16}%
This WD was observed once by \citet{Putney97}, who measured $\bz=4.1 \pm 9.5$\,kG using the strong H$\alpha$. It was observed once by us with FORS2, a non-detection with 1.4\,kG uncertainty (this work). Our FORS2 data also provide an upper limit of 300\,kG for \bs.

\subsubsection{WD\,0413$-$077  = 40\,Eri\,B (multiple system)}%
Observed first by \citet{AngLan70a} (with a Balmer line magnetograph, in the broad line wings of H$\gamma$, yielding $\bz = 20 \pm 5$\,kG), then by \citet{LanAng71} (broad-band visible light, $V/I=+0.04 \pm 0.05$), then in spectropolarimetric mode by \citet{SchSmi95} ($\bz = -4.5 \pm 3.1$\,kG and $1.2 \pm 0.9$\,kG), this star was considered for a long time as a weakly magnetic WD \citep{Fabetal03,Valetal03}. However, \citet{Lanetal15} obtained a number of highly precise measurements of the longitudinal field using both with ESPaDOnS and with ISIS. Although six of their measurements had uncertainty as low as 85--90\,G (the smallest $\bz$\ uncertainties ever obtained for a WD), no magnetic field was detected. The conclusion of \citet{Lanetal15} was that field detections previously reported in the literature were spurious, and the star is actually non-magnetic. In this paper we present a new ISIS field measurement, which is another high-precision non-detection. 

It is interesting to note that 40\,Eri was the object of a spectropolarimetric investigation by \citet{Thackeray47}, followed by a measurement made by \citet{Babcock48}. This star was the first WD ever observed with spectropolarimetric techniques.

\subsubsection{WD\,0415$-$594 = $\epsilon$\,Ret = HD\,27442B (VB: K2 at 13\arcsec)}%
This star is the stellar companion of planet-host star HD\,26442A (a K2 star at 13\arcsec) and was recognised as a WD by \citet{Chaetal06}. It was observed for the first time in spectropolarimetric mode in this work, with no field detection and $\sigma = 0.3$\,kG. Since we obtained only one measurement, it may be useful to re-observe it. 

\subsubsection{WD\,0433$+$270 = HD\,283750B (VM: K2 at 124\arcsec)} %
This is the secondary of a visual binary (the companion is a spectroscopic K2 binary with 124\arcsec\ separation). The early broadband polarimetric observations by \citet{LanAng71} (who had consider it as DC, and measured $V/I = 0.03 \pm 0.11$\,\%) do not set a strong constraint on the star's magnetic field. In this work we report one FORS2 ($\sigma = 1.5$\,kG) and one ISIS ($\sigma=5$\,kG)  measurement, with no field detection. From ISIS spectroscopy of H$\alpha$ we set the upper limit for \bs\ to 100\,kG. 

\subsubsection{{\bf WD\,0503$-$174} = {\bf LHS 1734} -- 4.3\,MG}%
A magnetic field was discovered in this star by \citet{Beretal92}.  The separation of the sigma components reveals a field of about $\bs = 4.3$\,MG. Based on their mass estimate of $0.38\,M_\odot$, \citet{Giaetal12} suggested that the star may be a unresolved DD. However, \citet{Bloetal19} derived for the stellar mass the higher value of $0.53\,M_\odot$. Both \citet{Holletal18} and \citet{Genetal19} provide H-rich mass values of about 0.50\,$M_\odot$.  We conclude that the star is no longer a strong DD candidate. The star should be re-observed to check for variability and possibly monitored.

\subsubsection{{\bf WD\,0553$+$053} = {\bf G\,99-47} -- 15\,MG}%
The magnetic field of WD\,0553$+$053 was discovered by \citet{AngLan72} on the basis of weak broad-band circular polarisation of about 0.3--0.4\,\%. A search for variability by those authors detected no statistically significant variations over any time scale from a few tens of seconds to 1\,yr. The star was re-observed by \citet{Lieetal75} and by \citet{PutJor95}. \citet{Lieetal75} interpreted the available spectroscopy and intermediate-band spectropolarimetry as revealing a field modulus of 15\,MG and a longitudinal field of 5.6\,MG. Re-analysis of the data of \citet{PutJor95} by \citet{BagLan20} yields $\bz = 6.2 \pm 1$\,MG and $\bs = 13.5 \pm 0.5$\,MG, essentially the same values found 20\,yr earlier. There is presently no evidence of variations on any time scale up to decades. Gaia does not provide parallax or proper motion, but these are reported by \citet{vanAetal95}.

\subsubsection{WD\,0642$-$166 = Sirius B (VB: A0 at 7.5\arcsec)}%
In 1844 F.W.Bessell identified invisible companions of Sirius and Procyon from meridian position observation of the wobbly proper motions of these stars. An orbit for Sirius B was computed in 1850 by C. A. F. Peters, with a semi-major axis of 2.4\arcsec, and a period of 50 years. In 1862 Sirius B was observed visually by Alvan Clark with his new 18" objective. Another 80 years would pass before S. Chandrasekhar would clearly explain the nature of this first discovered WD in the 1930s. 

H$\alpha$ profiles were observed at a spectral resolution of 0.56\,\AA\ per pixel with the STIS instrument of the HST by \citet{Joyetal18}; the absence of significant Zeeman splitting at H$\alpha$ (see their Fig.~7) allows us to set an upper limit on \bs\ of about 80-100\,kG. The difficulty of observing a WD of magnitude $V = 8.4$ only a few arcsec from a companion almost 10\,mag brighter is so severe that no spectropolarimetry has been attempted.

\subsubsection{WD\,0644$+$025 = G\,108-26}%
This high mass DA WD was observed in spectropolarimetric mode for the first time with ISIS in this work with no field detection ($\sigma \simeq 6$\,kG). Because it has been observed only once, and with low precision, so it should be re-observed.

\subsubsection{WD\,0644$+$375 = G\,108-26}%
Observed in broadband circular polarisation by \citet{Angetal81} ($V/I = 0.029 \pm 0.057$\,\%, $-0.050 \pm 0.038$\,\%). Observed twice in spectropolarimetric mode by \citet{SchSmi95}, who measured $\bz = 4.5 \pm 7.8$\,kG and  $\bz = 1.9 \pm 1.6$\,kG, and with ISIS by \citet{BagLan18} ($\bz = 0.75 \pm 0.91$\,kG). A new ESPaDOnS measurement presented in this work has $\sigma=0.35$\,kG. None of these measurements represents a field detection. ESPaDOnS data allow us to estimate the upper limit of 50\,kG to \bs.

\subsubsection{WD\,0655$-$390 = GJ\,2054}%
Identified as a WD by \citet{Subetal08} and observed for the first time in spectropolarimetric mode twice with FORS2 (this work). Both measurements are non detection with $\sigma \simeq 0.5$\,kG. FORS2 spectra set also the limit to $\bs \la 300$\,kG.

\subsubsection{WD\,0657$+$320 = GJ\,3420}%
Observed only once and for the first time with ISIS (this work), with no field detection. The spectrum shows only an extremely weak H$\alpha$ from which we set 300\,kG as the upper limit for \bs. Due to the weakness of H$\alpha$, the sensitivity of our \bz\ measurement is quite low ($\sigma \simeq 35$\,kG). 

\subsubsection{WD\,0727+482A = G\,107-70A (uDD)\label{Sect_0727}}%
This is a marginally resolved visual binary degenerate system with a separation of $\simeq 0.6"$, which is actually a member of a quadruple system -- the second pair (the CPM companion system G\,107-69) at about 103$"$ separation is an unresolved main sequence - main sequence binary of spectral type M. The orbital period of the DD has been found to be 20.5\,yr \citep{Stretal76,Haretal81,Haretal93,Tooetal17}. 

Gaia DR2 reports positions but no parallax or motions of this pair, but full astrometry of the CPM dM companions. Previous parallax and proper motion studies make it clear that these four stars from a physically associated system \citep{Tooetal17}. The separation of G\,107-69 from G\,109-70 in 2015.5 was 115\arcsec. \citet{Beretal01} discuss a model of the system based on parallax, spectroscopy and photometry, with two similar cool stars of masses around 0.6\,$M_\odot$ and effective temperatures of about 5000\,K. They observe a very weak, broad feature at H$\alpha$, so at least one of the two WDs is a DA. \citet{Neletal15} provide an accurate parallax of G\,107-70 and analyse the visual binary on the basis of HST fine guide sensor observations. They determine the astrometric masses of the two components: $0.634 \pm 0.01\,M_\odot$ and $0.599 \pm 0.01\,M_\odot$.  Unusually, the more massive component is also the brighter star of the WD pair, which therefore must be hotter than the somewhat larger secondary star. 

We adopt a hybrid model, taking the masses from \citet{Neletal15}, and the $\log g$ and radius values from the WD mass-radius relation. However, appropriate values of \teff\ need to be estimated. From \citet{Beretal01} we adopt a mean $\teff \approx 5000$\,K for the system.  However, because \citet{Neletal15} report that the more massive component is the brighter of the two stars, we cannot use the individual mass and temperature solution adopted by \citet{Beretal01}, which have the lower mass star as the larger and brighter member.  We try to identify a more consistent model as follows.

It appears widely agreed that the magnitude difference between the two components is about 0.3\,mag. This allows us to estimate that the integrated fluxes $f_1$ and $f_2$ from the two stars, proportional to $\teff^4 R^2$, are approximately in the ratio $f_1/f_2 \approx 1.32 \approx T_1^4 R_1^2/T_2^4 R_2^2$. Finding radii corresponding to the masses from the mass-radius relation, finally $T_1/T_2 \approx 1.09$. Starting from the system \teff\ value of about 5000\,K,  bracketing temperatures of $T_1 = 5225$\,K and $T_2 = 4775$\,K provide the necessary flux ratio, and the two WDs together, with similar fluxes,  would be expected to have a combined energy distribution roughly the same as that of a single WD of $\teff = 5000$\,K. We then estimate ages using the on-line Montreal cooling curves. 

Figure~18 of \citet{Beretal01} shows a very weak and broad H$\alpha$, quite different in shape from the normal profile of this line in cool stars, and very different from the line profile computed with their adopted model. Our ISIS Stokes $I$ spectrum is featureless in the blue but does show a weak feature at H$\alpha$. Like that illustrated by \citet{Beretal01}, it is bowl-shaped, about 40\,\AA\ wide and 2--3\,\% deep. If we assume that this feature is broadened by a very inhomogeneous magnetic field, the field would need to have a typical strength of several hundred kG. This would set a rather weak upper limit on the DA WD component of the DD.  Further constraint on possible magnetic fields can be obtained with polarimetry. WD\,0727+482AB was observed in polarimetric mode by \citet{Valetal03}, but their Fig.~7 shows a flat intensity spectrum around H$\alpha$. Although our polarised spectra show a very weak H$\alpha$ line, we are not able measure \bz, but we note that there is no evidence in our spectrum of significant circular polarisation within the H$\alpha$ feature. Another constraint may be provided by the lack of detection of polarisation in the continuum. The double degenerate was observed also by \citet{LanAng71}, who measured $V/I=-0.03 \pm 0.10$\,\% and $0.007 \pm 0.022$\,\% \citep[the same measurement is also reported by][]{Angetal81}. We estimate that neither component has a longitudinal magnetic field stronger than 1\,MG.

We are not able to assign a spectral classification to the companion WD\,0727+482B = G\,107-70B (uDD), which will be counted in the sample total, but not for detailed statistical purposes.

\subsubsection{WD\,0728$+$642 = G\,234-4 -- suspected magnetic}%
\citet{Putney97} obtained a $3\,\sigma$ detection: $\bz = 39.6 \pm 11.6$\,kG. We observed the WD with ISIS and measured $\bz=14 \pm 13$\,kG using H$\alpha$ (this work). We are not able to confirm the field detection, although our lack of detection could be simply do to geometry. At present we continue to consider the star as suspected magnetic, but the star should be re-observed in spectro-polarimetric mode. 

\subsubsection{WD\,0751$-$252 =	SCR\,J0753$-$2524 (VB: M0 at 400\arcsec)}%
Newly discovered as a WD within 20\,pc by \citet{Subetal08}, who originally classifed as a DC. Later, \citet{Giaetal12} obtained a spectrum that shows a weak H$\alpha$. The star has a visual M0 companion at 400\arcsec. We observed it in spectropolarimetric mode for the first time with FORS2, but with a poor choice of the instrument setting: we used grism 600B, while grism 1200R would have allowed us to check if the weak H$\alpha$ is polarised. There is no evidence of circular polarisation (this is not 100\% true, but we think it is spurious; otherwise we would see splitting in H$\alpha$). The most useful upper limits to the strength of its magnetic field comes from low {\it S/N} UVES archive data, that show an H$\alpha$ with no hint of Zeeman splitting, and no broadening. This allows us to set the limit to \bs\ to $\sim 0.5$\,MG. The star could be re-observed in spectropolarimetric mode around H$\alpha$, although, given the weakness of the spectral line, sensitivity will not be very high.

\subsubsection{WD\,0752$-$676 = GJ\,293}%
Observed in broadband circular polarisation by \citet{Angetal81} ($V/I=-0.018 \pm 0.064$\,\% and $0.147 \pm 0.066$\,\%). We observed it in spectropolarimetric mode using FORS2+1200R, with much higher sensitivity, and we obtained no \bz\ detection ($\sigma \simeq 0.4$\,kG). There is no hint of Zeeman splitting in archive UVES data, setting the limit of 50\,kG for \bs. Since the star was observed only once in spectropolarimetric mode, it should be re-observed. 

\subsubsection{{\bf WD\,0810$-$353} = {\bf UPM J0812-3529} -- 30\,MG}%
Discovered as strongly magnetic by \citet{BagLan20}, the star should be re-observed around H$\alpha$ for a better characterisation of its magnetic field. Note that the interpretation of the complex observed $I$ and $V/I$ spectrum by \citet{BagLan20} is very tentative, and the field strength is a rough estimate. 

\subsubsection{WD\,0821$-$669 = SCR\,J0821$-$6703}%
Spectroscopically confirmed DA WD by \citet{Subetal07}. We observed it for the first time in polarimetric mode with FORS2+1200R (this work), finding that the star is non magnetic with uncertainty of $\sim 4.5$\,kG. Our FORS2 data do not show obvious splitting in Stokes~$I$, but both spectral resolution and {\it S/N} are low, so we adopt $\bs \la 300$\,kG. 

\subsubsection{WD\,0839$-$327 = CD$-$32\,5613 (uDD?)}%
\citet{Kawetal07} comment that \citet{Braetal90} observed line profile variations, and suggest that this is a DD with a primary of $\teff = 9340$\,K and a secondary of $\teff = 7500$\,K.  \citet{Holletal18} and \citet{Genetal19} both give $log g = 7.79$ and $M = 0.48 M_\odot$, which suggests that the evidence of a double WDs may be not be strong. We are therefore treating it as a suspected DD, but still as a single star for statistical purposes.

\citet{Kawetal07} attempted to detect a field obtaining a null measurement with a 2.8\,kG uncertainty. A higher-precision non-detection was obtained by \citet{Aznetal04} ($\bz = 0.35 \pm 0.25$\,kG). From UVES archive data we suggest $\bz \la 50$\,kG.

\subsubsection{WD\,1019$+$637 = GJ\,1133}%
Observed in broadband circular polarisation by \citet{Angetal81} ($V/I=+0.02 \pm 0.05$\,\%). \citet{SchSmi95} measured $\bz = -0.1 \pm 5.1$\,kG. In this work we report a new measurement with ESPaDOnS with no detection (uncertainty of 1.2\,kG). For \bs\ we set the limit to 50\,kG.

\subsubsection{WD\,1121$+$216 = Ross\,627}%
\citet{SchSmi95} measured $\bz = 7.5 \pm 3.5$\,kG. We obtained one measurement with ESPaDOnS (this work) with non detection ($\sigma \simeq 0.8$\,kG). For \bs\ we set an upper limit of 50\,kG. 

\subsubsection{WD\,1134$+$300 = GD\,140 (uDD?)}%
Possibly this is a relatively close DD (relatively close because there is an apparent velocity difference between the two WDs). We note that from its proper motion anomaly, \citet{Keretal19} identified this star as a suspected binary system, although their results do not provide any guidance about whether the possible companion is a WD, a brown dwarf, or even a massive planet close to the primary. However, \citet{Giaetal12} show an excellent fit to the Balmer lines of this WD, from which they deduce a distance that is very close to the value provided by Gaia. If we suppose that the WD is actually an uDD, probably to be consistent with the single star model, we would need a system of two DA WDs, both with \teff very similar to that derived for the WD as a single star. For this system to have the same radiating surface area as the single star model, the two stars would need to be substantially smaller (and due to the mass-radius relation, more massive) than the single star model. We estimate that they would both need to have $M \approx 1.25 - 1.3 M_\odot$. This is a considerably larger mass than that of any star in the 20\,pc volume, and the value is dangerously close to the Chandrasekhar limit. In addition, we would expect such high mass (and corresponding gravity) to lead to Balmer lines significantly broader than those computed by \citet{Giaetal12} for the single star model with $M = 0.97 M_\odot$. Therefore we treat this system as a single WD, but note that it could be an uDD.

\citet{SchSmi95} measured $\bz = -0.6 \pm 2.2$\,kG, and \citet{Valetal06} measured $8.9 \pm 4.5$\,kG and $3.5 \pm 2.7$\,kG. Two consecutive non-detections were obtained with ISIS (one in the blue arm, one in the red arm) by \citet{BagLan18}: averaged together, the two measurements give $\bz = 0.3 \pm 0.4$\,kG. In this work we report a new non detection with ESPaDOnS ($\sigma = 0.8$\,kG). The core of H$\alpha$ in our ESPaDOnS exposure is rounded and shallower than those of WDs with similar fundamental parameters, and has a suggestively squarish shape. A similar feature is observed in our ISIS spectra. The non-detection of a field indicates that the observed broadening is not due to Zeeman effect. Assuming that the star is not an uDD, we may ascribe the rounding of the H$\alpha$ line core to stellar rotation. 

\subsubsection{WD\,1148$+$687 = PM\,J11508+6831}%
We have only one measurement with low {\it S/N} ($\bz = -2.2 \pm 2.6$\,kG) with ESPaDOnS (this work); $\bs\la 100$\,kG. H$\alpha$ and H$\beta$ have slightly rounded cores, possibly a symptom of a narrow core broadened by fast rotation.

\subsubsection{WD\,1236$-$495 = V* V886 Cen}%
Observed in broadband circular polarisation by \citet{Angetal81} ($V/I=0.018 \pm 0.070$\,\%). With spectropolarimetry. \citet{Kawetal07} measured  $\bz = 2.58 \pm 6.23$\,kG. We obtained a high-precision non-detection with FORS2 ($\bz=0.1 \pm 0.5$\,kG). From UVES and X-Shooter archive data we set the upper limit of \bs\ to 100\,kG.

\subsubsection{WD\,1257$+$037 = Wolf~457}%
The star was originally classified as DC. \citet{AngLan70Further} measured $V/I=+0.07 \pm 0.11$, and \citet{Angetal81} report the further measurement of $V/I=0.27 \pm 0.21$\,\%. \citet{Putney97} discovered that the star has H Balmer lines; from them she measured \bz= $11.2 \pm 18$\,kG and classified this as DAQZ star, but with no further comments. We could not find any reference to the presence of metal lines, and \citet{Zucetal03} classify it as DA. From our FORS2 polarisation spectrum (with grism 1200R) we obtained a non detection with $\sigma = 1$\,kG.

\subsubsection{{\bf WD\,1309$+$853} = {\bf G\,256-7} -- 5\,MG}%
From splitting of H$\alpha$, \citet{Putney95} measured a $\sim 5$\,MG field. We note that from the description of the method adopted for the field measurement, what is referred to as "effective field" by \citet{Putney95} probably corresponds to \bs; from the plots shown in Fig.~2 of that paper we derive $\bs = 5.4 +\pm 0.5$\,MG. From Putney's figure, it is clear that \bz\ is non-zero, but it was not measured. Because only one measurement is available, the star should be checked for variability.

\subsubsection{WD\,1316$-$215 = NLTT 33669}%
One of the faintest DA WDs in the local 20\,pc volume ($V=16.7$). \citet{KawVen12} used FORS1 to measure $\bz = -3 \pm 25$\,kG; we obtained two more accurate measurements using FORS2 with $\sigma \simeq 3$\,kG. FORS2 $I$ spectra set the upper limit of \bs\ to about 100\,kG.

\subsubsection{{\bf WD\,1315$-$781} = {\bf LAWD 45} -- 5.5\,MG}%
Originally classified as DC, \citet{BagLan20} discovered that its low resolution spectrum (obtained with FORS grism 300V) reveals H$\alpha$ and H$\beta$, and the star is magnetic, with no evidence of variability over a time scale of $\sim 1$ week. For a better characterisation of this magnetic field, the star should be re-observed with grism 1200R.

\subsubsection{WD\,1327$-$083 = BD-07\,3632 (VB: M4.5 at 503\arcsec)}%
Observed in broadband circular polarisation by \citet{AngLan70Further} ($V/I=+0.21 \pm 0.10$; the same measurement is reported also by \citeauthor{Angetal81} \citeyear{Angetal81}). Two FORS1 observations were published by \citet{Joretal07}, $\bz = -0.5 \pm 0.5$\,kG and $\bz=-0.4 \pm 0.5$\,kG; and one FORS2 measurement with grism 1200B and one ISIS measurement with grism 600B were published by \citet{BagLan18}: $\bz=-0.9 \pm 0.4$\,kG and $\bz=0.3\pm0.2$\,kG. From UVES archive spectra we estimate $\bs \la 50$\,kG. The star is member of a visual binary system. 

\subsubsection{WD\,1334$+$039 = Wolf~489}%
The star was in the past classified as DZ in older literature, then re-classified as DC by \citeauthor{Sioetal90} (\citeyear{Sioetal90}, see \citeauthor{Faretal09} \citeyear{Faretal09}).
Later, the star was again re-classifed, as a DA, by \citet{Giaetal12}. \citet{Angetal81} measured $V/I=-0.041 \pm 0.091$\,\%. Most recently we obtained one measurement with ISIS (non detection, $\sigma = 15$\,kG) and one with ESPaDOnS (from which, due to low S/N, we could not obtain any useful \bz\ measurement; hence it is not reported in Table~\ref{Table_Log}). Our ISIS spectrum shows indeed a (weak) H$\alpha$ line red-shifted by 5\,\AA with respect to absorption H$\alpha$ in the night sky background. \citet{Sioetal94} briefly discuss the strange motion of this star and place the star in the halo ellipse.
An upper limit from our ISIS spectra for \bs\ is 300\,kG. 

\subsubsection{WD\,1345$+$238 = LP\,380-5 (VB: M5 at 199\arcsec)}%
Cool WD in a binary system with an M5 star at 200\arcsec. \citet{Giaetal12} and \citet{Holbetal16} classify it as DA, although H$\alpha$ is extremely weak. The star was observed in broad-band circular polarimetric mode by \citet{LieSto80}, who measured $V/I=0.006 \pm 0.08$\,\%, and by us with ISIS (this work), with a non detection (a very weak H$\alpha$ is possibly visible in our ISIS intensity spectrum). Due to the fact that the spectrum is nearly featureless, the strongest constraint of \bz\ comes from the broadband circular polarisation measurements by \citet{LieSto80} ($\abz \la 1.5$\,MG).

\subsubsection{{\bf WD\,1350$-$090} = {\bf GJ 3814} -- 460\,kG, subtly variable}%
Discovered by \citet{SchSmi94}, who measured $\bz = 85 \pm 9$\,kG. We have obtained four polarised spectra with ESPaDOnS that show \bs\ $\simeq 450 - 465$\,kG. Data will be published in a forthcoming paper.

\subsubsection{WD\,1408$-$591 =  UCAC4 154$-$133995}%
Newly discovered WD by \citet{Holletal18}. Our two FORS2 observations (this work) show that it is a DA. Both measurements are 2\,$\sigma$ detections ($\sigma \simeq 0.8$\,kG). Therefore we currently consider the star non-magnetic, but further higher {\it S/N} measurements would be useful to check if the star is weakly magnetic.

\subsubsection{WD\,1544$-$377 = HD\,140901B (VB: G6 at 15\arcsec)}%
\citet{Kawetal07} presented two non-detections with FORS1 with 6--7\,kG uncertainty. We have obtained a new high-precision FORS2 measurement with grism 1200R (this work) which confirms non detection ($\sigma = 0.6$\,kG). From UVES archive data we set the upper limit for \bs\ to 50\,kG. The star had been observed in broadband circular polarisation by \citet{Angetal81} ($V/I= -0.043 \pm 0.027$\,\%).

\subsubsection{WD\,1620$-$391 = CD$-$38\,10980 (VB: G5 at 345\arcsec)}%
One (low-precision) measurement by \citet{Kawetal07}  ($\bz = -2.96 \pm 2.60$\,kG), one by \citet{KawVen12} ($\bz = 0.5 \pm 3.5$\,kG) and six high-precision measurements by \citet{Joretal07} (best uncertainty of 0.3\,kG), all non-detections.  \citet{Angetal81} had obtained broadband circular polarimetry ($V/I = 0.016 \pm 0.014$\,\%). UVES archive spectra set the upper limit of \bs\ to 50\,kG.

\subsubsection{WD\,1630$+$089 = G\,138-38}%
Discovered as DA WD within 20\,pc by \citet{Sayetal12}. Parameters for this star have been derived by \citet{Subetal17,Holletal18,Bloetal19}. The gravity and mass derived by \citet{Bloetal19} are quite discrepant from the other determinations. We prefer the results of \citet{Subetal17}, which are based on a spectrum and photometry from $B$ to $K$, and are concordant with those of \citet{Holletal18}. 

We obtained one ESPaDOnS measurement(this work), which is a non-detection with 2\,kG uncertainty.
The upper limit for \bs\ from H$\alpha$ spectroscopy is 50\,kG. Possibly more measurements would be useful to confirm that the star is non magnetic. 

\subsubsection{WD\,1647$+$591 = G\,226-29}%
\citet{Angetal81} obtained broadband circular polarimetry ($V/I = 0.046 \pm 0.022$\,\%). Three non-detections with ISIS published by \citet{BagLan18} each with 0.5\,kG uncertainty. ISIS H$\alpha$ spectroscopy set the upper limit for \bs\ to 100\,kG.

\subsubsection{{\bf WD\,1703$-$267} = {\bf UCAC4\,317-104829} -- 8\,MG variable}%
Newly discovered as a WD by \citet{Holletal18}. Discovered as a strongly magnetic star by \citet{BagLan20}, who showed that the star is variable on a time scale of some weeks or less.  Further observations are required for monitoring purpose.

\subsubsection{WD\,1756$+$827 = EGGR\,199}%
\citet{SchSmi95} measured $\bz = -2.8 \pm 4.2$\,kG. One ESPaDOnS measurement (this work) results in a non detection with 0.8\,kG uncertainty for \bz. H$\alpha$ spectroscopy sets the upper limit for \bs\ to 50\,kG.

\subsubsection{WD\,1814$+$124 = LSR J1817+1328}%
Discovered as a DA WD by \citet{Lepetal03}. We observed this star once with ISIS (this work) obtaining a non-detection with large uncertainty ($\sigma \simeq 10$\,kG) because only a weak H$\alpha$ is visible. From our ISIS spectra we also deduce a 200\,kG upper limit for \bs. 

\subsubsection{WD\,1820$+$609 = G 227-28}%
Cool DA. \citet{LieSto80} measured $V/I=0.066 \pm 0.11$\,\%, then \citet{Putney97} erroneously listed it as highly polarised \citep[see][Sect.~4]{Lanetal16}. ISIS observations presented in this paper (non-detection) show that H$\alpha$ is too weak to measure \bz, which is therefore better constrained by the broadband circular polarisation measurements of \citet{LieSto80} ($\abz \la 1.5$\,MG). From H$\alpha$ spectroscopy we deduce that $\bs \la 50$\,kG.

\subsubsection{WD\,1823$+$116 = UCAC4\,508-079937}%
Newly discovered as a WD within 20\,pc by \citet{Holletal18}. Originally classified as DC, it actually shows an extremely weak H$\alpha$ \citep{Treetal20}. We observed it with ISIS (this work) with no field detection, but based on loose constraints that come from our non detection of polarisation in the continuum we estimate  $\abz \la 2$\,MG. A stronger constraint comes from inspection of H$\alpha$: $\bs \la 0.5$\,MG.

\subsubsection{{\bf WD\,1829$+$547} {\bf = G\,227-35} -- 120\,MG}%
Strongly magnetic WD. Discovered as magnetic by \citet{Angetal75}, using both broadband polarimetry and low-resolution spectropolarimetry. It was observed again in spectropolarimetric mode by \citet{Cohetal93} who suggested the star has a dipolar field seen pole-on with a dipolar field strength of 130\,MG. \citet{PutJor95} observed the star again, and revised this estimate to a dipolar field strength of 170-180\,MG. It does not appear to vary. We observed the same star with ISIS twice in circular polarisation and once in linear polarisation, and will report on these data on a separate paper.

\subsubsection{{\bf WD\,1900$+$705} = {\bf Grw\,$+70\degr$\,8247} -- 200\,MG}%
First WD discovered to be magnetic \citep[by][]{Kempetal70}, it has a magnetic field of the order of hundreds MG, and shows little to nearly no variability. See \citet{BagLan19a} for a review and analysis of its polarimetric characteristics.  

\subsubsection{WD\,1919$+$061 = UCAC4\,482-095741}%
Discovered as WD by \citet{Holletal18}. Our new observation with ISIS (this work) confirms the star is a DA \citep[see also][]{Treetal20}, and that it does not show the presence of a magnetic field. The H$\alpha$ core is slightly rounded. The non detection of a mean longitudinal field (with a $\sigma \simeq 6$\,kG) suggests that this broadening is not due to the Zeeman effect. The upper limit to \bs\ is about 150\,kG.

\subsubsection{WD\,1919$+$145 = GD\,219}%
Observed by \citet{SchSmi95} ($\bz = 5.7 \pm 6.6$\,kG), then two times by \citet{Joretal07}, who measured $\bz = -1.5 \pm 0.8$\,kG and $\bz=-0.8 \pm 0.8$\,kG. UVES archive data allow us to set the upper limit for \bs to 50\,kG.

\subsubsection{WD\,1935$+$276 = G\,185-32}%
A pulsating star (ZZ Ceti type). One ISIS measurement was presented by \citet{BagLan18} ($\bz = -0.3 \pm 0.3$\,kG) and one additional ESPaDOnS measurement is reported in this work, also a non-detection with uncertainty of 0.6\,kG. Previously observed also by \citet{SchSmi95} but with lower precision ($\bz = -8.5 \pm 10.5$\,kG), and in broadband circular polarisation by \citet{Angetal81} ($V/I=0.00 \pm 0.09$\,\%). From ESPaDOnS H$\alpha$ spectroscopy we can estimate that the the upper limit for \bs\ is 50\,kG.

\subsubsection{{\bf WD\,1953$-$011} = {\bf GJ 772} -- 100 to 500\,kG variable and modelled}%
Observed spectropolarimetrically by \citet{SchSmi95} who did not detect a field ($\bz = 15.6 \pm 6.6$\,kG). \citet{Koeetal98} observed a Zeeman pattern in a noisy high-resolution SPY H$\alpha$ spectrum and suggested that a field of about 93\,kG was present. \citet{MaxMar99} obtained an H$\alpha$ spectrum which showed apparent Zeeman sigma components in $I$ corresponding to a field of about 500\,kG, and suggested that the star could actually be a double degenerate system with both components magnetic. \citet{Maxetal00} described a series of $I$ spectra that supported a model of WD\,1953$-$011 with a global dipolar field with polar field strength of order 100\,kG, together with a spot with a field of order 500\,kG that is sometimes visible and sometimes not, as the star rotates. Finally, a series of polarised spectra obtained with FORS1 and on the Russian 6-m telescope revealed a rotation period of 1.448 d and confirmed the basic idea of a global dipole with a single spot of much higher field \citep{Valetal08}.

\subsubsection{WD\,2007$-$303 = GJ\,2147}%
Two low-precision measurements by \citet{Kawetal07} ($\bz = 5.5 \pm 8.1$\,kG, $=3.7 \pm 3.$\,kG), one low-precision measurement with FORS1 by \citet{Lanetal12} ($\bz=1.1 \pm 2.8$\,kG) and two higher-precision FORS1 measurements by \citet{Joretal07}  ($\bz = 0.3 \pm 0.4$\,kG, $-0.5 \pm 0.4$\,kG) all non detections. From UVES archive data we set the upper limit for \bs to 50\,kG.

\subsubsection{WD\,2032$+$248 = HD 340611}%
Low precision from narrow-band polarimetry by \citet{AngLan70a} ($V/I=0.06 \pm 0.08$\,\%). One measurement by \citet{SchSmi95} ($\bz = 1 \pm 2.3$\,kG), one ISIS measurement published by \citet{BagLan18} ($\bz= 0.0 \pm 0.2$\,kG), and one ESPaDOnS measurements presented in this work ($\sigma = 0.3$\,kG), all non detections. ESPaDOnS spectroscopy of H$\alpha$ sets the upper limit for \bs\ to 50\,kG.

\subsubsection{WD\,2039$-$682 = EGGR\,140}%
\citet{Koeetal98} found that H$\alpha$ has a broadened core that could be explained either by a 50\,kG magnetic or a $\vsini = 80$\,\kms.
\citet{Kawetal07} measured $\bz = -6.0 \pm 6.4$\,kG, and suggested that the reason for broadening could be rotation rather than the Zeeman effect, but also the possibility that a magnetic spot is present at the surface of the star, but not visible at the time of their observation. We obtained four more high-precision \bz\ measurements with FORS2+1200R (this work), all non-detections, with $\sigma = 0.5-0.8$\,kG, which confirm the conclusion that the observed H$\alpha$ broadening is due to rotation. Formally, we still set the upper limit of \bs to 50\,kG.

\subsubsection{{\bf WD\,2047$+$372} = {\bf GJ\,4165} -- 60\,kG variable, modelled.}%
Discovered as magnetic by \citet{Lanetal16}, and monitored and modelled by \citet{Lanetal17} with ESPaDOnS data. Currently the weakest WD field that has been modelled in detail, based mainly on a series of 18 ESPaDOnS spectra. The rotation period, determined from the variation of \bz, is 0.243\,d. Originally, the star had been observed by \citet{SchSmi95}, who did not detect its weak and sign reversing field ($\bz = -42 \pm 59$\,kG and $-2.5 \pm 4.6$\,kG). 

\subsubsection{WD\,2048$+$263A  = G\,187-8A (uDD)}\label{Sect_DA2048}%
The suspected binary system WD\,2048$+$263 was earlier thought to be just outside the 20\,pc limit, but Gaia has made it a clear member of the 20\,pc volume. It is described by \citet{Beretal01} and \citet{Giaetal12} as a probable DD on the basis of a derived mass of $M = 0.24\,M_\odot$, and listed by \citet{Tooetal17} as unconfirmed but probable DD. The low derived mass has been confirmed by \citet{Holletal18}. The derived mass is so low (i.e. the star is so overluminous compared to other WDs) that the automated fitting routine of \citet{Genetal19} rejects it as a white dwarf. We consider this object to be a well-established uDD system.

The optical spectra published by \citet{Putney97} and \citet{Beretal01} show no sign of the absorption bands of a typical dM star, so the companion is almost certainly a DA or DC WD. Modelling by \citet{Beretal01} shows that the observed H$\alpha$ line is only about half as strong as it would be if both WDs are DAs of similar temperatures, so we conclude that the system probably contains one DA and one DC (the DC companion in listed in Sect.~\ref{Sect_DC2048}). The good fit of the single star model to the photometry suggests both stars have similar temperatures. 

We adopt the most recent model temperature from \citet{Bloetal19} and assign it to both stars. We use their $\log g$ and mass values to derive the single star radius from the mass-radius relation, and decrease the model radius to 0.707 of its initial value (to divide the observed luminosity equally between two stars). We then return to the mass-radius relation to obtain the new, larger mass ($0.53 M_\odot$) and gravity ($\log g = 7.90$) of the two smaller WDs.    This is of course only a rather rough first estimate of the parameters of the two stars, but is unlikely to seriously misrepresent either. Note that the two stars in the new model of the system both have masses within the normal range, rather than the very small mass estimated for the single star model of this system, so they may not have interacted very strongly during evolution. 

The system was observed in broad-band continuum polarisation by \citet{Angetal81} who report no detection with $V/I = 0.04 \pm 0.05$\,\%.
This pair was initially classified as a DC but reclassified by as DA9 by \citet{Putney97}, who detected a weak but clear H$\alpha$, and reported $\bz = 13.6 \pm 13.7$\,kG, a non-detection. It should definitely be re-observed with spectropolarimetry to be checked for magnetic field with higher precision.

\subsubsection{WD\,2057$-$493 = WT\,765 (VT: at 5$"$, 64\arcsec)}%
Newly identified as a nearby DA WD by \citet{Subetal17}. For this work we obtained one FORS2+1200R measurement which resulted in a non-detection with a 1.8\,kG uncertainty. Overall this is a triple system in which the WD is separated enough to be the only object in the slit. H$\alpha$ is quite visible but slightly asymmetric, with a feature at 6516\,\AA. The upper limit to \bs\ is 300\,kG. It should be re-observed again to confirm that it is not magnetic.

\subsubsection{WD\,2117$+$539 = GJ\,1261}%
Observed twice by \citet{SchSmi95}, who measured $\bz = -11.5 \pm 6.9$\,kG and $ 2.5 \pm 2.7$\,kG. One ISIS measurement was reported  by \citet{BagLan18} ($\bz = 0.0 \pm 0.2$\,kG) and one ESPaDOnS measurement is presented in this work ($\bz = 1.2 \pm 0.4$\,kG), all non detections. ESPaDOnS H$\alpha$ spectroscopy allows us to set the upper limit for \bs\ to 50\,kG. 

\subsubsection{WD\,2140$-$072 = PHL 1716?}%
Recently identified as a WD by \citet{Holletal18}. One ISIS measurement presented in this work is a non detection with 0.8\,kG uncertainty. ISIS spectroscopy of H$\alpha$ suggests an upper limit for \bs\ of 100\,kG.

\subsubsection{{\bf WD\,2150$+$591} = {\bf  UCAC4 747-070768} -- 800\,kG, variable (VB: M2 at 15\arcsec)}%
This star was discovered to be a strong WD candidate at a distance of only 8.47\,pc, and thus to be a probable member of the (supposedly already complete) 13\,pc volume-limited WD sample, by \citet{Schetal18}. It was spectroscopically confirmed as a WD, and found to have a magnetic field, by \citet{LanBag19a}, who reported  two ISIS measurements. These ISIS spectra showed clearly that the field is variable with a period of hours or days.  We have monitored the star with one ESPaDOnS observation and several more ISIS spectra. These observations and a model of the star's magnetic field will be presented in a forthcoming paper.  According to \citet{Schetal18} the star is member of a wide binary system including a M2 star, but was inadvertently omitted from the study of Sirius-like systems with a magnetic WD member by \citet{LanBag20}.

\subsubsection{WD\,2159$-$754 = V* CD Oct}%
Observed twice by \citet{Kawetal07} who obtained $\bz = -7.8 \pm 8.6$\,kG and $ -11.8 \pm 7.1 $\,kG.
The one FORS2+1200R measurement presented in this work is another non detection with 0.8\,kG uncertainty. From UVES archive data we set the upper limit for \bs to 50\,kG.

\subsubsection{WD\,2211$-$392 = LEHPM 4466 -- suspected magnetic}%
Observed in polarimetric mode for the first time in this work, with three FORS2 measurements: one $3\,\sigma$ detection and two non detections ($\sigma \simeq 1$\,kG). We cannot consider this star as a magnetic one, but it may well be a weakly and variable MWD. The situation of WD\,2211$-$392 may be somewhat similar to that of WD\,1105$-$084, a DA star outside of the local 20\,pc volume, in which the (weak) magnetic field is detected only in a fraction of the observations \citep{Aznetal04,BagLan18}. From the FORS2 intensity spectrum we estimate an upper limit on \bs\ of 300\,kG.

\subsubsection{WD\,2246$+$223 = EGGR 155}%
Observed in polarimetic mode for the first time in this work, with two ESPaDOnS measurements (non-detections with 1.4\,kG  uncertainty). Upper limit from H$\alpha$ from ESPaDOnS spectroscopy is $\bs \la 50$\,kG.

\subsubsection{WD\,2248$+$293A = WD\,2248$+$294A = GJ 1275A (uDD)}%
The system WD\,2248$+$294 was recently classified as WD by \citet{Holbetal08}. The parameters derived by \citet{Giaetal12}, $\teff = 5592$\,K, $\log g = 7.55$ and  $M = 0.35\,M_\odot$, strongly suggests that this object may be an uDD. The parameters obtained by \citet{Holletal18}, \citet{Genetal19}, and \citet{Bloetal19} are all very close to 5615\,K, 7.71 and 0.43\,$M_\odot$. As the derived mass is always found to be  too small to be the result of single star evolution, it appears that this object is an uDD \citep[a view supported by][]{Holletal18}. Accordingly, we count this system as two objects in the statistical discussion. Because the fit obtained by \citet{Giaetal12} to the observed H$\alpha$ line is essentially perfect, we conclude that the system is composed of two DAs with very similar parameters and ages. We adopt the \teff\ value of \citet{Bloetal19}; we assign it to both stars; we take the single star model radius and reduce it to 0.70 of its initial values, and then retrieve the new mass and $\log g$ values of the two stars using the mass-radius relation. Finally we interpolate new Montreal cooling ages as described in the text. 

This system was observed polarimetric for the first time in this work, with only one ISIS measurement (non detection with 4\,kG uncertainty, but the upper limit for \abz\ should be increased by a factor of two because of binarity). From H$\alpha$ spectroscopy we estimated the upper limit for \bs\ to be 100\,kG. We see no sign of a companion in our ISIS spectra. The star should be re-observed. 

\subsubsection{WD\,2248$+$293B = WD\,2248$+$294B = GJ 1275B (uDD)}%
See section above.

\subsubsection{WD\,2307$+$548 = LSPM J2309+5506E (VB: K3 13\arcsec)}%
Observed for the first time in polarimetric mode in this work, with one ISIS measurement (non detection with 5\,kG uncertainty), the spectrum shows both H$\alpha$ and a very weak H$\beta$. From H$\alpha$ spectroscopy we set our upper limit for \bs\ to 150\,kG. The star is a member of a visual binary system with a K3 star at 13.1\arcsec. The star should be re-observed.

\subsubsection{WD\,2336$-$079 = GD\,1212}%
This star was observed in polarimetric mode for the first time in this work: twice with ISIS, once with FORS2+1200B and once with ESPaDOnS: all these measurements are non-detections, with lowest uncertainty of 0.3\,kG. \bs\ upper limit from ESPaDOnS spectroscopy is the usual 50\,kG. 

\subsubsection{WD\,2341$+$322 = LAWD 93 (VB: M3 at 175\arcsec)}%
The star was observed in broadband circular polarisation by \citet{AngLan70a} ($V/I= 0.11 \pm 0.18$\,\%), and four additional observations were published by \citet{Angetal81}. Two much higher precision non-detections were obtained by \citet{BagLan18} with 0.3\,kG uncertainty, and in this work we have presented another ESPaDOnS non-detection ($\sigma = 0.9$\,kG). From ESPaDOnS H$\alpha$ spectroscopy we set the upper limit for \bs\ to 50\,kG.

\subsection{DAZ stars}\label{Sect_DAZ}
\subsubsection{WD\,0141$-$675 = LTT\,934}
Observed two times by \citet{Kawetal07} at the Mount Stromlo Observatory with the Steward CCD spectropolarimeter (results $\bz = 2.93 \pm 5.44$\,kG, and $\bz = -1.68 \pm 4.65$\,kG), and three times by us with FORS2 and grism 1200R (this work): one measurement was obtained with accuracy similar to those by \citet{Kawetal07}, while two more recent measurements had a ten times smaller uncertainty (0.3--0.4\,kG). None of these measurements suggests the presence of a detectable field.

\subsubsection{WD\,0208$+$396 = EGGR\,168}
Observed by us once with ISIS and once with ESPaDOnS (this work), with uncertainties of $\sim 0.5$ and 1\,kG, respectively. Both \bz\ measurements are non-detections. Sharp Ca\,{\sc ii} H and K lines apparent in Espadons spectrum, Mg not visible.

\subsubsection{WD\,0245$+$541 = G\,174-74}
Observed twice with ISIS (this work), no field detection with $\sigma \simeq 12$\,kG. It is classified as DAZ by \citet{Zucetal03}, but Ca\,{\sc ii} lines are very weak in our ISIS spectra.  
It was originally measured in broadband circular polarisation by \citet{Angetal81} ($V/I = -0.14 \pm 0.09$\,\%).

\subsubsection{{\bf WD\,0322$-$019} = {\bf G77-50} -- 120\,kG}
The star was discovered to be a MWD by \citet{Faretal11} from the analysis of UVES spectroscopy. The mean field modulus \bs\ has been measured from Zeeman splitting of weak metal lines using co-added spectra and is found to be $\bs \approx 120$\,kG.  From the UVES times series, \citet{Faretal11} suggests that the rotation period could be either $\sim 1$\,d or $\sim 30$\,d, the longer period being more likely than the former.  Two FORS2 measurements by \citet{Faretal18} obtained during two consecutive nights ($\bz=-5.4 \pm 3.0$\,kG and $-16.5 \pm 2.3$\,kG) suggest that the longitudinal field may actually change at a time scale much shorter than a month. 

\subsubsection{WD\,0856$-$007 = LP\,606-32 (uDD?)}
This star could be considered to be a uDD? because the mass derived by \citet{Bloetal19} is $0.41 M_\odot$ with $\teff = 4655$\,K. 
(We note that this WD appears in Table 3 of \citeauthor{Bloetal19} \citeyear{Bloetal19} with the wrong sign for $\delta$.) However, 
\citet{Subetal17} find $\teff = 5240$\,K,$\log g = 8.10$, and $M = 0.64 M_\odot$; 
\citet{Holletal18} find $\teff = 4963$\,K and $\log g = 7.92$; and \citet{Genetal19} find $\teff = 4967$\,K, $\log g = 7.93$, and $M = 0.54 M_\odot$. We consider this WD as a possible but weakly supported uDD? candidate, and count it as a single star for the statistics. The star was observed for the first time in spectropolarimetric mode using FORS2  with grism 300V. This star was originally classified as DC by \citet{Subetal08}, but our spectra clearly show the presence of H$\alpha$ and Ca\,II H\&K lines. Our \bz\ measurement from H$\alpha$ has low precision (12\,kG uncertainty) but is a $3\,\sigma$ detection;  the star should be re-observed with higher resolution around H$\alpha$, for instance using FORS2 grism 1200R. 

\subsubsection{WD\,1202$-$232 = LP\,852-7 }
One ISIS measurement published by \citet{BagLan18} ($\bz = 0.43 \pm 0.33$\,kG) and three FORS1 measurements by \citet{Joretal07} ($\bz = 0.69 \pm 0.52$\,kG; $-0.19 \pm 0.35$\,kG, $-0.06 \pm 0.37$\,kG, as revised by \citet{Bagetal15}). From UVES SPY spectra we estimate $\bs \la 50$\,kG.

\subsubsection{WD\,1208$+$576 = G\,197-47}
 Classified as DAZ by \citet{Zucetal98} \citep[see also][]{Zucetal03}. In this work we report one \bz\ measurement with the ISIS instrument that turned to be a non detection with $\sigma \simeq 2.5$\,kG.  From non split weak H Balmer lines we deduce that the upper limit for \bs\ is 200\,kG. To confirm that is not magnetic, the star should be re-observed.

\subsubsection{WD\,1223$-$659 = WG\,21}
\citet{Wegner73} found weak Ca\,{\sc ii} H\&K lines in the spectrum of this star. Later, \citet{Kawetal07} could not see them in their spectra, and classified the star as DA. However, ESO Archive X-Shooter spectra clearly show the presence of Ca\,{\sc ii} lines, therefore we consider this as a DAZ WD. In this work we report three measurements obtained with FORS2 + grism 1200R. They are all non-detections with a typical 0.3\,kG uncertainty. From the X-Shooter archive spectra we set 200\,kG\ as the upper limit of \bs.

\subsubsection{WD\,1633$+$433 = GJ\,3965}
\citet{SchSmi95} measured $\bz= 3.6 \pm 2.9$\,kG. We obtained a new ESPaDOnS \bz\ measurement (this work), which is a non detection with $\sigma = 1.4$\,kG. From ESPaDOnS H$\alpha$ spectroscopy we obtain $\bs \la 50$\,kG.

\subsubsection{WD\,1821$-$131 = EGGR\,176}
A DAZ star with very weak metal lines: \citet{Zucetal03} have measured a Ca\,{\sc ii} K line with an equivalent width of 46\,m\AA. We obtained one measurement with ISIS and one measurement with FORS2+1200R, both non detections, with lowest $\sigma \simeq 1.2$\,kG. From ISIS spectra we estimate that the upper limit for \bs\ is 200\,kG. From UVES spectra archive we set $\bs \la 50$\,kG.

\subsubsection{{\bf WD\,2105$-$820} = {\bf LAWD 83} -- 40\,kG}
This star was suspected to have a weak magnetic field by \citet{Koeetal98}, who observed that the core of H$\alpha$ was abnormally broad, but they could not decide whether this was due to rapid rotation of $v \sin i \approx 65$\,km/s or to a magnetic field of about 43\,kG. Five FORS1 polarised spectra of the star by \citet{Lanetal12} revealed an apparently nearly constant magnetic field of $\bz \approx 10$\,kG. Later FORS2 polarised spectra by \citet{BagLan18} and \citet{Faretal18} reveal that the field sometimes decreases to $\bz \approx 4$\,kG. No period has been established yet. Our FORS2 1200B spectra show a very weak Ca\,{\sc ii} K line, so the star really is a DAZ.

\subsubsection{WD\,2326$+$049 = V* ZZ Psc }
\citet{Holbetal16} erroneously lists this 0.4\,Gyr old WD as magnetic, but actually \citet{SchSmi95} had measured $\bz = 2.8 \pm 12.8$\,kG. The FORS2 measurements of \citet{Faretal18} ($\bz = -0.7 \pm 0.5$\,kG) and our ESPaDOnS measurement (this work, $\bz = 0.0\pm 0.5$\,kG) confirm that this star is not magnetic. The ESPaDOnS spectrum provides 50\,kG as the upper limit for \bs.

\subsection{DB stars}\label{Sect_DB}
There are no DB stars in the local 20\,pc volume, except for a DBQA star (WD\,1917$-$077) which is listed in Sect.~\ref{Sect_DQs} among the DQ stars.
\subsection{DC stars}\label{Sect_DC}
\subsubsection{WD\,0000$-$345 = LAWD 1}
Based on the detection of a broad absorption feature between 4500 and 4700\,\AA\ via low-resolution spectroscopy, \citet{Reietal96} claimed that the star has a magnetic field with dipolar strength of 86\,MG. However, \citet{Schetal01} observed the same star in circular spectropolarimetric mode, and failed to detected the polarisation signal that would be expected from a magnetic star with a 86\,MG field. Furthermore, \citet{Schetal01} failed to detect the feature around 4500-4700\,\AA, (there are UVES and FORS1 spectra in the archive to check this). The spectrum by \citet{Giaetal12} (available through the MDWD) is clearly featureless, confirming the hypothesis by \citet{Schetal01} that the feature seen by \citet{Reietal96} was spurious. For these reasons, this star is considered to be non magnetic. 
Integrated over their optical spectra, \citet{Schetal01} measured once $-0.05$\,\% and once  $+0.14$\,\%, from which we obtain an upper limit of $\bz \la 2$\,MG.

\subsubsection{{\bf WD\,0004$+$122} = {\bf NLTT\,287} -- 100\,MG}%
Identified as a DC star by \citet{KawVen06}, WD\,0004$+$122 was found to exhibit a strong signal of circular polarisation by \citet{BagLan20}, who estimated $\bz \sim +30$\,MG, suggesting that the \bs\ value may be between 60 and 200\,MG. (In Table~\ref{Tab_Stars} we adopt $\bs=100$\,kG.) The star was observed only once and should be re-observed to check for variability.

\subsubsection{WD\,0121$-$429B = LP 991-16B}\label{Sect_DC0121}
This star is a member of a uDD system composed of a magnetic DA and a DC WD. Since the system was never observed in polarimetric mode, we cannot reach any conclusion about the magnetic nature of the DC component, and is not be considered in our statistical analysis. See Sect.~\ref{Sect_DA0121} for more details about the system and the DA component.

\subsubsection{WD\,0123$-$262 = EGGR\,307}%
The star was never observed in spectropolarimetric mode before our FORS2 measurement that shows $\vmax \la 0.03$\,\%, for $\abz \la 0.5$\,MG (this work).

\subsubsection{WD\,0233$-$242B = NLTT 8435B (uDD)}\label{Sect_DC0233} %
This star is a  member of a uDD system composed of a magnetic DA and a DC WD. No circular polarisation is detected in the continuum away from H$\alpha$ (which is polarised by the magnetic DA companion), with $\vert V/I \vert \la 0.05$\%. We conclude that the DC component has $\abz \la\ 1.5$\,MG. See Sect.~\ref{Sect_DA0233} for more details about the system and the DA component.

\subsubsection{WD\,0423$+$120 = G\,83-10 (uDD?)}%
 \citet{Holbetal08} considers the star as too bright for its parallax and classify it as an unresolved DD. \citet{Giaetal12} discussed WD\,0423$+$120 as an object without H$\alpha$ but their spectral energy distribution is better fit by a H-rich model with $\teff \simeq 6100$\,K. They do not mention DD models or assign a very low mass to WD, but suggest that a large magnetic field may be present that washes out the H$\alpha$ profile. They note, however, that \citet{Putney97} did not detect any polarisation signal from the star. \citet{Holletal18} give $\log g = 8.20$ and \citet{Genetal19} give the WD mass as $0.76 M_\odot$ (H) or $0.70 M_\odot$ (He).  We note the star as a possible uDD system, but we consider it as a single star. 

The star was observed by \citet{LanAng71}, who measured $V/I=0.03 \pm 0.11$\,\% and \citet{Putney97}, who measured  $V/I = 0.038 \pm 0.034$\,\% in the region 3800-4600\,\AA, and $V/I = -0.060 \pm 0.020$\,\% in the region 6000-8000\,\AA. \citet{Putney97} considers the star as having a magnetic field $\la 20$\,MG, but if we transform the measurement by \citet{LanAng71} and the average of \citet{Putney97} measurements in the two bands into $\bz = 0.45 \pm 1.65$\,MG and $\bz = -0.33 \pm 0.45$\,MG, respectively, we obtain $\abz \la 0.6$\,MG. If this is a uDD system, a polarisation signal coming from one of the stars would be diluted by a factor of roughly two; therefore we double our upper limit and set $\abz \la 1.2$\,MG. Perhaps it would be worth while to observe the star again in spectropolarimetric mode with higher precision.

\subsubsection{WD\,0426$+$588 = G\,175-34B (VB: dM at 9\arcsec)}%
This bright DC WD was observed by \citet{LanAng71}, who measured $V/I=-0.12 \pm 0.13$\,\% and by \citet{Angetal81}, who measured  $V/I=0.006\pm0.013$\,G. The star was observed in spectropolarimetric mode with ISIS by \citet{BagLan18}; in the $B$ filter we report $V/I = 0.00 \pm 0.01$\,\%, and in the red $0.01 \pm 0.01$\,\%. Altogether, these measurements set the upper limit for \abz\ to 0.17\,MG. 

\subsubsection{{\bf WD\,0708$-$670} = {\bf SCR\,J0708$-$6706} -- At least 200\,MG}%
Identified as a DC WD by \citet{Subetal08}, it was discovered to be magnetic by \citet{BagLan20}, who estimated the star to have $\bz \sim 200$\,MG.

\subsubsection{WD\,0743$-$336 = GJ 288 B (VM: F9 at 870\arcsec)}%
Identified as a cool WD by \citet{Kunetal84}. Observed in polarimetric mode only once (with FORS2, this paper, using grism 300V), the star shows no evidence for circular polarisation. We set 1.5\,MG as the upper limits of the star's longitudinal field. The star is member of a triple system.  We note that the star has $G=15.3$, not $V=16.7$ as tabulated in Simbad and reported by \citet{Kunetal84}.

\subsubsection{WD\,0743$+$073.2 = GJ\,1102\,A} %
Observed only once in polarimetric mode (with FORS2, this paper, using grism 300V), no signal was detected.  We set the upper limit of \abz\ to 1.5\,MG. This is a member of a resolved DD system with WD\,0743$+$073.1 at 16\arcsec\ (see below).

\subsubsection{WD\,0743$+$073.1 = GJ\,1102\,B}%
Observed only once in polarimetric mode with FORS2 (this paper), using grism 300V. No evidence of circular polarisation.  We set the upper limit of \abz\  to 0.45\,MG. This is a member of a resolved DD system with WD\,0743$+$073.2 at 16$"$ (see above).
 
\subsubsection{WD\,0810$+$489 = G\,111-64}%
Listed as probable WD by \citet{Luyten70}, this star was spectroscopically confirmed as a DC WD by \citet{Kawetal04}. Observed with ISIS (this paper) with bad seeing conditions, this WD shows $V/I \simeq 0.25$\,\% in the blue, but this signal may well be spurious. We set the upper limit for \abz\ to 3\,MG.

\subsubsection{{\bf WD\,0912$+$536} = {\bf G\,195-19} -- 100\,MG, variable}%
This is the second WD ever discovered magnetic \citep{AngLan71-Second}, and the first one to be discovered variable \citet{AngLan71-Periodic}. An improved ephemeris was provided by \citet{Angetal72-eph}. With circular polarisation reaching 2\,\% we infer $\bz \sim 30$\,MG and $\bs \sim 100$\,MG. There has been almost no further interest in this star since the 1970s. 

\subsubsection{WD\,0959$+$149 = G\,42-33}%
Observed by \citet{LanAng71} in broadband ($V=0.02 \pm 0.11$\,\%), and by \citet{Putney97} ($V=0.080 \pm 0.033$\,\%). Neither detect circular polarisation. We set an upper limit of 2\,MG for the mean longitudinal field.

\subsubsection{WD\,1033$+$714 = LP\,37-186}%
Observed by \citet{LieSto80}, who measured $V/I=-0.039 \pm 0.16$\,\%.  For \abz\ we set an upper limit of 1.5\,MG, but the star should be re-observed in circular spectropolarimetric mode.

\subsubsection{WD\,1055$-$072 = LAWD\,34}%
Observed with ISIS (this paper) with gratings R600B and R1200R in less than optimal sky conditions, we did not detect a convincing signal of circular polarisation, with upper limits of 0.2\,\% in the blue and 0.4\,\% in the red (corresponding to $\abz \la 6$\,MG). 

\subsubsection{WD\,1116$-$470 = SCR\,J1118$-$4721 -- suspected magnetic}%
Discovered as a DC WD by \citet{Subetal08}, we have observed it twice with FORS2 (this work) with grism 300V. Both observations show a similar signal of circular polarisation at $-0.2$\,\%, close to the instrumental detection limit. Because of the consistency between the two observations, we suspect that this small signal is real. In this case the star would have a longitudinal field of about 3\,MG.

\subsubsection{WD\,1132$-$325 = HD 100623B  (VB: K0 at 16\arcsec)}%
A Sirius like system with a K0 at 16$"$, firmly classified as DC by \citet{Holbetal16}. We observed it with FORS2 (this work) with grism 300V. The result was a non detection, and we set the upper limit to \abz\ as 1.5\,MG. 

\subsubsection{WD\,1145$-$747 = SSSPM\,J1148-7458 (uDD?)}%
This is the coolest and faintest WD of our sample, discovered to be a DC WD by \citet{Schetal04}, who estimated its distance to $36 \pm 5$\,pc, but according to the Gaia parallax ($\pi 50.10 \pm 0.08$\,mas) it is just within the local 20\,pc volume.  Assuming a H-rich atmosphere, \citet{Holletal18} give \teff = 3711\,K, $\log g$ = 7.67;  \citet{Genetal19} gives 3761\,K 7.70, and $M = 0.42 m_\odot$. Since this mass is well below the lower limit for production by single star evolution, this result suggests that this object may be an uDD. However, assuming an He-rich atmosphere leads to the parameters  4109\,K, 7.81, and $M = 0.47 M_\odot$ \citep{Genetal19}. If this is the case, this object could be produced by single-star evolution. Because of the ambiguity of the mass of WD\,1145--747, we classify it as uDD? 

We observed the object with FORS2 (this work) with grism 300V and (quasi-simultaneously) with grism 600B, with no detection of circular polarisation. We set $\vert \bz \vert \la 1.5$\,MG, but if the system is shown to be an uDD, this upper limit should be multiplied by two.

\subsubsection{WD\,1310$-$472 =  GJ\,3770}%
One of the faintest WDs in the local 20\,pc volume. We observed it with FORS2+300V (this work), detecting $V/I \simeq 0.1$\,\%. This signal is consistent with instrumental polarisation, but the star should be re-observed. We set $\abz \la 1.5$\,MG . 

\subsubsection{WD\,1338$+$052 = LSPM\,J1341$+$0500}%
Identified as a new DC WD of the local 20\,pc volume by \citet{Sayetal12}, this star was observed in polarimetric mode only once with FORS2+300V (this work), which resulted in the upper limit $\abz \la 1.5$\,MG.

\subsubsection{WD\,1444$-$174 = LP\,801-9}%
Observed by \citet{LieSto80} who measured $V/I = -0.233 \pm 0.09$\,\% (but note that in their paper, the Vilanova identification was erroneously given as WD\,1414$-$175). \citet{Putney97} detected circular polarisation in the blue ($0.36 \pm 0.002$\,\%), but not in the red ($V=0.044 \pm 0.042$\,\%); her conclusion was that the detection in the blue arm was spurious, and data were consistent with null polarisation. This star should be re-observed. We set the upper limit for \abz\ to 5\,MG.

\subsubsection{WD\,1917$+$386 = G\,125-3}%
Observed once by \citet{Angetal81} ($V/I=-0.02 \pm 0.03$)\,\% and twice by \citet{Putney97}. In one of her measurements \citet{Putney97} detected a small amount of polarisation in the continuum ($V/I=-0.49 \pm 0.03$\,\% in the blue and $+0.062 \pm 0.001$\,\% in the red), while a second observation was consistent with zero ($V/I = -0.07 \pm 0.02$\,\% in the blue and $V/I = -0.05 \pm 0.03$ in the red). The star was not reported as magnetic. If we assume 0.1\,\% as the upper limit for $\vert V/I \vert$, we obtain $\abz \la 1.5$\,MG.

\subsubsection{WD\,2002$-$110 = EGGR 498}%
Observed quasi-simultaneously with FORS2 with grisms 300V and 600B (this work). We obtained $\vmax \la 0.1$\,\% and $\vmax \la 0.2$\,\%, respectively, leading to $\abz \la 3$\,MG.

\subsubsection{WD\,2008$-$600 = SCR\,J2012-5956}%
The star was spectroscopically confirmed to be a DC WD by \citet{Subetal07}. Observed using FORS2 with grism 300V (this work) for an upper limit of $\abz \la 1.5$\,MG.

\subsubsection{WD\,2017$-$306 = EC\,20173$-$3036}%
Discovered to be a WD by \citet{Holletal18}, who assumed that it is H-rich. One measurement using FORS2 (this work) with grism 1200\,B shows no features and $\vert V/I \vert \la 0.02$\,\%, for $\abz \la 0.3$\,MG. Since it has $\teff \simeq 10500$, we conclude that the star is clearly an He-dominated rather than H-dominated DC WD.

\subsubsection{WD\,2048$+$263B  = G\,187-8B (uDD)}\label{Sect_DC2048}%
This is the probable DC companion of WD\,2048$+$263A, with which we believe it forms a uDD system. See Sect.~\ref{Sect_DA2048} for more details.
From the broad-band polarisation measurement by \citet{Angetal81} ($V/I = 0.04 \pm 0.05$\,\%),  assuming a dilution by a factor of two, we estimate $\abz \la 1.5$\,MG

\subsubsection{{\bf WD\,2049$-$253} = {\bf UCAC4 325-215293} -- 20\,MG}%
Identified as a WD by \citet{Holletal18}. Observed with FORS2 with grism 1200B by \citet{BagLan20}, who found that the star exhibits  continuum circular polarisation of $\sim 0.5$\,\%. Since this discovery observation was obtained during dark time, when the background was essentially unpolarised, we believe that the star is very probably magnetic. However, this should be confirmed by further observations.

\subsubsection{WD\,2054$-$050 = VB\,11 (VB + uDD?)}%
\citet{Giaetal12} classify this WD as DC, and provide a mass estimate of $0.37 M_\odot$. Both they and \citet{Tooetal17} consider it likely that the star is actually an uDD. However, \citet{Holletal18} and \citet{Genetal19} agree on $\log g$ values near 7.87, implying  $M = 0.50 M_\odot$, which could result from single star evolution, and \citet{Holletal18} does not even mention it as a possible uDD. We consider that this object is not a strong candidate uDD, and we count it as a single WD in the statistics. It is also a VB with an M3 companion at 15$"$.

\citet{LieSto80} measured $V/I=0.23 \pm 0.13$\,\%, \citet{Angetal81} $V/I = -0.15 \pm 0.12$\,\%. This allows us to set an upper limit to \abz\ of 2\,MG.

\subsubsection{WD\,2226$-$754 = SCR J2230$-$7513 (DD)}%
Reported as visual (resolved VB) DD by \citet{Schetal02} (see WD\,2226$-$755). Observed with FORS2+300V (this work), no detection, which sets the upper limit of \abz\ to 0.75\,MG.

\subsubsection{WD\,2226$-$755 = SCR J2230$-$7515 (DD)}%
Reported as visual DD by \citet{Schetal02} (see WD\,2226$-$754 above). Observed with FORS2+300V (this work), no detection, which gives us an upper limit to \abz\ of 1.2\,MG.

\subsection{DZ and DZA stars}\label{Sect_DZ}
\subsubsection{WD\,0046$+$051 = vMA\,2}%
This is the prototype of metal-enriched WDs -- its strong Ca\,{\sc ii} H+K lines were observed more than one century ago by \citet{van17}. With ISIS, \citet{BagLan18} obtained a $2.4\,\sigma$ detection (with $\sigma \sim 1$\,kG), but a new more accurate FORS2 measurement presented in this work, obtained with ($-0.5 \pm 0.3$\,kG), suggests that the star is not magnetic. UVES archive data obtained with $R\sim 20\,000$ show no hint of Zeeman splitting in the Ca\,{\sc ii} H+K lines, which allows us to set the upper limit of 200\,kG for \bs. 
The star is very bright, and perhaps one or two additional high-precision field measurements could be useful to check if the star has a very weak field.
JDL

\subsubsection{WD\,0552$-$106 = UCAC4\,398$-$010797}%
Newly identified as a WD of the local 20\,pc volume by \citet{Holletal18}, our two ISIS spectra (this work) show weak Ca\,II lines H\&K and no Balmer lines, consistent with the classification as a DZ star by \citet{Treetal20}. We obtained no field detection, although with low precision  ($\sigma \simeq 40$\,kG). From intensity spectra we set an upper limit of 300\,kG for \bs.

\subsubsection{WD\,0552$-$041 = EGGR 45}%
According to  \citet{Giaetal12} and \citet{Subetal17}, WD\,0552$-$041 is the only known DZ star in the local 20\,pc volume with a H-rich atmosphere, however \citet{Couetal19} prefers to describe it as He-rich, and suggest that there is no H at all.  We observed it with ISIS and FORS2+1200B (this work), and we do not see H$\alpha$ in our ISIS spectrum, but at $\teff \approx 4500$, H$\alpha$ would not be expected, and selecting the dominant element in the atmosphere must be done using the energy distribution. We found no magnetic field (our most accurate measurement is $\bz = 5.9 \pm 4.6$\,kG). From ISIS intensity spectra we set the upper limit of \bs\ to 300\,kG.

\subsubsection{WD\,0738$-$172 = L745-46A (VB: M6.5 at 21\arcsec)}%
The stellar spectrum shows no sign of higher Balmer lines, in spite of $\teff = 7600$\,K, but H$\alpha$ is clearly visible with very shallow broad wings and deep line core. \citet{BagLan19b} considered this star as a candidate (weakly) magnetic DZ star on the basis of FORS1 measurements around H$\alpha$ by \citet{Frietal04}, who found that the star is magnetic, although at a low significance level ($\bz = -6.9 \pm 2.1$\,kG). In the FORS1 archive we have retrieved the original polarisation spectra obtained on 2000-12-03 and on 2000-12-19 with grism 1200R, and using  H$\alpha$ we measure $\bz=+5.1 \pm 1.3$\,kG and $\bz=-3.8\pm 2.2$\,kG, respectively \citep[note that we obtained the opposite sign for \bz\ than that found by][]{Frietal04}. None of our new ISIS \citep[][and this work]{BagLan19b}, ESPaDOnS and FORS2+1200B observations confirms the presence of a magnetic field. Because of the very sharp core of its H$\alpha$ (FWHM $\la 1$\,\AA\ wide), the upper limit on \bs\ from ESPaDOnS data is as small as 20\,kG. The star should still be considered as a candidate magnetic star, but at present we treat it as non-magnetic. 

\subsubsection{{\bf WD\,0816$-$310} = {\bf SCR J0818-3110}  -- 90\,kG, variable} %
Discovered as a magnetic WD by \citet{BagLan19b} who obtained one FORS2 and one ISIS measurement. The field is too weak to split the star's metal lines observed at low resolution, but using X-Shooter spectra with resolution $\sim 10000$, \citet{Kawetal21} measured $\bs 92 \pm 1$\,kG. 

\subsubsection{WD\,0840$-$136 = LP\,726-1}%
The star was discovered as a DZ WDs in the local 20\,pc volume by \citet{Subetal07}. Two FORS2+1200B observations presented in this work \citep[but see also][]{BagLan19b} did not lead to any field detection with uncertainties of $\sim 2.5$\,kG. From FORS2 Stokes $I$ spectra we estimate $\bs \la 200$\,kG.

\subsubsection{{\bf WD\,1009$-$184} = {\bf WT 1759} -- 150\,kG, variable (VB: K7 at 400\arcsec)}%
The star was classified as DZ by \citet{Subetal09}, and was discovered to be a  MWD by \citet{BagLan19b}, who published one measurement obtained with FORS2 and one obtained with ISIS, showing that the star is variable.  (The most extreme value was measured with FORS2, $\bz = -47 \pm 1.5$\,kG.) As is the case for WD\,0816$-$310, the field is too weak to split the star's metal lines, hence $\bs \la 200$\,kG. 

\subsubsection{{\bf WD\,1532$+$129} = {\bf G\,137-24} -- 50\,kG, variable}%
Originally classified as DZ WD by \citet{Kawetal04}, the star was discovered to be a MWD by \citet{BagLan19b}, who published two FORS2+1200B measurements and one ISIS measurement. The star is variable, with \bz\ values from  FORS2 measurements of $ -21 \pm 1$\,kG and $-4 \pm 1$\,kG, while \bs\ is not strong enough to split spectral lines, leading to  $\bs \la 300$\,kG.  
In revising our data we have realised that the uncertainty of $\simeq 2$\,kG in our ISIS measurement reported by \citet{BagLan19b} was actually seriously underestimated, and should have been $\sigma \simeq 5.5$\,kG.

\subsubsection{WD\,1626$+$368 = Ross 640}%
The star is a DZA. An early broad-band circular polarisation measurement was reported by \citet{Angetal81} of  $V/I = -0.16 \pm 0.11$\,\%.  In this work we present one measurement with ESPaDOnS and two measurements with ISIS. none of which show evidence of a magnetic field (with typical 1--2\,kG uncertainty). ESPaDOnS data (this work) gives us an upper limit for \bs\ of 50\,kG.

\subsubsection{WD\,1705$+$030 = EGGR 494 }%
The only \bz\ measurement available, obtained with ISIS (this work) reveals no field, with a $\simeq 8$\,kG uncertainty. The width of the core of the Ca\,{\sc ii} H line at 3968\,\AA\ gives us a much weaker limit of $\bs \la 300$\,kG. 

\subsubsection{WD\,1743$-$545 = PM\,J17476$-$5436}%
This star was identified as a DC WD by \citet{Subetal17}, but the star shows very weak Ca {\sc ii} H\&K lines, and an extremely weak H$\alpha$. Observed using FORS2 with grism 300V (this work), we obtain an upper limit $\vmax \la 0.2$\,\% that corresponds to $\abz \la 3$\,MG; from the 20\,\AA\ full width at half maximum of the weak H-line we deduce $\bs \la 1$\,MG.

\subsubsection{{\bf WD\,2138$-$332} = {\bf L 570-26} -- 50\,kG variable}%
Magnetic according to two FORS2 measurements and one ISIS measurements published by \citet{BagLan18} and \citet{BagLan19b}. A new ISIS and a new ESPaDOnS field measurement are presented in this work. The longitudinal field of this star is certainly variable and the star should be monitored for modelling.

Very close examination of our single ESPaDOnS spectrum and of our 1200R ISIS spectrum strongly suggests that a very weak H$\alpha$ line is present in both $I$ spectra. This feature is so weak (less than 100 m\AA) that its reality is not certain. It should be confirmed with a spectrum at higher \snr. In any case, if real, this feature indicates that the He-dominated atmosphere also contains a very small amount of H. 

\subsubsection{WD\,2251$-$070 = EGGR 453}%
This is the second coolest WD of the local 20\,pc volume. We have only one FORS2+1200B observation (this work), which shows strong Ca\,{\sc i} (4226\,\AA) and Ca\,{\sc ii} (3933, 3968\,\AA) resonance lines. The polarisation spectrum of these lines shows no magnetic field, but with large uncertainty ($\sim 5$\,kG).

\subsection{DQ stars}\label{Sect_DQs}
\subsubsection{WD\,0038$-$226 = LHS\,1126}%
Peculiar DQ. \citet{LieSto80} measured  $V/I=-0.15 \pm 0.20$\,\% with no filter (integrating over the interval $3200-8600$\,\AA).  \citet{Angetal81} obtained three BBCP measurements, including one with a marginal detection ($V/I=0.120 \pm 0.041$). Spectropolarimetric data by \citet{Schetal95} show that $V/I=-0.04 \pm 0.03$\,\%. \citet{Schetal95} estimated $\bs \la 3$\,MG. 

\subsubsection{WD\,0115$+$159 = GJ\,1037}%
Cool DQ WD. \citet{AngLan70Further} obtained one measurement ($V/I = -0.12 \pm 0.1$\,\%) and \citet{Angetal81} reported a best measurement of $V/I = 0.003 \pm 0.019$\,\%. \citet{Voretal13} published circular spectropolarimetry that they obtained in 2008 with FORS1 using grism 600B. These new data show no signal of non-zero circular polarisation at about the 0.1--0.2\,\% level, and we assume $\abz \la 0.5$\,MG.

\subsubsection{WD\,0208$-$510 = HD\,13445B  (VB: K0 at 1.9\arcsec)}%
The star has a K0 companion at 1.9\arcsec. Optical STIS spectroscopy by \citet{Faretal13} shows that the star is a helium-rich white dwarf with C2 absorption bands and $\teff = 8180$\,K, thus making the binary system rather similar to Procyon. Its spectral features in intensity are too broad to estimate any useful upper limit to \bs, and we cannot observe it in polarimetric mode with FORS2 because of its proximity to a much brighter object. For our statistical analysis, the star is considered as not observed, hence it is ignored.

\subsubsection{WD\,0341$+$182 = GJ\,151}%
\citet{LanAng71} measured $V/I=0.06 \pm 0.10$\,\%, \citep[the same observations were later reported by][]{Angetal81}, and by \citet{Voretal13} with FORS1 using grism 600B, with no field detection, for $\abz \la 2$\,MG.

\subsubsection{WD\,0435$-$088 = GJ\,3306}%
\citet{Angetal81} obtained two measurements: $V/I = -0.038 \pm 0.057$\,\% and $V/I = -0.017 \pm 0.021$\,\%. Observed by \citet{Voretal13} with FORS2 using grism 600B with no detection. We assume an upper limit of 0.5\,MG for the magnetic field \bs.

\subsubsection{{\bf WD\,0548$-$001} = {\bf GJ\,1086} -- 5\,MG, constant}%
Magnetic field discovered by \citet{LanAng71}, and re-observed and modelled by \citet{AngLan74} who deduced the value of \bz\ to be about 3.6\,MG from the polarisation signature of the CH G-band.  Newer, higher-resolution FORS1 polarised spectra were obtained by \citet{Berdetal07} and \citet{Voretal10} from which $\bz \sim 2.5$\,MG is deduced.
The signal of circular polarisation seems constant between 2005 and 2008, and overall there is no evidence of change compared with the earlier broadband polarisation measurements by \citet{AngLan74}.

\subsubsection{WD\,0736$+$053 = Procyon B (VB: F5 at 5.3\arcsec)}%
\citet{Proetal02} observed the star with STIS of the HST, revealing a DQZ spectrum with carbon Swan bands and Mg\,{\sc i} and Mg\,{\sc ii}) UV lines. Their Fig.~9 shows that Mg\,{\sc i} and Mg\,{\sc ii} lines around 2800\,\AA do not appear split by Zeeman effect, setting a limit to 300\,kG for \bs. Dues to its proximity ($\sim 5\arcsec$) to a much brighter F-type companion, the star has never been observed in polarimetric mode.

\subsubsection{WD\,0806$-$661 = L97-3}%
Our FORS2 measurements with grism 300V (this work) do not show any polarisation signal, with a signal smaller than 0.02\,\%. Our optical spectrum is featureless, but the star is classified as DQ on the basis of IUE spectra \citet{Beretal97,Subetal09}. The star was observed twice by \citet{Angetal81} in BBCP, for $0.060 \pm 0.058$\,\% and $0.031 \pm 0.055$\,\%. There is no evidence that the star is magnetic, and we  estimate $\abz \la 0.3$\,MG.

\subsubsection{{\bf WD\,1008$+$290} =  {\bf LHS\,2229} -- 100\,MG}%
Cool peculiar DQ star discovered to be magnetic by \citet{Schetal99}, who from a $V/I \sim 10$\,\% estimate a field of the order of $\abz \sim 100$\,MG. We have not found other polarimetric observations since its discovery, so the star should be re-observed and checked for variability and possibly monitored.

\subsubsection{{\bf WD\,1036$-$204} = {\bf LP\,790-29} -- 200\,MG, constant}%
Magnetic field discovered via spectropolarimetry by \citet{Lieetal78} and observed also by \citet{Schetal95} \citep[see also][for further analysis]{Schetal99}. \citet{BeuRei02} have monitored the star with EFOSC in spectropolarimetric mode to search for short-term variability, without finding any significant variations.
From the measured signal of circular polarisation of $\sim 10$\,\% we estimate a longitudinal field of the order of 100\,MG. 

\subsubsection{WD\,1043$-$188 = BD$-$18\,3019B (VB: M3 at 8\arcsec)}%
Visual binary with an M3 at 8\arcsec. Observed by us using FORS2 with grism 600B (this work): we did not detect any signal, and we set the upper limit for \abz\ to 0.3\,MG.

\subsubsection{WD\,1142$-$645 = GJ\,440}%
Observed only by \citet{Angetal81}, who did not detect any signal of circular polarimetry ($V/I=0.010 \pm 0.026$\,\% and $V/I=-0.014 \pm 0.016$\,\%). From these measurements we set an upper limit of $\vert \bz \vert \la 0.5$\,MG. The star is quite bright ($V=11.5$) but there are no spectropolarimetric measurements available in the literature, and not having observed it with FORS was an oversight on our part.

\subsubsection{WD\,1633$+$572 = EGGR 258 (MS: M-type eclipsing binary at 26\arcsec)}%
Observed twice by \citet{Angetal81} ($V/I = -0.08 \pm 0.14$\,\% and $-0.09 \pm 0.12$\,\%), and by \citet{Schetal95} ($V/I \la 0.03$\,\%). From these measurements we deduce $\abz \la 0.3$\,MG. We note that this star is also called WD\,1633$+$571/2/3, as well as G\,226-17 and G\,225-58. 

\subsubsection{{\bf WD\,1748$+$708} = {\bf G\,240-72} -- 300\,MG, very slowly variable}%
Intrinsic broadband circular and linear polarisation of this MWD were discovered by \citet{Angetal74b}. It was later observed in linear polarisation by \citet{West89} and by \citet{BerPii99}. \citet{BerPii99} found evidence of a variation of linear polarisation in a time scale of 20 years. We have observed the star with ISIS both in circular (twice) and in linear polarisation; our observations will be presented in a later paper. 

The intensity spectrum of this star is quite unique, with a huge “sag” in energy distribution between 4300 and 6700\,\AA. \citet{Beretal97} relate this feature to a group of C$_2$H stars which have molecular bands in the same general region and similar \teff\ (see their Fig.~30), and sugggest that the feature observed in  WD\,1748$+$708 is C$_2$H band distorted by the presence of a strong magnetic field, so they classify the star as a DQ. \citet{Kowalski10} suggests that the same feature is actually an extremely pressure shifted Swan band, implicitly agreeing on the star's classification as DQ. Nevertheless it is still somewhat uncertain whether this classification is correct, and if correct, exactly what it means.

\subsubsection{WD\,1917-077 = LAWD\,74 (VB: M6 at 27.3\arcsec)}  %
This star is classified as DBQA. It has an extremely weak H$\alpha$ \citep[see Fig.~8 of][]{Giaetal12} but $\teff \simeq 11000$\,K, which leads to the conclusion that the atmosphere is probably He dominated. The UV shows a single line of C\,{\sc i}. There is no sign of the C$_2$ Swan bands in the optical \citet{Oswetal91}. 
This DBQA star was observed by \citet{Angetal81}, who obtained $V/I = 0.05 \pm 0.05$\,\%, and with FORS2 by \citet{BagLan18}, who found $\vert V/I \vert \la 0.03$\,\%. We have also re-observed the star with ISIS both in the blue and in the red, but because of a instrument setup error, we cannot measure and subtract the sky contribution (hence there is no corresponding entry in Table~\ref{Table_Log}). However, from ISIS spectroscopy of the broad, shallow H$\alpha$ we can still set the upper limit for \bs\ to 1\,MG, while from circular polarimetry we deduce $\abz \la 0.5$\,MG. In our FORS2 spectrum we do not see evidence for He lines.

\subsubsection{WD\,2140$+$207 = EGGR\,148}%
Observed in BBCP mode by \citet{AngLan70Further} ($V/I= -0.01 \pm 0.04$\,\%), then observed again in BBCP two times by \citet{Angetal81} ($V/I=0.025 \pm 0.017$\,\%, $-0.032 \pm 0.077$\,\%, and once in spectropolarimetric mode by \citet{BagLan18} ($V/I \la 0.02$\,\%). No lines are visible in the optical spectrum. The DQ classification comes from \citet{Koeetal98}, who analysed IUE spectra of luminous WDs.  The non detection of circular polarisation allows us to place the $1\,\sigma$ uncertainty for field detection at 200\,kG.

\subsubsection{{\bf WD\,2153$-$512} = {\bf WG\,39}  -- 1.3\,MG (VB: M2 at 28\arcsec)}%
Magnetic field discovered by \citet{Voretal10}, who derived the $\bz = 1.3 \pm 0.5$\,MG value from modelling their polarisation observations. These data do not constrain \bs. The star is a member of a wide binary system \citep[see][]{LanBag20}.vThe star should be checked for variability in spectropolarimetric mode. Note that this star has also been named WD\,2154$-$511 and WD\,2154$-$512. 

\subsection{DX}\label{Sect_DX}
\subsubsection{WD\,0211$-$340}%

This star is too close to a background star, and could not be characterised.

\bsp 
\label{lastpage}
\end{document}